\documentclass[aps,prd,showpacs]{revtex4-1}

\usepackage{amsmath,amssymb,amsfonts,amsthm} 
\usepackage{color}
\usepackage{graphicx}
\usepackage{pstricks,pstricks-add}
\usepackage{pst-plot}
\usepackage[hang,nooneline]{subfigure}
\usepackage{slashbox}

\definecolor{dark-green}{rgb}{0,0.7,0}
\definecolor{dark-blue}{rgb}{0,0.2,0.5}
\definecolor{med-blue}{rgb}{0,0.7,1}
\definecolor{mblue}{rgb}{0,0.2,1}
\definecolor{cnc}{rgb}{0.8,0,0}
\definecolor{light-red}{rgb}{1,0.8,0.8}
\definecolor{dark-yellow}{rgb}{1,0.8,0}
\definecolor{light-blue}{rgb}{0.8,0.9,1}
\definecolor{grey}{rgb}{0.211,0.211,0.211}
\definecolor{verylight-blue}{rgb}{0.93,0.95,1}
\definecolor{light-yellow}{rgb}{1,0.9,0.8}

\begin{document}

\title{Detection of cosmic superstrings by geodesic test particle motion}

\author{Betti Hartmann $^{(a)}$ }
\email{b.hartmann@jacobs-university.de}

\author{Claus L{\"a}mmerzahl $^{(b),(c)}$}
\email{laemmerzahl@zarm.uni-bremen.de}

\author{Parinya Sirimachan $^{(a)}$}
\email{p.sirimachan@jacobs-university.de}

\affiliation{
$(a)$ School of Engineering and Science, Jacobs University Bremen, 28759 Bremen, Germany\\
$(b)$ ZARM, Universit\"at Bremen, Am Fallturm, 28359 Bremen, Germany\\
$(c)$ Institut f\"ur Physik, Universit\"at Oldenburg, 26111 Oldenburg, Germany}
\date\today

\begin{abstract}

(p,q)-strings are bound states of p F-strings and q D-strings and are predicted
to form at the end of brane inflation. As such these cosmic superstrings should be detectable 
in the universe.
In this paper we argue that they can be detected by the way that massive and massless test particles
move in the space-time of these cosmic superstrings, in particular we study solutions 
to the geodesic equation in the space-time of field theoretical (p,q)-strings. The geodesics can be classified according to the test particle's
energy, angular momentum and momentum in the direction of the string axis. 
We discuss how the change of the magnetic fluxes, the ratio between the symmetry breaking scale
and the Planck mass, the Higgs to
gauge boson mass ratios and the binding between the F- and D-strings, respectively,
influence the motion of the test particles. While massless test particles
can only move on escape orbits, a new feature as compared
to the infinitely thin string limit is the existence of bound orbits for massive test
particles. In particular, we observe that - in contrast to the space-time of a 
single Abelian-Higgs string - bound orbits for 
massive test particles in (p,q)-string space-times
are possible if the Higgs boson mass is larger than the gauge boson mass. 
We also compute the effect of the binding between the p- and the q-string on observables such as the 
light deflection and the perihelion shift. While light deflection can also be caused
by other matter distributions, the possibility of a negative perihelion shift seems
to be a feature of finite width cosmic strings that could lead to the unmistakable 
identification of such objects.
In Melvin space-times, which are asymptotically non-conical, 
massive test particles have to move on bound orbits, while massless test particles can only escape to
infinity if their angular momentum vanishes.
\end{abstract}

\maketitle

\section{Introduction}
Cosmic strings are topological defects that are predicted to have formed via the Kibble mechanism
\cite{kibble}
during one of the phase transitions in the early universe and in the field theoretical
description \cite{no}  can be considered to be an example of a topological soliton. Due to the fact that these
objects can be extremely heavy (up to $10^{12}$ kg/m) they were believed to be 
a possible source of the density 
perturbations that led to structure formation and the anisotropies in the 
cosmic microwave background (CMB) \cite{vs}. However, the detailed measurement of the 
CMB power spectrum
as obtained by COBE, BOOMERanG and WMAP showed
that cosmic strings cannot be the main source for these anisotropies.

In recent years cosmic strings gained renewed 
interest due to the possible connection to the fundamental entities of
String Theory  \cite{polchinski}. Brane inflation is a popular inflationary model that
can be embedded into String Theory and 
predicts the formation of cosmic string networks at the end of inflation \cite{braneinflation}.
E.g. in the framework of type IIB String Theory the inflaton field corresponds to the
distance between two Dirichlet branes with 3 spatial dimensions (D$3$-branes) and inflation ends
when these two branes collide and annihilate. The production of strings (and lower dimensional branes)
then results from the collision of these two branes.
Each of the original D3-branes has a U(1) gauge symmetry that gets broken when
the branes annihilate.
If the gauge combination is Higgsed, magnetic
flux tubes of this gauge field carrying Ramond-Ramond (R-R) charge are D-branes with one spatial dimension, 
so-called
D-strings. When the gauge combination is confined the field is condensated into 
electric flux tubes carrying Neveu Schwarz-Neveu Schwarz (NS-NS) charges and these objects are fundamental strings (F-strings) \cite{dvali_vilenkin}. 
D-strings and F-strings are so-called cosmic superstrings \cite{polchinski} 
which seem to be a generic prediction of supersymmetric 
hybrid inflation \cite{lyth} and grand unified based inflationary models \cite{jeannerot}. 
D- and F-strings, however, have different properties than the usual (solitonic) cosmic strings. 
The probability of intercommutation 
of solitonic strings is equal to one but less than one in the case of cosmic superstring.
Therefore, solitonic strings do not merge, while cosmic superstrings tend to form bound states. 
When p F-strings and 
q D-strings interact, they can merge and 
form bound states, so-called (p,q)-strings \cite{copeland_myers_polchinski} whose
properties have been investigated \cite{bulk}. 
Even though the origin of (p,q)-strings is type IIB string theory, 
their properties can be investigated in the framework of field theoretical models
\cite{saffin,rajantie,salmi,urrestilla}. 
The influence of gravity on field theoretical (p,q)-strings has been studied in 
\cite{hartmann_urrestilla}.

Since there are good reasons for cosmic superstrings to be a consequence of String Theory
it is very exciting to search for observational consequences of their existence. There has 
been considerable effort in 
numerically modeling cosmic string networks to obtain
CMB power and polarization spectra \cite{cmb}. Comparison with observations has
shown that cosmic strings might well contribute considerably to the
energy density of the universe.
There is another way to detect cosmic strings in the universe, namely through the motion of test bodies in
such string space-times. The test particle motion in different space-times containing
cosmic strings has been investigated in \cite{ag,gm,cb,Ozdemir2003,Ozdemir2004}, while the complete set of
orbits of test particles in the space-time of black hole pierced by an infinitely thin cosmic string has been given
for a Schwarzschild black hole in \cite{hhls1} and for a Kerr black hole in \cite{hhls2}.

In this paper we follow the latter approach and use the field theoretical model 
discussed in \cite{hartmann_urrestilla} to describe (p,q)-strings by two coupled
Abelian-Higgs models in curved space-time. For vanishing coupling between the two sectors,
the model corresponds to the Abelian-Higgs model coupled minimally to gravity. This model
has solutions describing strings with finite core width that 
have been investigated in \cite{clv,bl}. Geodesics in this space-time have been
studied recently \cite{hartmann_sirimachan} and can only be given numerically.
Here we would like to extend this investigation
to the field theoretical description of cosmic superstrings.

Our paper is organized as follows: in Section II, we discuss the field theoretical
model that possesses (p,q)-string solutions and we also work out the geodesic equation.
In Section III we discuss our numerical results, in particular we give examples
of orbits and demonstrate how the ratio between the symmetry breaking scale and the
Planck mass, the ratios between Higgs and gauge boson masses, the magnetic fluxes  and the binding
between the F- and D-string influence our results.
We conclude in Section IV.

\section{The Model}

\subsection{The space-time of a (p,q)-string}
The field theoretical model to describe gravitating (p,q)-strings 
reads \cite{hartmann_urrestilla}
\begin{eqnarray}
S&=&\int d^4x\sqrt{-g}\left(\frac{1}{16\pi G}R+\mathcal{L}_{\rm m}\right)  \ ,
\label{action}
\end{eqnarray}
where $R$ is the Ricci scalar and $G$ is Newton's constant. 
The matter Lagrangian $\mathcal{L}_{\rm m}$ is given by \cite{saffin}
\begin{eqnarray}
\mathcal{L}_{\rm m}&=&D_{\mu}\phi(D^{\mu}\phi)^*-
\frac{1}{4}F_{\mu\nu}F^{\mu\nu}+D_{\mu}\xi(D^{\mu}\xi)^*-\frac{1}{4}H_{\mu\nu}H^{\mu\nu}-u(\phi,\xi)
\end{eqnarray}
with the covariant derivatives  
$D_{\mu}\phi$ = $\nabla_{\mu}\phi$ - $i e_1 A_{\mu}\phi$, 
$D_{\mu}\xi$ = $\nabla_{\mu}\xi$ - $i e_2 B_{\mu}\xi$ of the two complex scalar 
fields (Higgs fields)
$\phi$ and $\xi$ and the field strength tensors 
$F_{\mu\nu}=\nabla_{\mu}A_{\nu}-\nabla_{\nu}A_{\mu}=\partial_{\mu}A_{\nu} - \partial_{\nu}A_{\mu}$, 
$H_{\mu\nu}=\nabla_{\mu}B_{\nu} - \nabla_{\nu}B_{\mu}= \partial_{\mu}B_{\nu} - \partial_{\nu}B_{\mu}$ 
of two U(1) gauge potential $A_{\mu}$, $B_{\nu}$ with coupling constants $e_1$ 
and $e_2$. $\nabla_{\mu}$ denotes the gravitational covariant derivative. 
Finally, the potential $V(\phi,\xi)$ reads:
\begin{eqnarray}
u(\phi,\xi)&=&\frac{\lambda_1}{4}(\phi\phi^*-\eta_1^{2})^2+\frac{\lambda_2}{4}(\xi\xi^*
-\eta_2^{2})^2-\lambda_3(\phi\phi^*-\eta_1^{2})(\xi\xi^*-\eta_2^{2})  \ ,
\label{Vpq}
\end{eqnarray}
where $\lambda_1$ and $\lambda_2$ are the self-couplings of the two scalar fields,
while $\lambda_3 > 0$ is the coupling between the two sectors. $\eta_1$ and $\eta_2$ are the
vacuum expectation values of the scalar fields.

In order for both U(1) symmetries to spontaneously break which then leads to the formation
of strings we have to require that the (absolute) minimum of the potential (\ref{Vpq}) 
is at non-vanishing
values of $\vert\phi\vert$ and $\vert\xi\vert$. This leads to the requirement \cite{saffin}
\begin{eqnarray}
\lambda_1\lambda_2 &>&4\lambda_3^2   \ .
\end{eqnarray}

The most general static cylindrically symmetric line element invariant under boosts along the $z$-direction is
\begin{eqnarray}
\label{metric}
ds^2=N(\rho)^2dt^2-d\rho^2-L(\rho)^2d\varphi^2-N(\rho)^2dz^2  \ .\label{cysymmetric}
\end{eqnarray}
For the matter and gauge fields, we apply the Ansatz \cite{no}
\begin{eqnarray}
\phi(\rho,\varphi)&=&\eta_1h(\rho)e^{in\varphi}\quad,\quad A_{\mu}dx^{\mu}=\frac{1}{e_1}(n-P(\rho))d\varphi\\
\xi(\rho,\varphi)&=&\eta_2f(\rho)e^{im\varphi}\quad,\quad B_{\mu}dx^{\mu}=\frac{1}{e_2}(m-R(\rho))d\varphi  \ ,
\end{eqnarray}
where $n$ and $m$ are integers indexing the vorticity of the two Higgs fields around the $z$-axis
and correspond to the degree of the map from $S^1\rightarrow S^1$, where the homotopy group
is $\pi_1(S^1)=\mathbb{Z}$. In our field theoretical model of (p,q)-strings the p corresponds
to the winding $n$ and the q to the winding $m$.

We can then do the following rescaling
\begin{equation}
\label{rescaling}
\rho\rightarrow \frac{\rho}{e_1\eta_1} \ \ \ , \ \ \ L\rightarrow \frac{L}{e_1\eta_1}\ 
\end{equation}
such that the total Lagrangian only depends on the following dimensionless coupling constants
\begin{eqnarray}
\gamma=8\pi G\eta_1^2\quad,\quad g= \frac{e_2}{e_1}\quad,
\quad q=\frac{\eta_2}{\eta_1}\quad,\quad\beta_i=\frac{\lambda_i}{e_1^2}  \ \ , \ \ \label{pqparameters}
\end{eqnarray}
where $i = 1, 2, 3$. $\gamma$ is proportional to the ratio between the Planck mass $M_{\rm Pl}=G^{-1/2}$
and the symmetry breaking scale $\eta_1$. Moreover, $\sqrt{\beta_1}$ is proportional to the ratio
between the Higgs mass $M_{\rm H,1}=\sqrt{\lambda_1} \eta_1$ and the corresponding
gauge boson mass $M_{\rm W,1}=\sqrt{2}e_1\eta_1$, while $\sqrt{\beta_2}/g$ is proportional
to the ratio
between the Higgs mass $M_{\rm H,2}=\sqrt{\lambda_2} \eta_2$ and the corresponding
gauge boson mass $M_{\rm W,2}=\sqrt{2}e_2\eta_2$. Each of the strings
possesses a scalar core with width $\rho_{{\rm H},i}\sim  M_{{\rm H},i}^{-1}$ and
a gauge field core with width $\rho_{{\rm W},i}\sim  M_{{\rm W},i}^{-1}$, $i=1,2$. 
Note that with the rescaling (\ref{rescaling}) the width of the gauge field cores is $\rho_{{\rm W},1}\sim 1/\sqrt{2}$, 
$\rho_{{\rm W},2}\sim 1/(gq\sqrt{2})$ while
the widths of the scalar cores is given by $\rho_{{\rm H},i}=1/\sqrt{\beta_i}$, $i=1,2$.

The variation of the action (\ref{action}) with respect to the matter fields leads to the 
following equations \cite{hartmann_urrestilla}
\begin{eqnarray}
\frac{(N^2Lh')'}{N^2L}&=&\frac{P^2h}{L^2}+\frac{1}{2}\frac{\partial u}{\partial h}\label{EL1} \ ,\\
\frac{(N^2Lf')'}{N^2L}&=&\frac{R^2f}{L^2}+\frac{1}{2}\frac{\partial u}{\partial f} \ ,\\
\frac{L}{N^2}\left(\frac{N^2P'}{L}\right)'&=&2h^2P \ ,\\
\frac{L}{N^2}\left(\frac{N^2R'}{L}\right)'&=&2g^2f^2R  \ ,
\end{eqnarray}
where the prime denotes the derivative with respect to $\rho$ and the potential $u$ reads
\begin{eqnarray}
u(h,f)=\frac{\beta_1}{4}(h^2-1)^2+\frac{\beta_2}{4}(f^2-q^2)^2-\beta_3(h^2-1)(f^2-q^2) \ .
\end{eqnarray}

The variation of (\ref{action}) with respect to the metric leads to the Einstein equations
\begin{eqnarray}
R_{\mu\nu}=-\gamma\left(T_{\mu\nu}-\frac{1}{2}g_{\mu\nu}T\right)  \ , 
\end{eqnarray}
where $T$ is the trace of the energy momentum tensor. Using our Ansatz these read 
\cite{hartmann_urrestilla}
\begin{eqnarray}
\label{einstein1}
\frac{(LNN')'}{N^2L}&=&\gamma\left[\frac{(P')^2}{2L^2}+\frac{(R')^2}{2g^2L^2}-u\right],\\
\frac{(N^2L')'}{N^2L}&=&-\gamma\left[\frac{2h^2P^2}{L^2}+\frac{2R^2f^2}{L^2}
+\frac{(P')^2}{2L^2}+\frac{(R')^2}{2g^2L^2}+u\right]\label{ES2}  \ .
\end{eqnarray}
In addition there is a constraint equation that is not independent. This reads
\begin{equation}
 2 \frac{N' L'}{NL} + \frac{ (N')^2 }{N^2} = \gamma \left[ (h')^2  + (f')^2 + \frac{ (P')^2}{2 L^2} +
\frac{(R')^2}{2g^2 L^2} - \frac{h^2 P^2}{L^2} - \frac{R^2f^2}{L^2} - u\right]   \ .
\end{equation}

The set of differential equations can be solved only numerically subject to an appropriate 
set of boundary conditions. The requirement of regularity at $\rho = 0$ leads to the 
following conditions
 \begin{eqnarray}
 h(0) = 0,\quad f(0)= 0,\quad P(0) = n, \quad R(0) = m
 \end{eqnarray}
for the matter fields and
 \begin{eqnarray}
 N(0) = 1,\quad N'(0)= 0,\quad L(0) = 0, \quad L'(0) = 1
 \end{eqnarray}
for the metric fields, while the requirement of finiteness of the energy per unit length leads to
\begin{equation}
h(\infty)=1, \ f(\infty)=q \ , \ P(\infty)=0 \ , \ R(\infty)=0  \ .
\end{equation}

The inertial energy per unit length $E_{\rm in}^{(n,m)}$ of the (p,q)-string is given by
\begin{eqnarray}
E_{\rm in}^{(n,m)}&=&\int\sqrt{-g_3}T^0_0d\rho d\varphi\label{Eperunitlength}\\
&=&2\pi\int^{\infty}_0
NL\left((h')^2+(f')^2+\frac{(P')^2}{2L^2}+\frac{(R')^2}{2g^2L^2}+\frac{h^2P^2}{L^2}+\frac{R^2f^2}{L^2}+u\right)d\rho  \ ,  
\end{eqnarray}
where $g_3$ is the determinant of the 2+1-dimensional space-time given by ($t,\rho,\varphi$). 
Note that there is also another notion of energy in this space-time, namely
that of the Tolman energy \cite{clv,bl}. This defines the gravitationally active mass.

In the Bogomolnyi-Prasad-Sommerfield (BPS) limit \cite{bps} given by $\beta_1=\beta_2\equiv \beta=2$, $\beta_3=0$ and with the choice
$q=g=1$ we have that $T_{\rho}^{\rho}=T^{\varphi}_{\varphi}=0$ such that it follows from (\ref{einstein1}) that $N(\rho)\equiv 1$. The
remaining BPS equations are 
\begin{eqnarray}
 h'=\frac{Ph}{L} \ \ & , & \ \ f'=\frac{Rf}{L} \ , \\
\frac{P'}{L}=h^2 -1 \ \ & , & \ \ \frac{R'}{L}=f^2 -1 \ ,
\end{eqnarray}
for the matter fields and 
\begin{equation}
 \frac{L''}{L}=-\gamma\left[ \frac{2 h^2 P^2}{L^2} + \frac{2 R^2 f^2}{L^2} + (h^2-1)^2 + (f^2-1)^2\right]
\end{equation}
for the non-trivial metric function.
The solutions fulfill an energy bound such that
\begin{equation} 
E_{\rm in}^{(n,m)}=2\pi (n+m) \ .
\end{equation}
Note that in this limit the widths of the scalar cores $\rho_{{\rm H},i}$ become equal to the widths of
the respective gauge field cores $\rho_{{\rm W},i}$, $i=1,2$. 

The binding energy per unit length of a (p,q)-string $E_{\rm b}^{(n,m)}$ can be defined as
\begin{eqnarray}
E_{\rm b}^{(n,m)}&=&E_{\rm in}^{(n,m)}-nE_{\rm in}^{(n,0)}-mE_{\rm in}^{(0,m)} \ .\label{Ebindperunitlength}
\end{eqnarray}
Finally the (p,q)-string possesses magnetic fields in $z$-direction 
$\vec{B}_1=B_1 \vec{e}_z$ and $\vec{B}_2=B_2 \vec{e}_z$ with
\begin{equation}
 B_1=-\frac{P'}{L} \ \ , \ \ B_2=-\frac{R'}{L} \ ,
\end{equation}
where $B_1$ and $B_2$  are given in units of $M_{W,1}^2$.
The magnetic fluxes then read
\begin{equation}
 \Phi_{M,1}= 2\pi n \ \ , \ \ \Phi_{M,2}=2\pi m
\end{equation}
and are obviously quantized. Hence, changing the winding numbers $n$ and $m$ changes the magnetic
fluxes along the (p,q)-string.

\subsection{The geodesic equation}
The Lagrangian $\mathcal{L}_g$ describing geodesic motion of a test particle 
in the static cylindrically symmetric space-time (\ref{cysymmetric}) reads
\begin{eqnarray}
\label{geolag}
\mathcal{L}_{\rm g}&=&g_{\mu\nu}\frac{dx^{\mu}}{d\tau}\frac{dx^{\nu}}{d\tau}=\varepsilon
=N^2\left(\frac{dt}{d\tau}\right)^2-\left(\frac{d\rho}{d\tau}\right)^2
-L^2\left(\frac{d\varphi}{d\tau}\right)^2-N^2\left(\frac{dz}{d\tau}\right)^2 \ ,
\label{lag}
\end{eqnarray}
where $\varepsilon = 0, 1$ for massless or massive test particles, respectively 
and $\tau$ is an affine parameter that corresponds to the proper time for massive test particles
moving on time-like geodesics. The space-time has three Killing vectors 
$\frac{\partial}{\partial t}$, $\frac{\partial}{\partial \varphi}$ and
$\frac{\partial}{\partial z}$ which lead to the following constants of motion: the energy $E$, 
the angular momentum $L_z$ along the string axis ($z$-axis) 
and the momentum $p_z$
\begin{eqnarray}
N^2\frac{dt}{d\tau}=:E \ \ , \ \ 
L^2\frac{d\varphi}{d\tau}=:L_z \ \ , \ \ 
N^2\frac{dz}{d\tau}=:p_z  \ .
\end{eqnarray}
Using the rescaling (\ref{rescaling}) the constants of motion must be rescaled 
according to $E$ $\rightarrow$ $E/(e_1\eta_1)$, 
$p_z$ $\rightarrow$ $p_z/(e_1\eta_1)$, $L_z$ $\rightarrow$ $L_z/(e_1\eta_1)^2$.  
We then find from (\ref{lag})
\begin{eqnarray}
\varepsilon=N^2\left(\frac{dt}{d\tau}\right)^2-\left(\frac{d\rho}{d\tau}\right)^2
-L^2\left(\frac{d\varphi}{d\tau}\right)^2-N^2\left(\frac{d z}{d\tau}\right)^2
=\frac{E^2-p_z^2}{N^2}-\left(\frac{d\rho}{d\tau}\right)^2-\frac{L_z^2}{L^2}  \label{udot}  \ .
\end{eqnarray}
Using the constants of motion we find from (\ref{geolag})
\begin{eqnarray}
\frac{1}{2}\left(\frac{d\rho}{d\tau}\right)^2 &=&
\frac {E^2-\varepsilon}{2}-\frac{1}{2}
\left[E^2\left(1-\frac{1}{N^2}\right)+\frac{p_z^2}{N^2}+\frac{L_z^2}{L^2}\right]\ . \label{Nrdotsqr}
\end{eqnarray}
The left hand side of (\ref{Nrdotsqr}) is always positive 
and $E^2-\varepsilon$ is a constant of motion. Following \cite{kkl} we can then rewrite this equation as
\begin{eqnarray}
\label{eq1}
\frac{1}{2}\left(\frac{d\rho}{d\tau}\right)^2 &=& \mathcal{E}-V_{\rm eff}(\rho) \label{afromstring}  \ ,
\end{eqnarray}
where
\begin{eqnarray}
\label{effective_pot}
V_{\rm eff}(\rho)&=&\frac{1}{2}\left[E^2\left(1-\frac{1}{N^2}\right)+\frac{p_z^2}{N^2}+\frac{L_z^2}{L^2}\right] \label{Veff}
\end{eqnarray}
is the effective potential 
and $\mathcal{E}$ = $(E^2-\varepsilon)/{2}$. Note that with this definition the effective
potential becomes explicitly energy-dependent. 

In the following, we would like to find $t(\rho)$, $\varphi(\rho)$ and $z(\rho)$. For this,
we rewrite the geodesic equation in the form
\begin{eqnarray}
d\varphi&=&\pm \frac{L_z d\rho}{L(\rho)^2\left(\frac{E^2-p_z^2}{N(\rho)^2}
-\frac{L_z^2}{L(\rho)^2}-\varepsilon\right)^{1/2}}\label{phiint}  \ , \\
dz&=&\pm \frac{p_z d\rho}{N(\rho)^2 \left(\frac{E^2-p_z^2}
{N(\rho)^2}-\frac{L_z^2}{L(\rho)^2}-\varepsilon\right)^{1/2}}\ , \label{zint}\\
dt&=&\pm \frac{E d\rho}{N(\rho)^2 \left(\frac{E^2-p_z^2}{N(\rho)^2}-\frac{L_z^2}{L(\rho)^2}-\varepsilon\right)^{1/2}}\label{tint}  \ .
\end{eqnarray}

The solution for each component can then be calculated as a 
function of $\rho$ by using numerical integration methods.

\section{Numerical results}
We have solved the set of differential equations (\ref{EL1}) - (\ref{ES2}) numerically using the ODE solver COLSYS 
that uses a Newton-Raphson adaptive grid method
\cite{colsys}. The relative error of the solutions is on the order of 10$^{-13}$ - 10$^{-10}$. 
Each component of the geodesic equation can then be integrated numerically by using the 
integrating function \verb!quad!, i.e. a recursive adaptive Simpson quadrature in MATLAB  
with an absolute error tolerance 10$^{-8}$. 
However the numerical profiles of the metric functions $N(\rho)$ and $L(\rho)$  must first be interpolated. 
This was done using a piecewise cubic Hermite interpolating polynomial, i.e. with \verb!pchip! 
in MATLAB. With this procedure it is possible to obtain a smooth curve for the effective potential.

In the following we will distinguish between bound orbits, escape orbits and terminating orbits.
Note that when we talk about bound, escape and terminating orbits we are referring to the motion
in the $x$--$y$--plane. The particles can, of course, move along the full $z$--axis
from $-\infty$ to $+\infty$ for $p_z\neq 0$.

Bound orbits are orbits on which test particles move 
from a minimal value of $\rho$, $\rho_{\rm min} > 0$ to a maximal value of
$\rho$, $\rho_{\rm max} < \infty$ and back again. These orbits have hence two turning points
with $(d\rho/d\tau)^2=0$. On escape orbits, on the other hand, particles come from $\rho=\infty$,
reach a minimal value of $\rho$, $\rho_{\rm min} > 0$ and move back to $\rho=\infty$, which means that escape
orbits have only one turning point with $(d\rho/d\tau)^2=0$.
Looking at (\ref{eq1}) it is obvious that turning points are located at those $\rho$ at
which  ${\cal E} - V_{\rm eff}(\rho) = 0$.  
Finally, terminating orbits are orbits that end at the string axis $\rho=0$. 

For all our calculations we have chosen $q=1$ and $g=1$. 

\subsection{Generalities}
Solutions to the model (\ref{action}) have been extensively studied previously. The Table \ref{SumSolns} summarizes the
particular cases. 

\begin{table}[h!]
\begin{center}
\begin{tabular}{|l|c|c|c|c|}
  \hline
  Solution & $\beta_1$, $\beta_2$ & $\beta_3$ & $\gamma$ & Studied in \\
  \hline
  Abelian-Higgs string in flat space-time& $\beta_1=\beta_2\neq 0$ & $\beta_3=0$ & $\gamma=0$& \cite{no}\\
  Abelian-Higgs string in curved space-time & $\beta_1 =\beta_2\neq 0$  & $\beta_3=0$ & $\gamma\neq0$ & \cite{clv}, \cite{bl} \\
  (p,q)-string in flat space-time & $\beta_1 = \beta_2 = 2$ & $\beta_3\neq 0$ & $\gamma=0$ & \cite{saffin}\\
  (p,q)-string in curved space-time & $\beta_1=\beta_2 = 2$  & $\beta_3\neq 0$ & $\gamma\neq 0$& \cite{hartmann_urrestilla} \\
  \hline
\end{tabular}
\end{center}
\caption{Known string solutions of the model (\ref{action}).}  \label{SumSolns}
\end{table}

It has been observed in \cite{clv} that there are two types of solutions if one couples the Abelian-Higgs model
minimally to gravity: 
string solutions and Melvin solutions which exist for the same values of the parameters in the model.
These differ by their asymptotic behaviour of the metric functions at infinity.

\subsubsection{String solutions}
The string solution behaves like
\begin{eqnarray}
\label{stringinfinity}
N(\rho\rightarrow\infty)=c_1,\quad L(\rho\rightarrow\infty)=c_2\rho+c_3, \quad c_2 > 0  \ ,
\end{eqnarray}
where $c_1$, $c_2$ and $c_3$ are constants depending on $n$, $m$, $g$, $\gamma$ and $\beta_i$, $i = 1,2,3$.
For $\beta_3=0$ it has been found \cite{clv,bl} that $c_1 > 1$ for $\beta_1=\beta_2\equiv \beta < 2$, 
$c_1 < 1$ for $\beta_1=\beta_2\equiv \beta > 2$ and $c_1=1$ in the BPS limit $\beta_1=\beta_2\equiv \beta = 2$.

A solution with the asymptotics (\ref{stringinfinity}) describes a conical space-time with deficit angle $\delta$ given by
\begin{equation}
\label{melvininfinity}
\delta = 2\pi(1-c_2)
\end{equation}
In linear order the deficit angle $\delta$ is given by the product of the coupling $\gamma$ and the
inertial energy per unit length $E_{\rm in}^{(n+m)}$ with $\delta\sim \gamma E_{\rm in}^{(n+m)}$.
As such the constant $c_2=1$ for $\gamma=0$ (or $E_{\rm in}^{(n+m)}=0$) and $c_2$ decreases
for either $\gamma$ or $E_{\rm in}^{(n+m)}$ increasing. If the coupling $\gamma$ or
the energy per unit length is too large then $c_2 < 0$ and the 
deficit angle $\delta > 2\pi$. In this case the solution would have a 
singularity at a finite, parameter-dependent value of $\rho=\rho_0$ with $L(\rho=\rho_0)=0$, while
$N(\rho_0)$ stays finite. These solutions are the so-called super-massive string solutions \cite{gl} 
(or inverted string solutions). We will not consider these kind of solutions in this paper and will
always assume the deficit angle to be smaller than $2\pi$.

The ``force'' exerted on a test particle corresponds to the right hand side of 
\begin{eqnarray}
\label{force}
\frac{d^2\rho}{d\tau^2}&=&-\left(\frac{E^2-p_z^2}{N^3}\right)N'+\left(\frac{L_z^2}{L^3}\right)L'    \ .
\end{eqnarray}
Note that for string solutions the effective potential tends asymptotically to a constant with $V_{\rm eff}(\rho\rightarrow\infty)\rightarrow
\frac{E^2(c_1^2-1)+ p_z^2}{2c_1^2}$  and hence there is no force exerted on test particles far from the string. 
While the force associated to the angular momentum
$L_z$ is always repulsive, the total force close to the string can either be attractive or
repulsive. Since $E^2 - p_z^2 \geq 0$ and $N >0$ this depends on the sign of $N'$ (see more details below).

For $\rho \ll 1$ the string solutions behave like
\begin{equation}
\label{zero}
 N(\rho \ll 1)\sim 1 + O(\rho^2)\ \ , \ \  L(\rho \ll 1) \sim \rho \ .
\end{equation}
Hence there is an infinite potential barrier at $\rho=0$ for test particles with non-vanishing angular momentum $L_z$, i.e.
these test particles can never reach the string axis at $\rho=0$ since their is no force to counterbalance the
repulsive centrifugal force. On the other hand, for $L_z=0$ the effective potential
tends to a constant $V_{\rm eff}(\rho\rightarrow 0) \rightarrow p_z^2/2$. Hence particles
with $E^2 -\varepsilon < p_z^2$ can reach the string axis. Since $E^2 > p_z^2$ these terminating
orbits are 
only possible for massive test particles with $\varepsilon =1$. \\

{\it Infinitely thin cosmic strings} \ The infinitely thin limit corresponds to the
case where both the width of the scalar core as well as that of the gauge field core
tend to zero. The string is hence a 1-dimensional object that can e.g. be described by the
Nambu-Goto action. In this case the metric function $N(\rho)\equiv 1$ (or some other constant that can
be absorbed into the definition of $t$) and $L(\rho)\equiv c_2\rho$ for
$\rho > 0$. In this case, the only component in the force (\ref{force}) exerted on a particle is the
repulsive angular momentum contribution. Hence, bound orbits are not possible in this
space-time. This can also easily be understood when noting that the space-time
of an infinitely thin cosmic string is locally flat \cite{vs} and geodesics are just straight lines. 
The fact that bound orbits are possible in a finite width cosmic string space-time 
is related to the fact that close to the string axis the conical space-time is smoothed
on scales comparable to the width of the string. The existence of bound orbits in
``pure'' cosmic string space-times \cite{remark1} is hence 
a new feature when considering cosmic strings with finite width. 

\subsubsection{Melvin solutions}

The Melvin solutions exist for the same parameter values as the string solutions, but have a different asymptotic behaviour:
\begin{equation}
\label{melvin}
 N(\rho\rightarrow \infty) \rightarrow a_1 \rho^{2/3}  \ \ , \ \  L(\rho\rightarrow \infty) \rightarrow a_2 \rho^{-1/3} \ \ ,
\end{equation}
where again $a_1$ and $a_2$ are parameter dependent positive constants. This space-time 
is not asymptotically flat and the proper length of a curve with $t=const.$, $\rho=const.$, $z=const$ and
$\varphi=0\rightarrow 2\pi$ is $s=2\pi a_2 \rho^{-1/3}$. This tends to zero for $\rho\rightarrow \infty$. 
For the Melvin space-time with the asymptotic behaviour (\ref{melvin}) the effective potential
tends to infinity asymptotically with $V_{\rm eff}(\rho\rightarrow \infty) \propto \rho^{2/3}$ for $L_z\neq 0$.
Hence there is an infinite potential barrier at infinity for test particles with non-vanishing angular momentum, i.e.
these particles can never reach infinity. This is related to the fact that the total force (\ref{force}) on a test particle
is always attractive at large $\rho$ in Melvin space-times. 
For $L_z=0$, the effective potential tends to $E^2/2$ for $\rho\rightarrow\infty$. Hence, the
asymptotic value of the effective potential is always larger than (for massive test particles)
or equal to (for massless test particles) $\mathcal{E}$. Massive
test particles moving on radial geodesics can thus not reach infinity, while massless
test particles have a turning point at infinity. 

For $\rho \ll 1$ the Melvin solutions behave like the string solutions (\ref{zero}).

\subsection{Geodesic motion in (p,q)-string space-times: string solutions}
We will mainly discuss the geodesic motion in space-times with the asymptotic behaviour (\ref{stringinfinity}) since
we believe this to be the physically relevant case. However, since the Melvin solution is a solution
to the Abelian-Higgs model coupled minimally to gravity, we will also comment on this below.

\subsubsection{The effective potential}
The case $\beta_3=0$, $\beta_1=\beta_2\equiv \beta$ has been discussed for $n=m=1$ in \cite{hartmann_sirimachan}.
It was found that bound orbits are only possible for $\beta < 2$ and for massive particles. In fact, in order to
have bound orbits we need (at least) two turning points of the motion, i.e two intersection points between $V_{\rm eff}$ and 
$\mathcal{E}$. Note that for $\mathcal{E}$ finite and larger than the minimal value  of the effective potential
we will always have
one intersection point for $L_z\neq 0$ due to the infinite potential barrier at small $\rho$ such that escape orbits always exist.
However, bound orbits are only possible if in addition the effective potential has 
local minima and maxima with $\frac{dV_{\rm eff}}{d\rho}= 0$. 
At these local extrema we should then have
\begin{eqnarray}\label{dv0}
\frac{E^2-p^2_z}{L_z^2}&=&\frac{N(\rho)^3}{N'(\rho)}\frac{L'(\rho)}{L(\rho)^3} \ .
\end{eqnarray}
Since $E^2-p_z^2  > 0$, $N(\rho) > 0$, $L(\rho) > 0$, $L'(\rho) > 0$ this equation has only solutions for
$N'(\rho) > 0$. 
For $\beta_3 = 0$ it has been observed \cite{hartmann_sirimachan} that the metric function $N(\rho)$ 
is either monotonically decreasing
(for $\beta > 2$) or monotonically increasing (for $\beta < 2$), while $N(\rho)\equiv 1$ in the BPS limit $\beta=2$. 
Hence the sign of $N'(\rho)$ doesn't change and in particular,
bound orbits are only possible for $\beta < 2$. In this case 
the energy-momentum part of the force (\ref{force})
becomes attractive for $\beta < 2$, i.e. if the width of the scalar
core is larger than the width of the gauge field core and can balance the repulsive part associated to the angular momentum.
On the other hand for $\beta=2$ ($\beta >2$) the width of the scalar core is equal (smaller) than the width of the
gauge field core. We observe that this leads to a vanishing (repulsive) energy-momentum part in the force (\ref{force})
and only escape orbits are possible.

\begin{figure}[h!]
  \begin{center}
    \subfigure[ $\beta_1= 1$, $\beta_2=2$     ]{\includegraphics[scale=0.32]{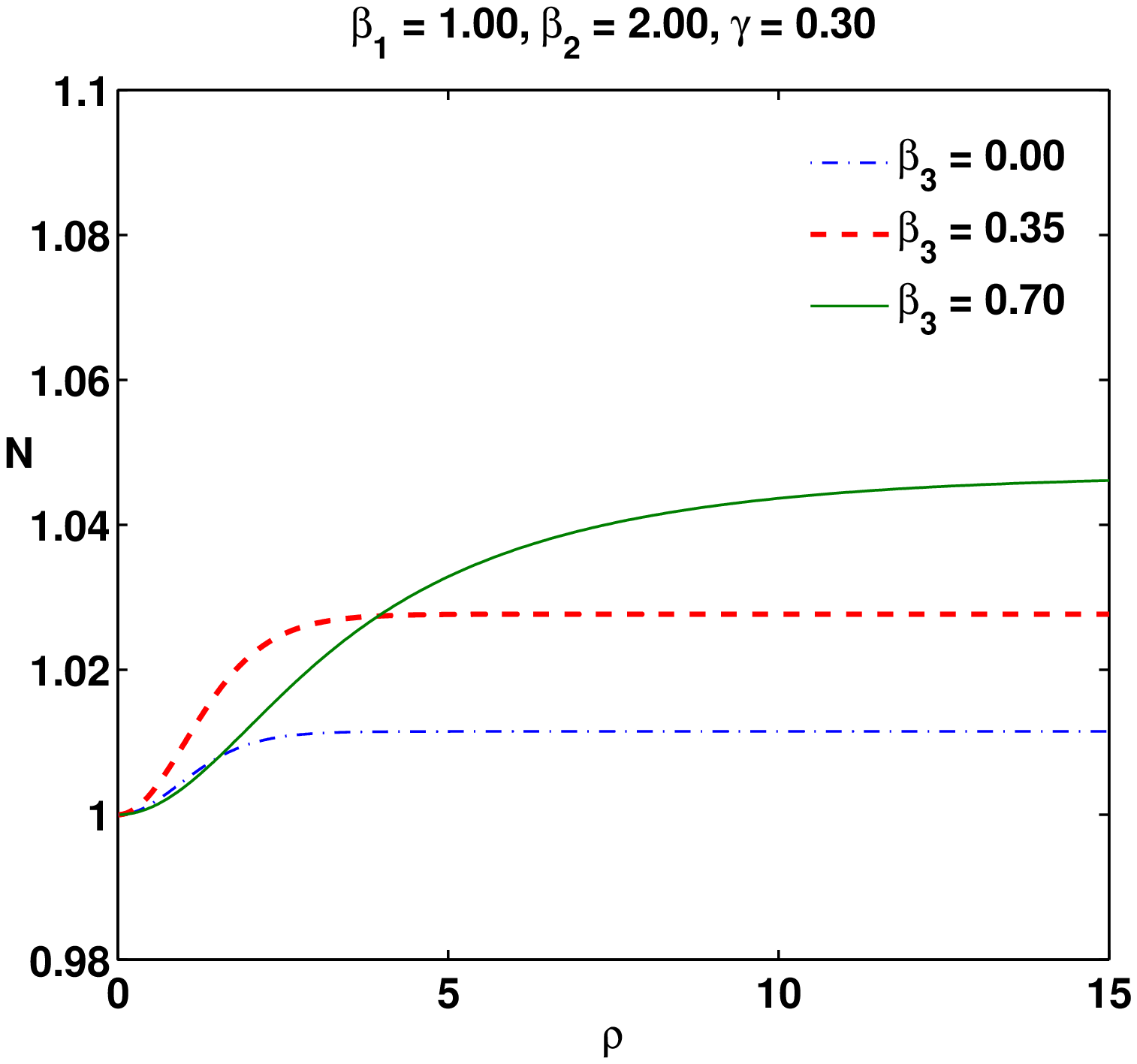}}
    \subfigure[$\beta_1=\beta_2=2$       ]{\label{NprofB}\includegraphics[scale=0.32]{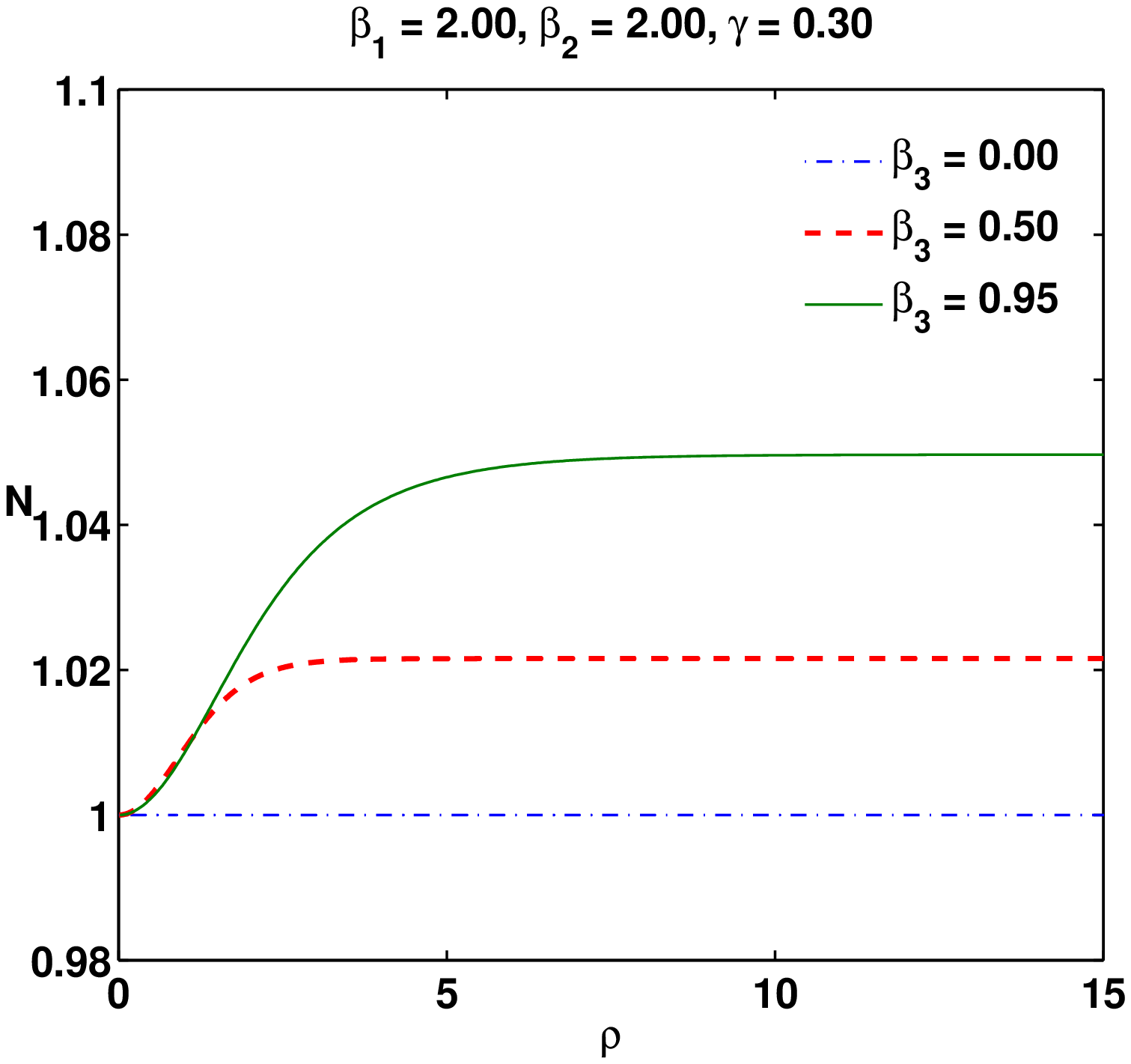}}
    \subfigure[ $\beta_1= 1$,  $\beta_2= 1$       ]{\label{NprofC}\includegraphics[scale=0.32]{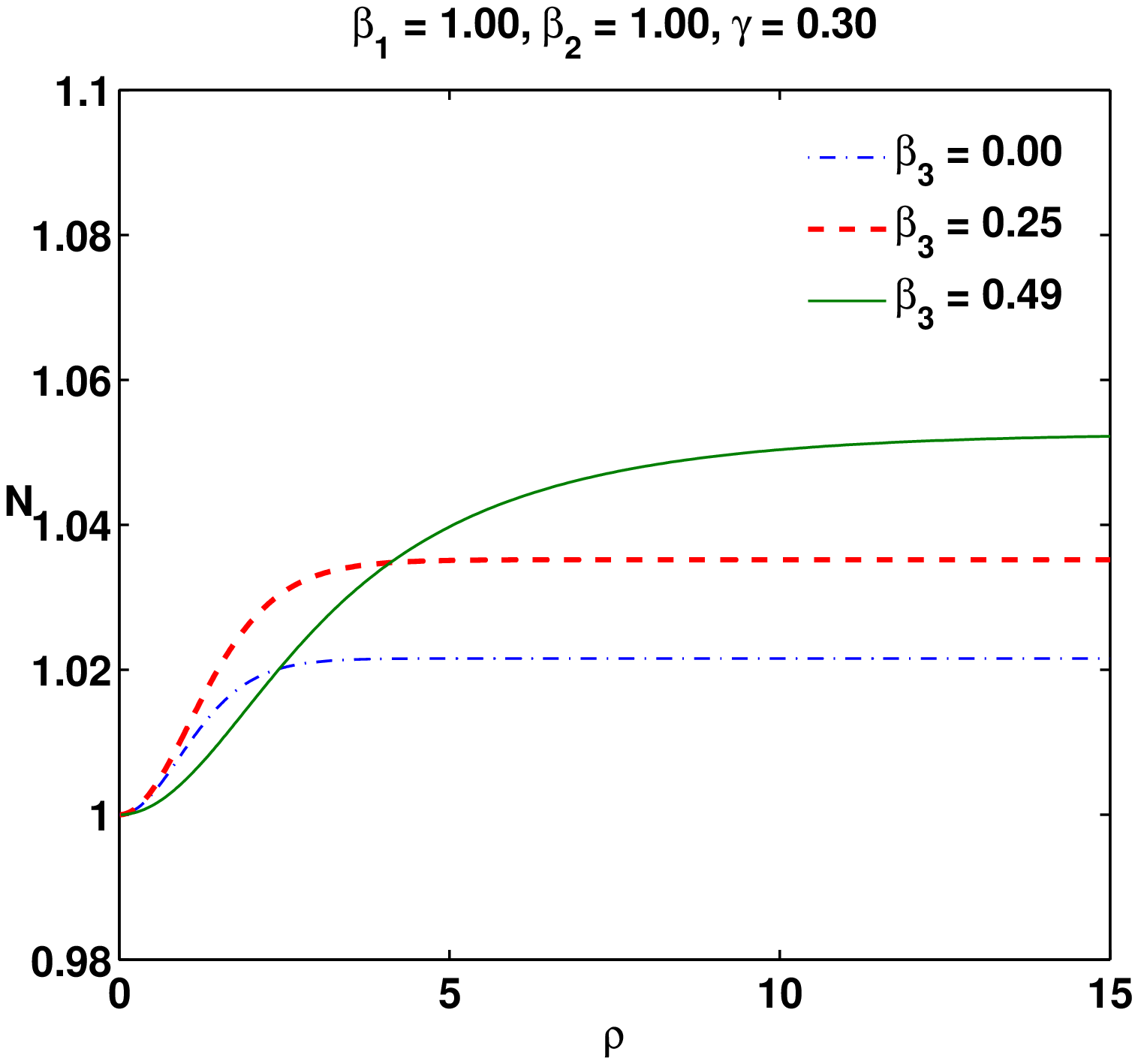}}\\
    \subfigure[$\beta_1=1.5$,  $\beta_2 =6$        ]{\label{NprofD}\includegraphics[scale=0.32]{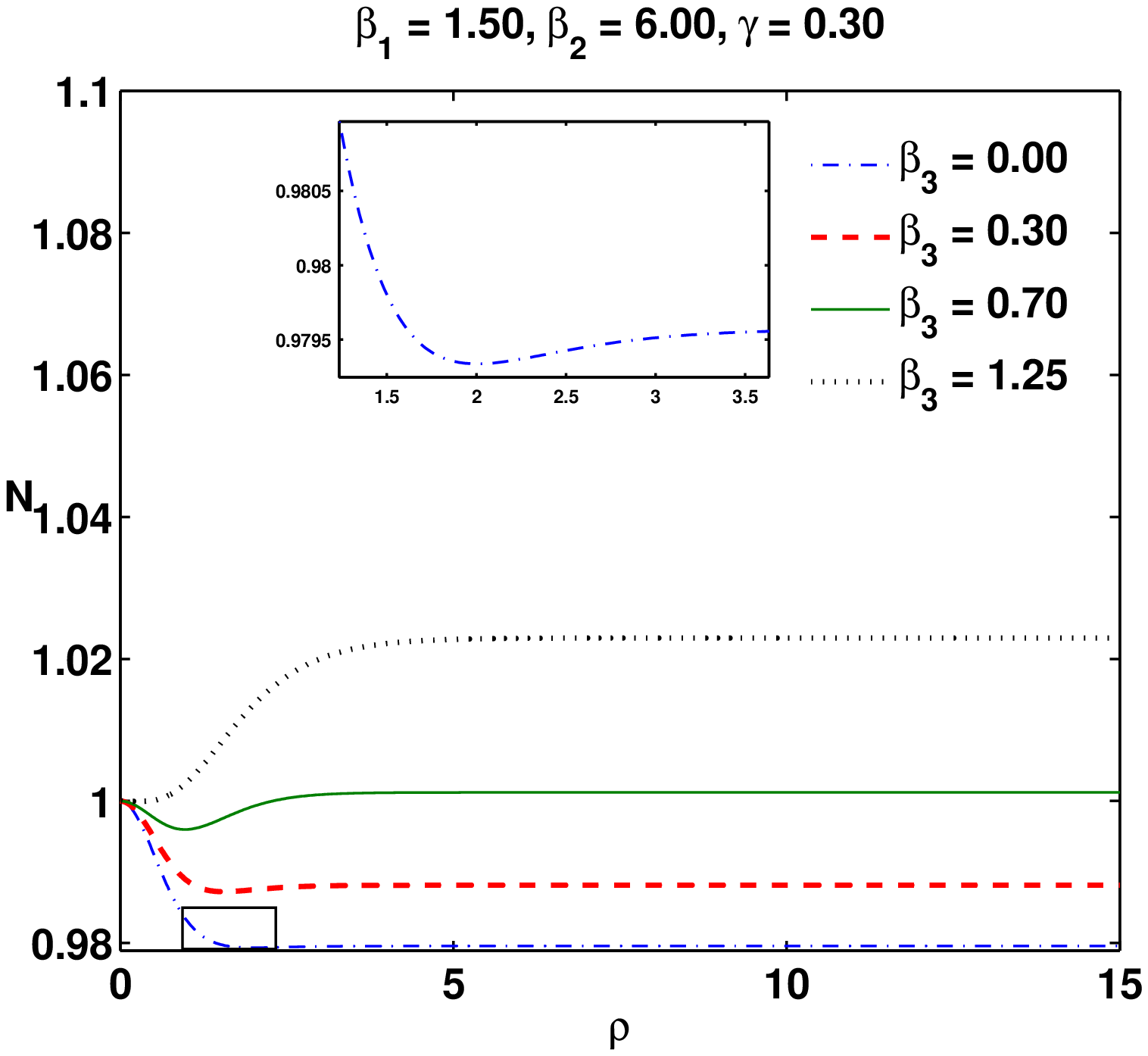}}
    \subfigure[$\beta_1 = 2$,  $\beta_2 = 4.5$       ]{\label{NprofE}\includegraphics[scale=0.32]{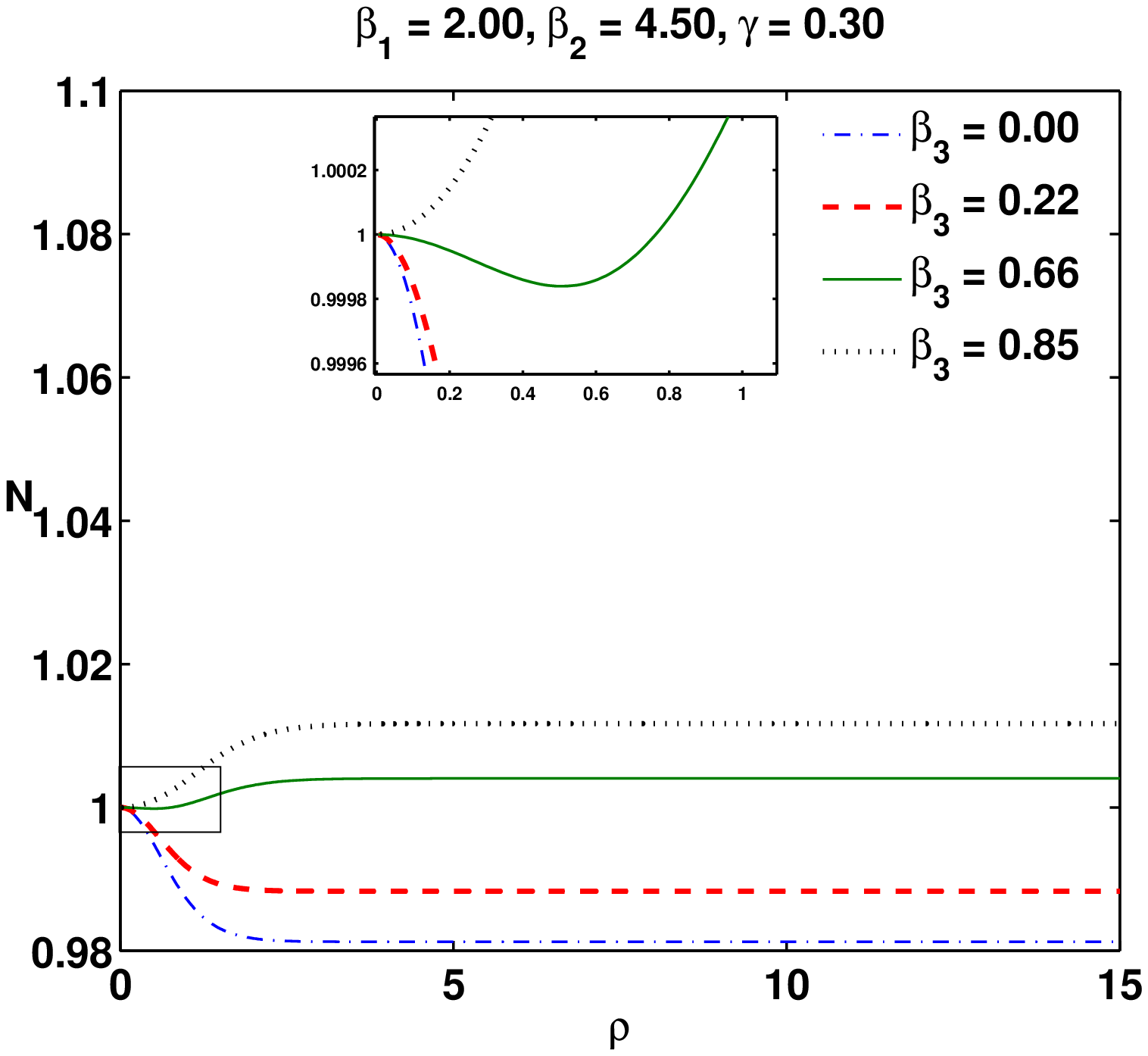}}
    \subfigure[ $\beta_1 =2.25$,  $\beta_2 = 4$     ]{\label{NprofF}\includegraphics[scale=0.32]{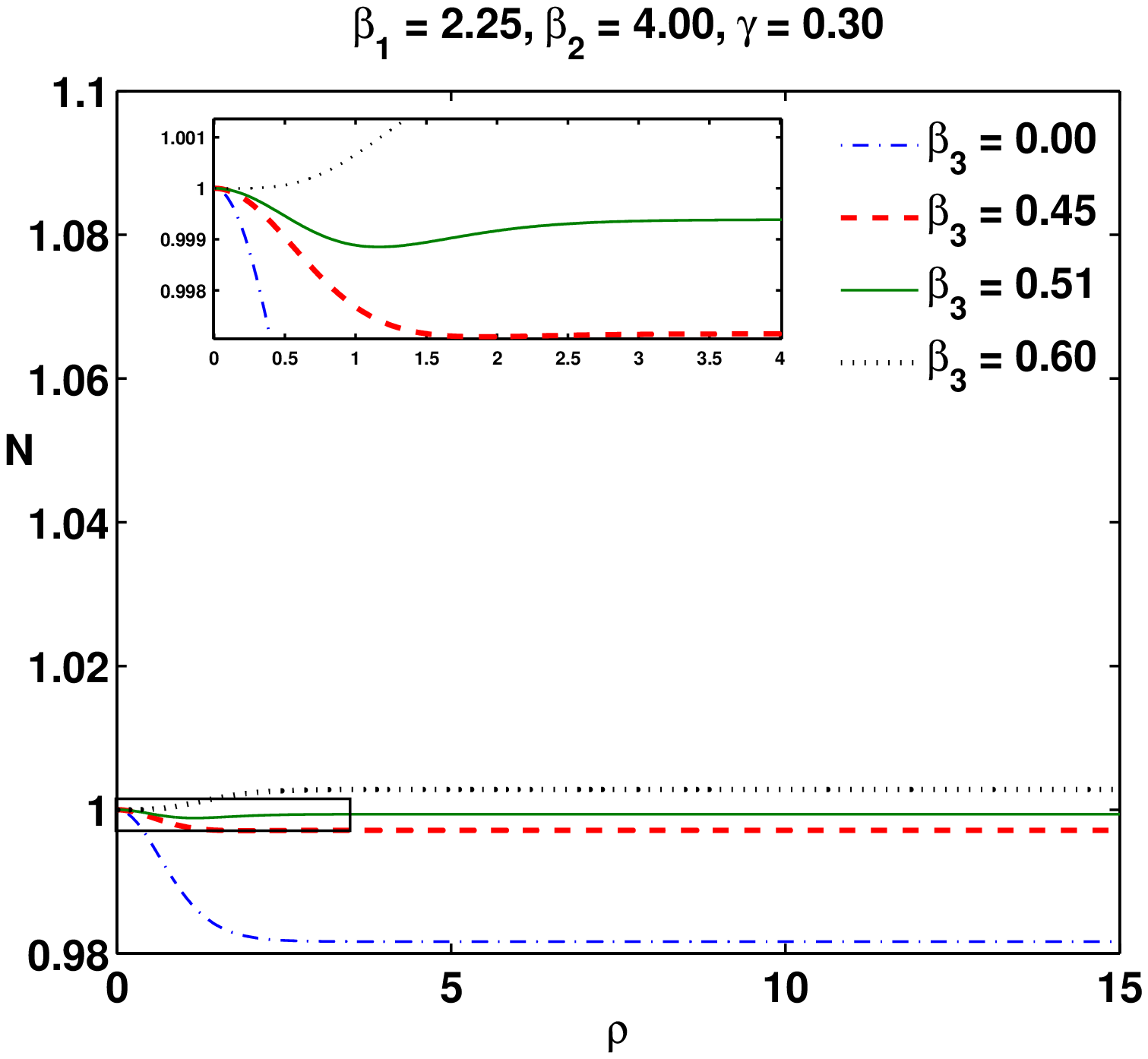}}
   \end{center}
   \caption{The metric function $N(\rho)$ of a (1,1)-string is shown for $\gamma=0.3$ and different choices of $\beta_1$, $\beta_2$ and $\beta_3$.
                    }  \label{Nprof}
  \end{figure}

This is different when $\beta_3 > 0$. We will first discuss the case $n=m=1$. 
The behaviour of the metric function $N(\rho)$ of a (1,1)-string for $\gamma=0.3$ and different choices of
$\beta_i$, $i=1,2,3$ is shown in Fig. \ref{Nprof}. In all cases the blue dotted-dashed line corresponds to $\beta_3=0$ and for
cases (a), (b) and (c) the green solid line corresponds to  $\beta_3\approx \beta_3^{\rm (max)}$ with $\beta_3^{\rm (max)}\equiv
\sqrt{\beta_1\beta_2}/2$ the maximally allowed value for a given choice of $\beta_1$ and $\beta_2$.
For (a) $\beta_1=1$, $\beta_2=2$, (b) $\beta_1=\beta_2=2$ and (c) $\beta_1=\beta_2=1$ 
the increase of $\beta_3$ leads to an increase
of the asymptotic value of $N(\rho)$ for all choices of the $\beta_i$, $i=1,2,3$. 
Hence, the increased binding between the p- and the q-string pronounces the effect already observed
in the $\beta_3=0$ limit. Note that while bound orbits are not possible in the BPS limit $\beta_1=\beta_2=2$ for $\beta_3=0$
bound orbits do exist for $\beta_3 >0$ and $\beta_1=\beta_2=2$ (which, of course, no longer corresponds to a BPS limit).
For (d) $\beta_1=1.5, \beta_2=6$, (e) $\beta_1=2, \beta_2=4.5$ and (f) $\beta_1=2.25, \beta_2=4$ the metric function
$N(\rho)$ can have a local minimum if $\beta_3 < \beta_3^{\rm (cr)}(\beta_1,\beta_2)$, i.e. if the binding
between the strings is not too large. This is new as compared to the $\beta_3=0$ limit. We find that
$\beta_3^{\rm (cr)}(1.5,6)\approx 0.7$, $\beta_3^{\rm (cr)}(2,4.5)\approx 0.66$,  $\beta_3^{\rm (cr)}(2.25,4)\approx 0.51$.

\begin{figure}[h!]
  \begin{center}
\subfigure[$\beta_1= 2$, $\beta_2=2$     ]{\includegraphics[scale=0.32]{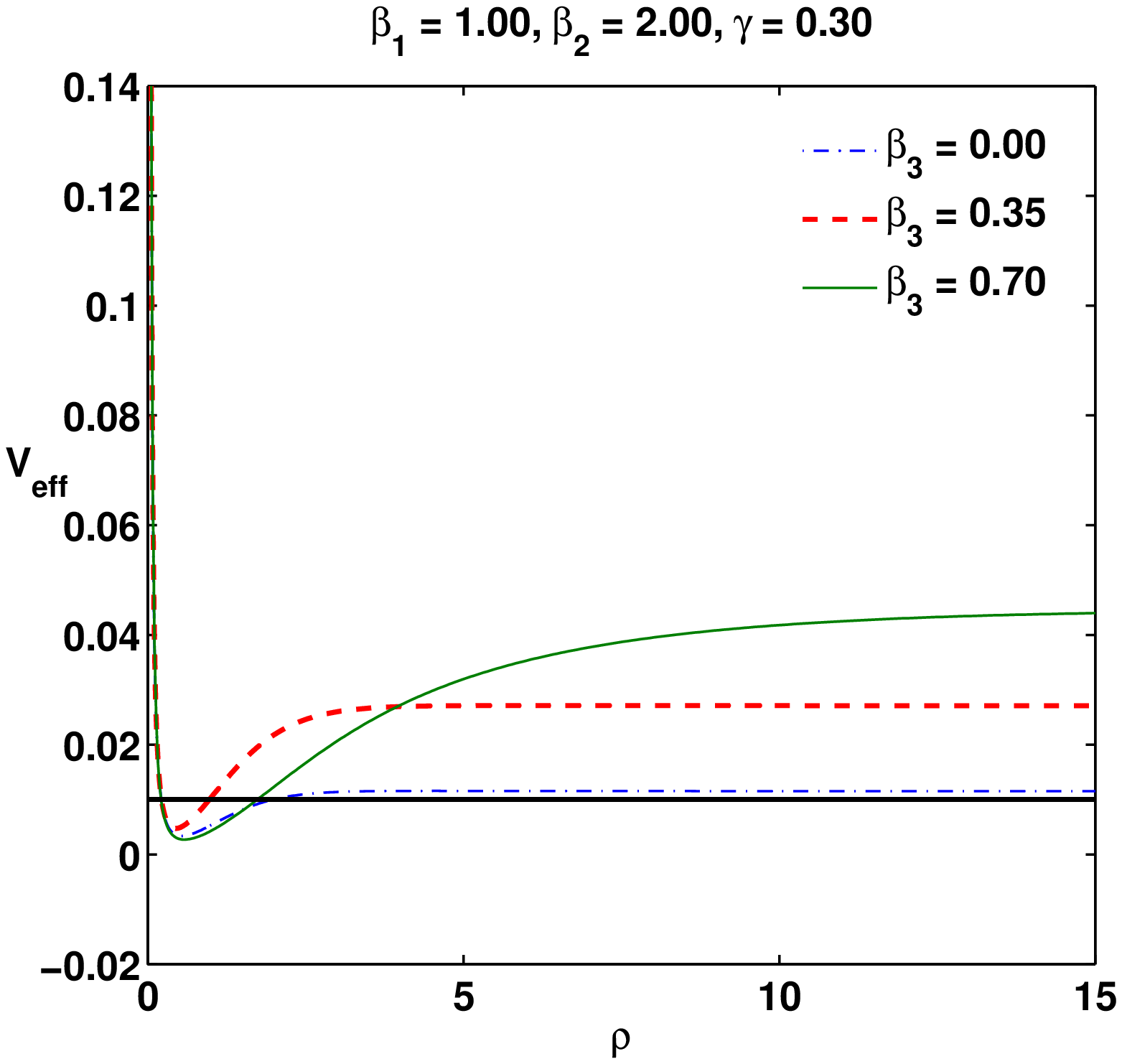}}
    \subfigure[ $\beta_1=\beta_2=2$        ]{\label{NprofB}\includegraphics[scale=0.32]{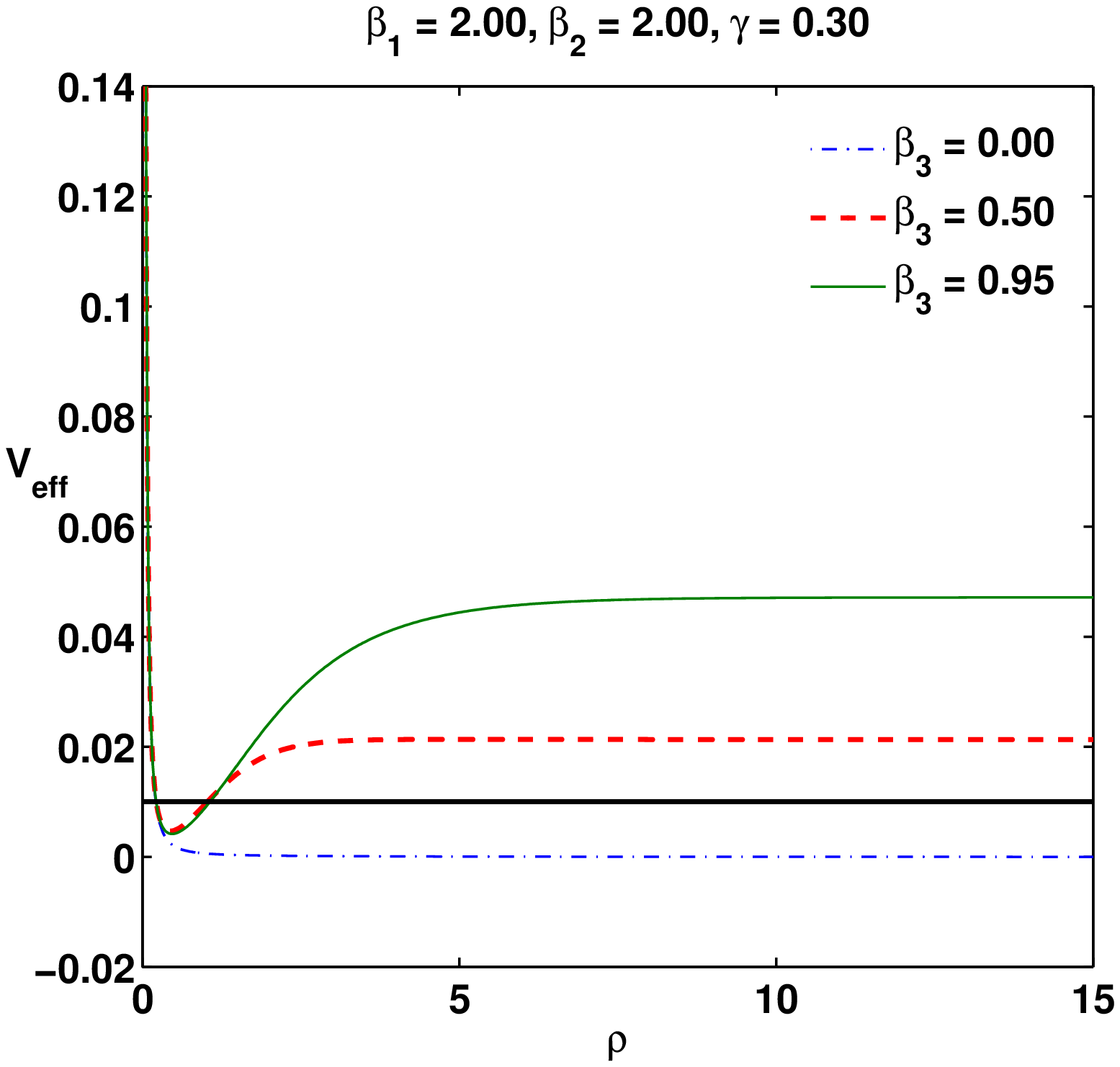}}
    \subfigure[$\beta_1= 2$,  $\beta_2= 1$      ]{\label{NprofC}\includegraphics[scale=0.32]{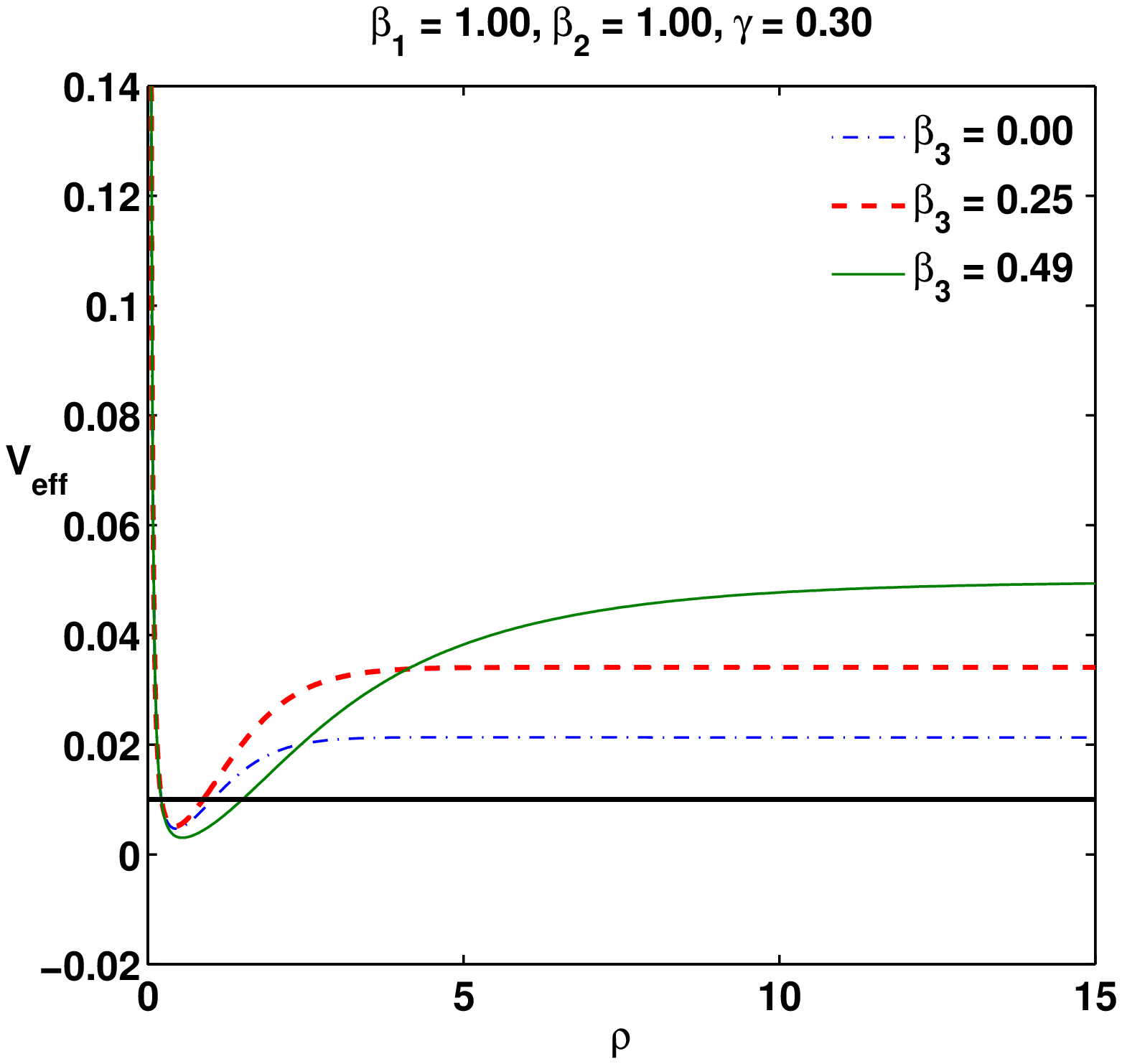}}\\   
    \subfigure[ $\beta_1= 1.5$,  $\beta_2 = 6$        ]{\label{NprofD}\includegraphics[scale=0.32]{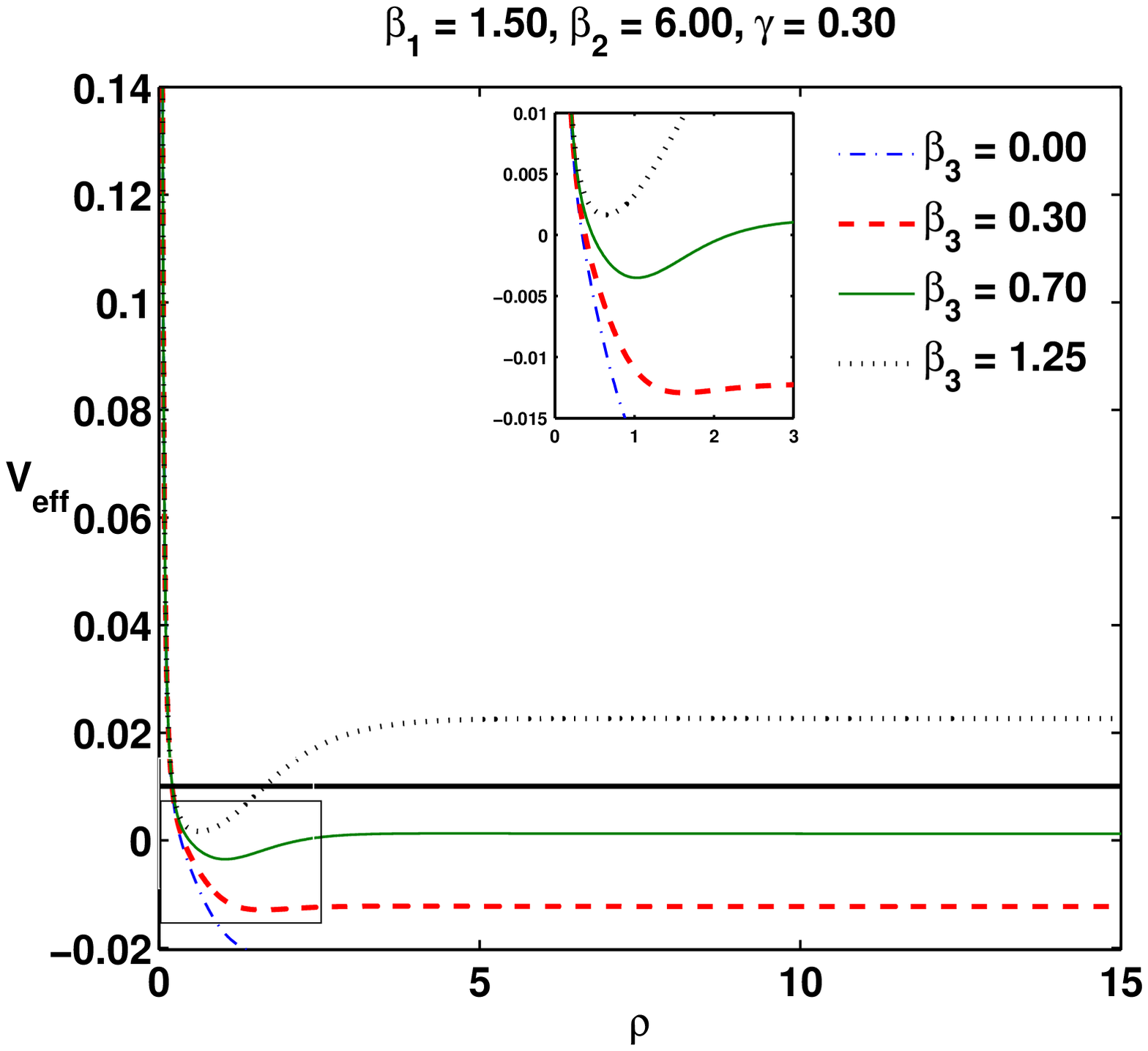}}
    \subfigure[$\beta_1 = 2$,  $\beta_2 = 4.5$         ]{\label{NprofE}\includegraphics[scale=0.32]{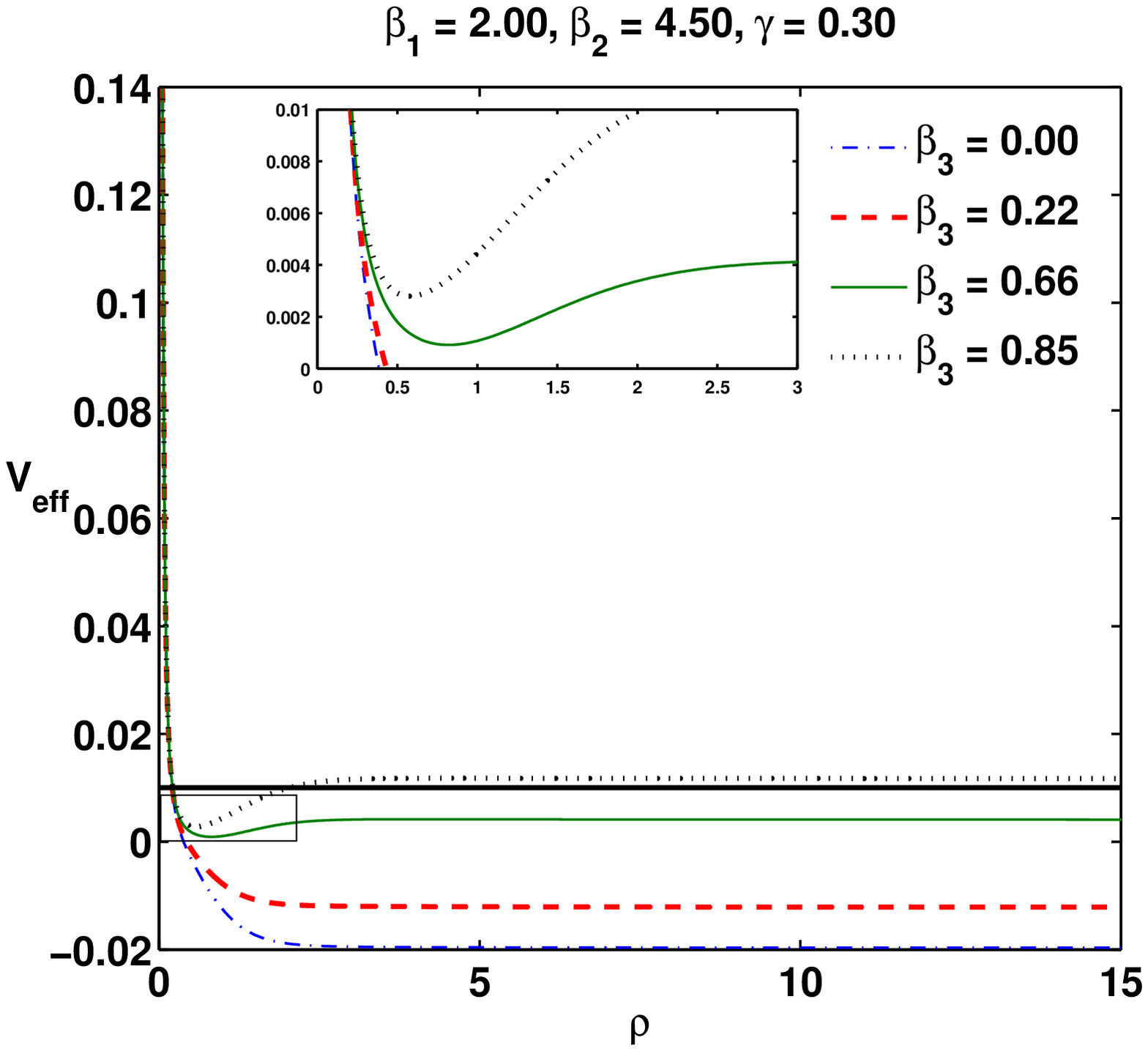}}
    \subfigure[ $\beta_1 = 2.25$,  $\beta_2 = 4$          ]{\label{NprofF}\includegraphics[scale=0.32]{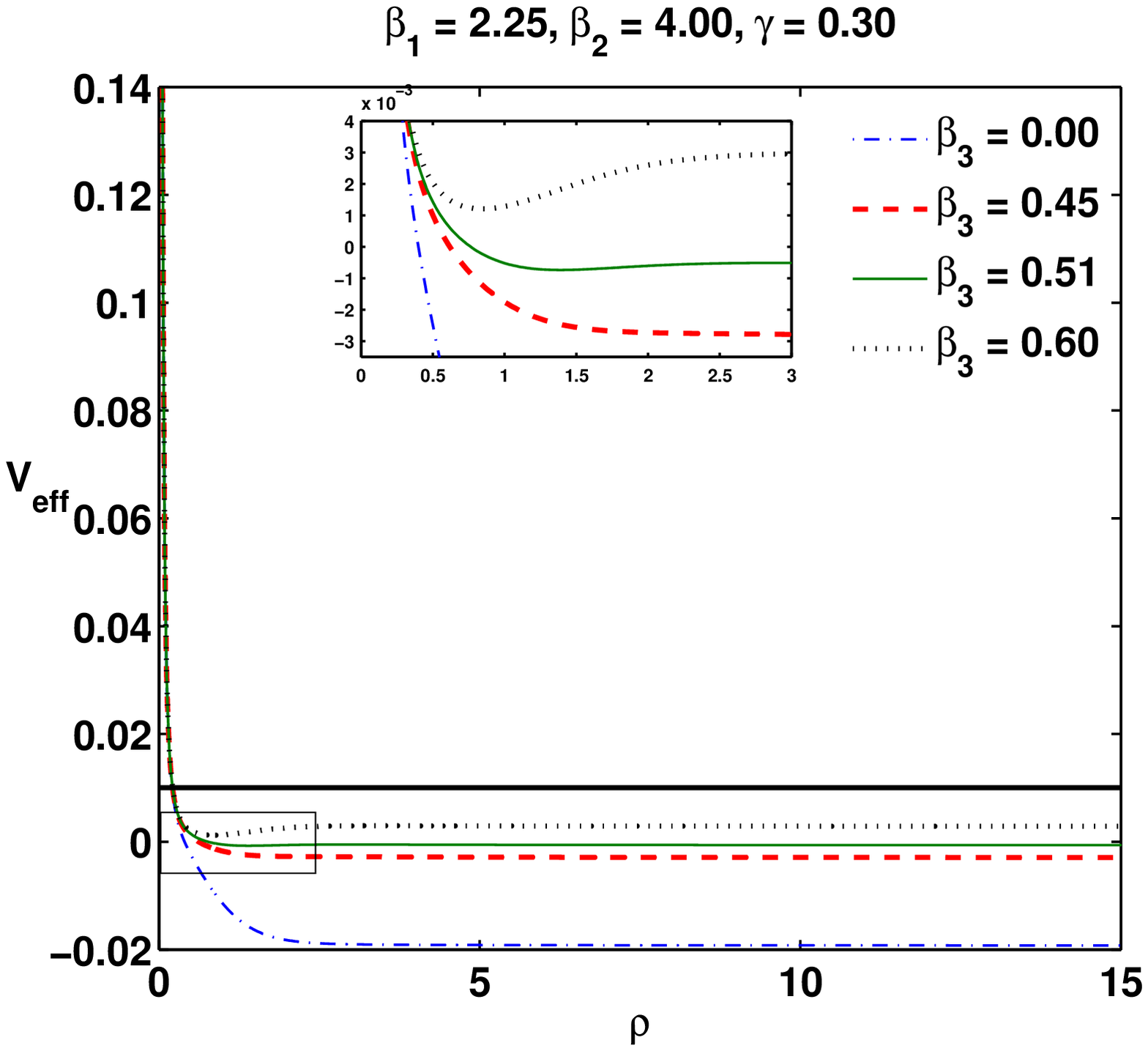}}
   \end{center}
   \caption{The effective potential $V_{\rm eff}(\rho)$ in the space-time of a (1,1)-string 
     is shown for $\gamma=0.3$ and different choices of $\beta_1$, $\beta_2$ and $\beta_3$. Here $E=0.01$, $L_z=0.03$ and $p_z=0$.
                            }  \label{Vprof}
  \end{figure}

Obviously, for (a) $\beta_1< 2$, $\beta_2=2$,  (b) $\beta_1=\beta_2=2$ and (c) $\beta_1< 2$,  $\beta_2< 2$ the metric
function $N(\rho)$ increases monotonically, while for the other cases $N(\rho)$ can first decrease from $N(\rho=0)=1$, have a local
minimum at $\rho=\rho_{\rm min}$ with $N(\rho_{\rm min}) < 1$ and then increase again to $N(\rho \gg 1) > 1$. 
This has important consequences for the shape of the effective potential as discussed below and can be understood
as follows: consider the (1,1)-string to be a superposition of a (1,0)-string and a (0,1)-string.
Now, for $\beta_3=0$, these two strings do not interact. In this case, we know that for $\beta_i > 2$, $i=1,2$ the metric
function $N(\rho)$ would monotonically decrease, while for $\beta_j < 2$, $j=1,2$ the metric
function $N(\rho)$ would monotonically increase. Superposing a string with $\beta_i>2$ and one with $\beta_j < 2$
leads than to a metric function $N(\rho)$ that first decreases and than increase again. Note that the opposite is not possible
since the scalar core
of a string with $\beta_i > 2$ is smaller than that of a string with $\beta_j < 2$.

This behaviour of the metric function $N(\rho)$
leads to the observation that the effective potential can have a negative minimum which
for $p_z=0$ is located exactly at $\rho_{\rm min}$. In the case $\beta_3=0$, the
effective potential can have a local minimum for $\beta < 2$, but this will always be positive valued since $N'(\rho) > 0$ 
means $N(\rho) > 1$.
This is shown in Fig.\ref{Vprof} for a particle with $E=0.01$, $L_z=0.03$ and $p_z=0$. For cases (a), (b) and (c) the effective potential is 
positive for all values of $\rho$ for non-vanishing $L_z$ or $p_z$, while it can become negative for the other cases. In fact, the potential
becomes positive everywhere for $\beta_3\approx\beta_3^{(cr)}$. This will have 
influence on the existence of bound orbits as discussed below. In particular if the potential 
has a negative valued minimum as is e.g. the case for $\beta_1=1.5$, $\beta_2=6$ and $\beta_3=0.7$, particles
with ${\cal E} <0$, i.e. $E^2 < 1$ can move on bound orbits. 

\begin{figure}[p!]
  \begin{center}
   \subfigure[$\beta_1 =\beta_2 =4$, $\beta_3 =1.9$]{\label{NII-II}\includegraphics[scale=0.29]{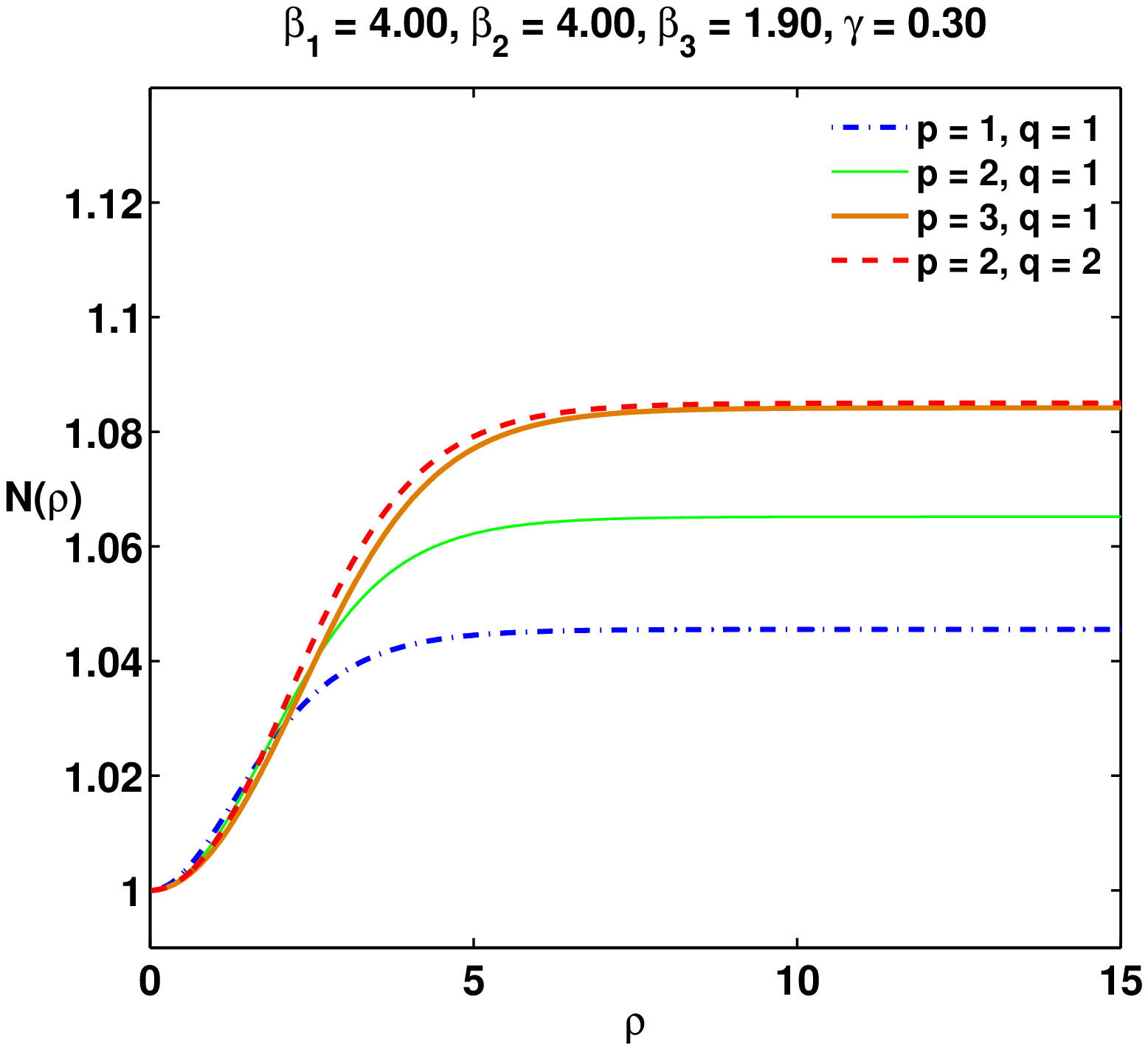}}
    \subfigure[$\beta_1=\beta_2=2$, $\beta_3=0.7$]{\label{NB-B}\includegraphics[scale=0.29]{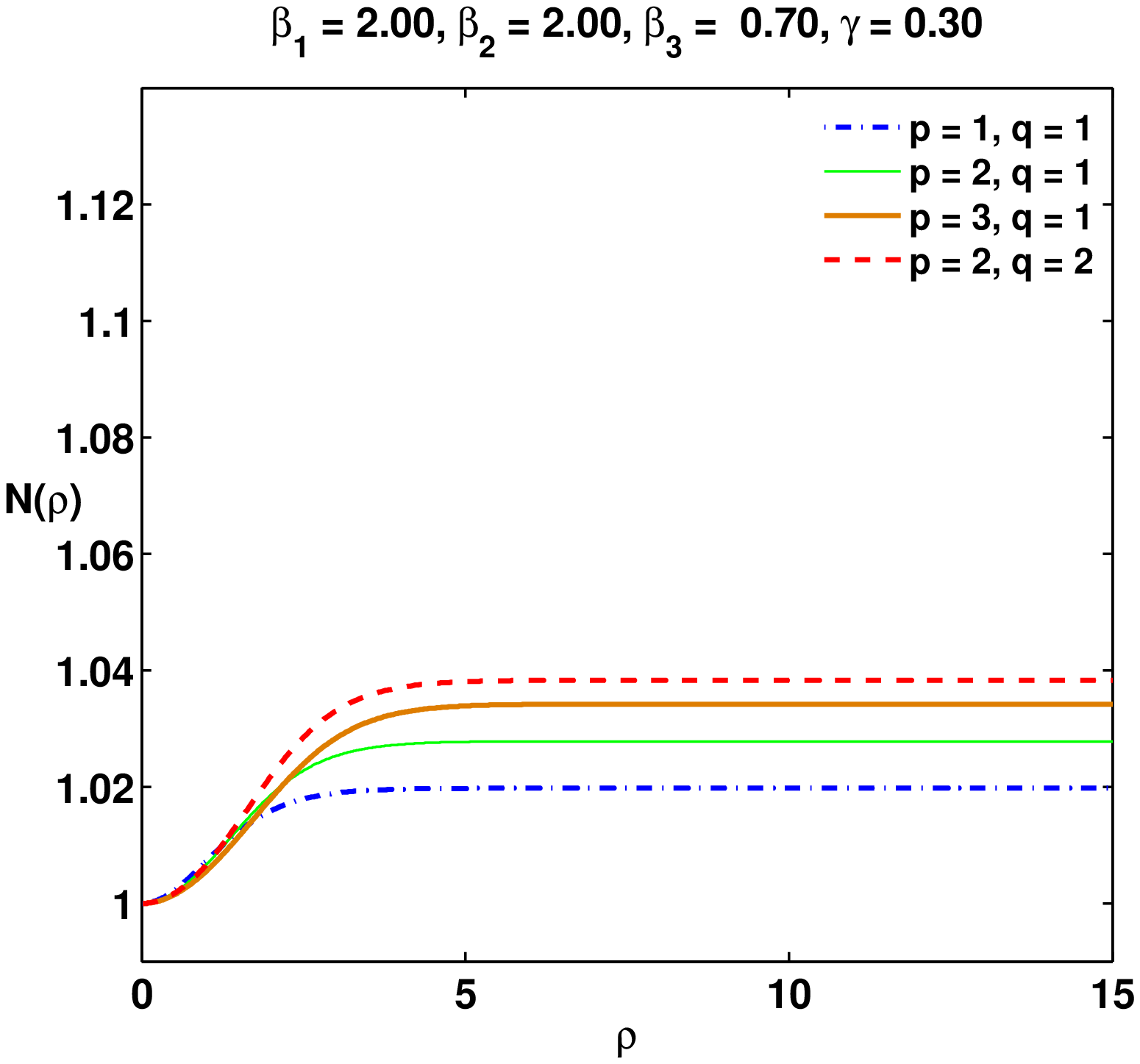}}
    \subfigure[$\beta_1 =\beta_2=0.5$, $\beta_3 =0.24$]{\label{NI-I}\includegraphics[scale=0.29]{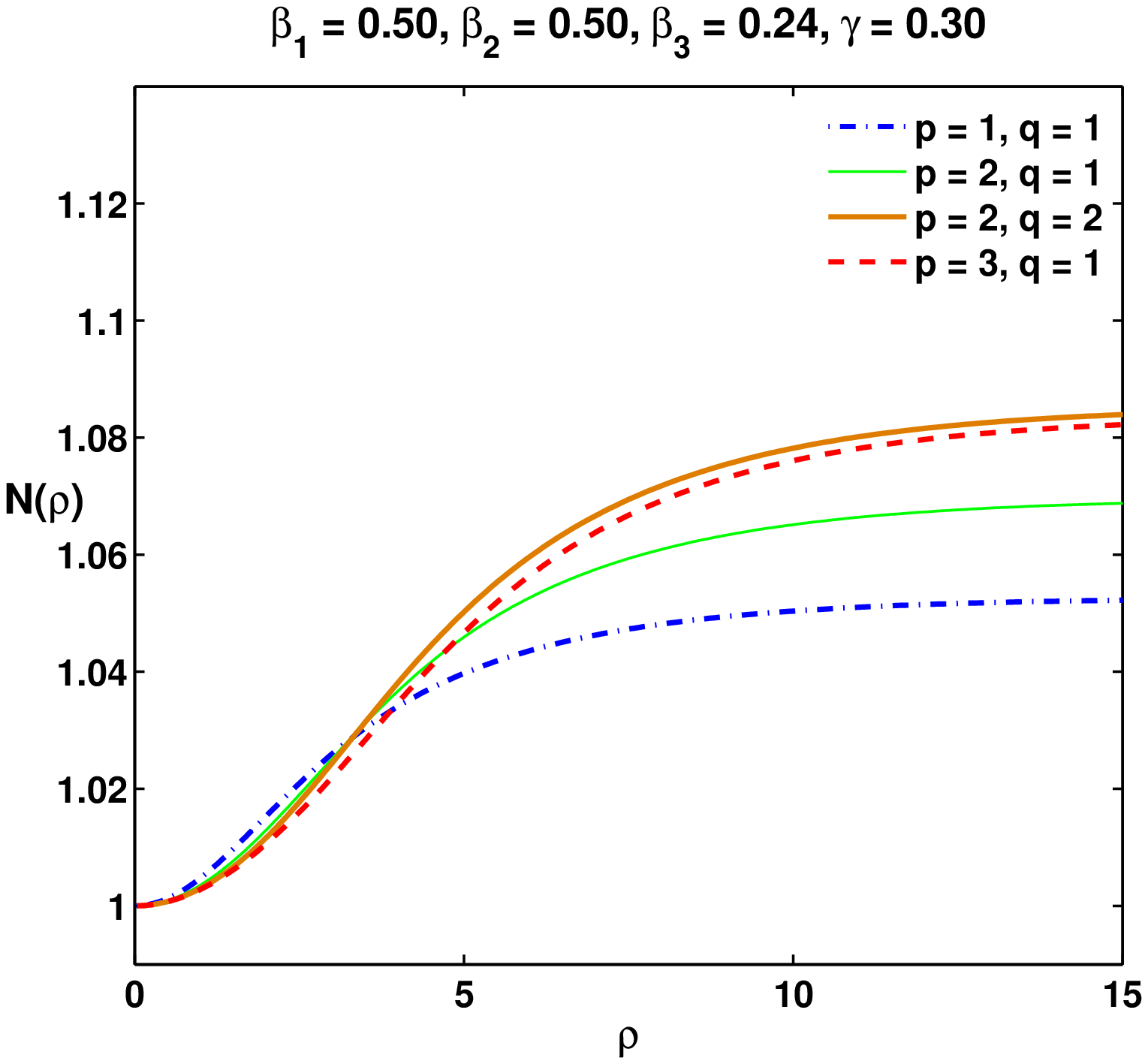}}\\
    \subfigure[$\beta_1 = 1$, $\beta_2 = 4$, $\beta_3=0.99$]{\label{NI-II}\includegraphics[scale=0.29]{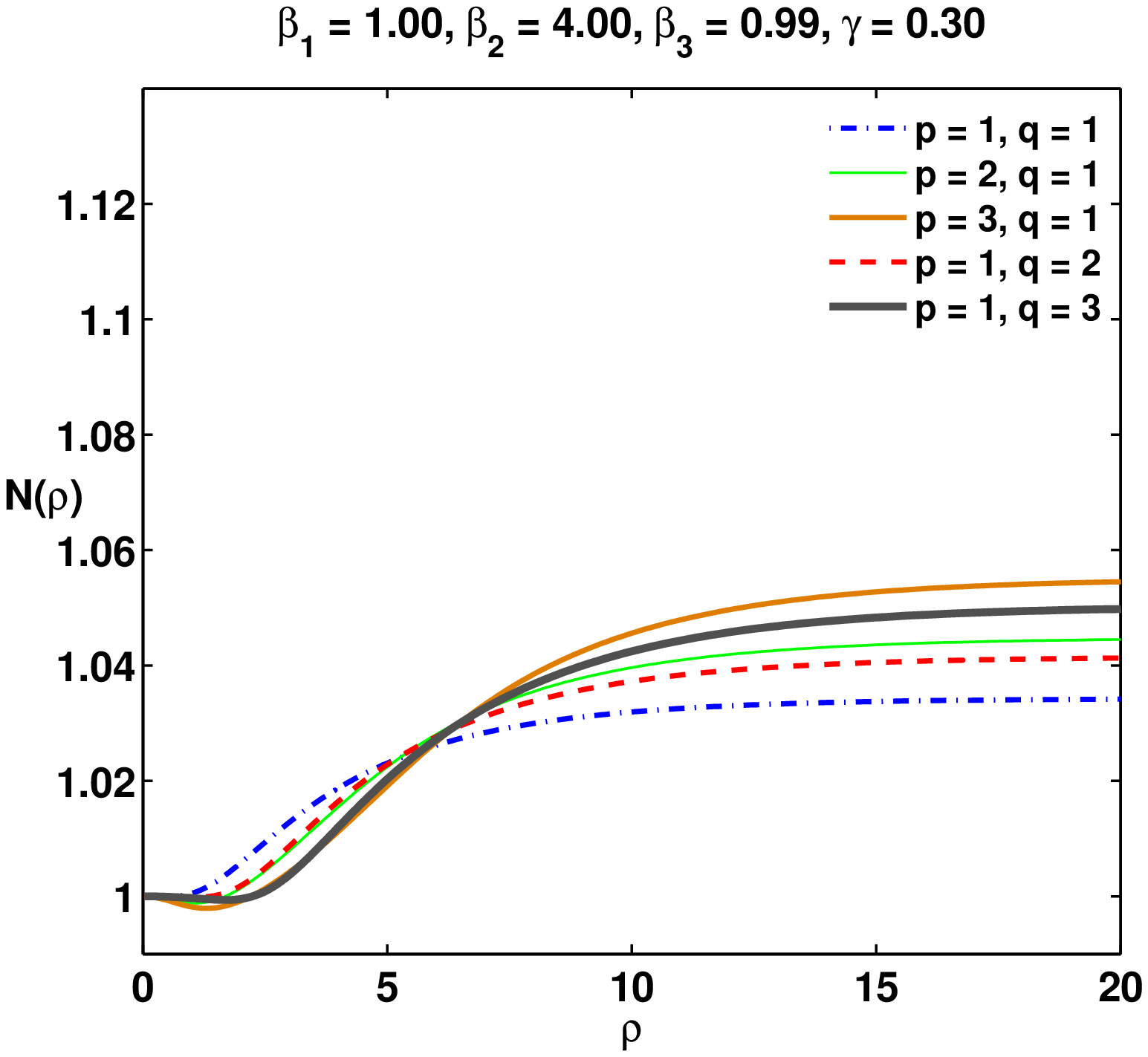}}
    \subfigure[$\beta_1 = 2$, $\beta_2=4.5$, $\beta_3=1.45$]{\label{NII-B}\includegraphics[scale=0.29]{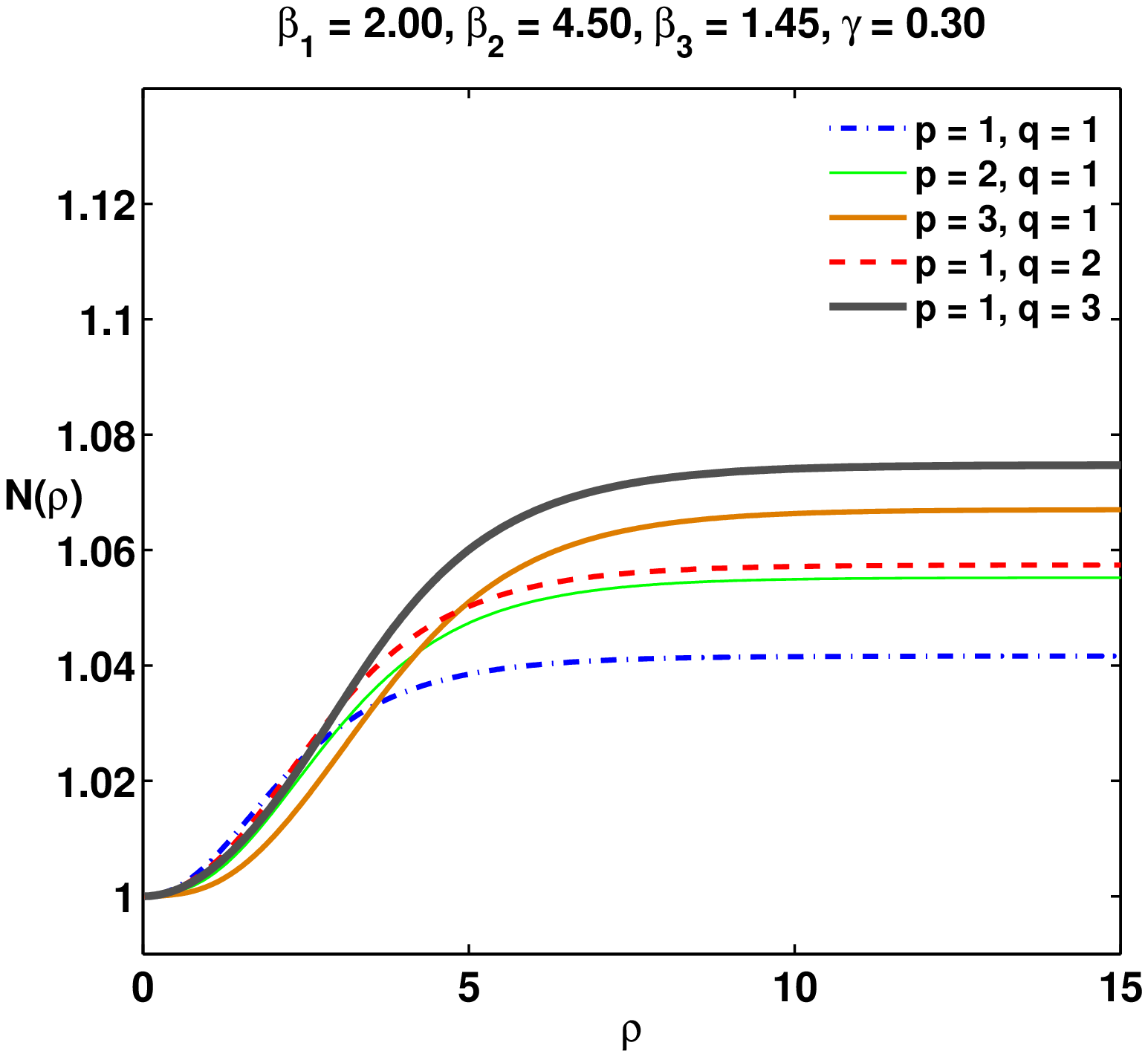}}
     \subfigure[$\beta_1 = 0.5$, $\beta_2=2$, $\beta_3=0.45$]{\label{NI-B}\includegraphics[scale=0.29]{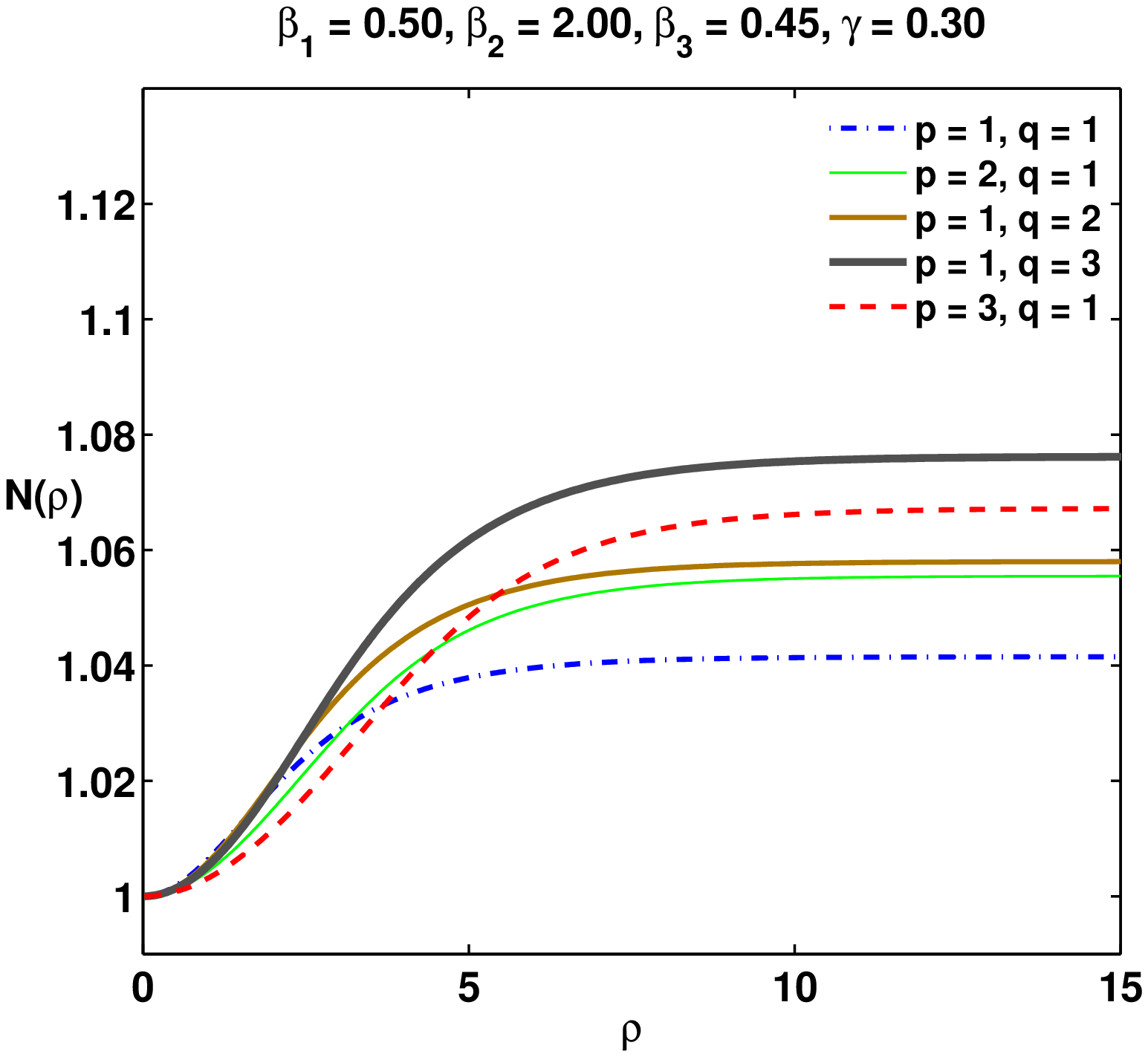}}
   \end{center}
   \caption{The metric function $N(\rho)$ of a (p,q)-string for $\gamma=0.3$, different
choices of $\beta_i$, $i=1,2,3$ and different choices of (p,q)$\equiv (n,m)$.} \label{Nprofvpq}
  \end{figure}

We have also investigated how the metric function $N(\rho)$ changes
when changing the winding numbers $n$, $m$ and hence the magnetic fluxes along the string. Our results are shown
in Fig.\ref{Nprofvpq} for $\gamma=0.3$.

\begin{figure}[p!]
  \begin{center}
    \subfigure[$\beta_1 =\beta_2 =4$, $\beta_3 =1.9$]{\label{LII-II}\includegraphics[scale=0.29]{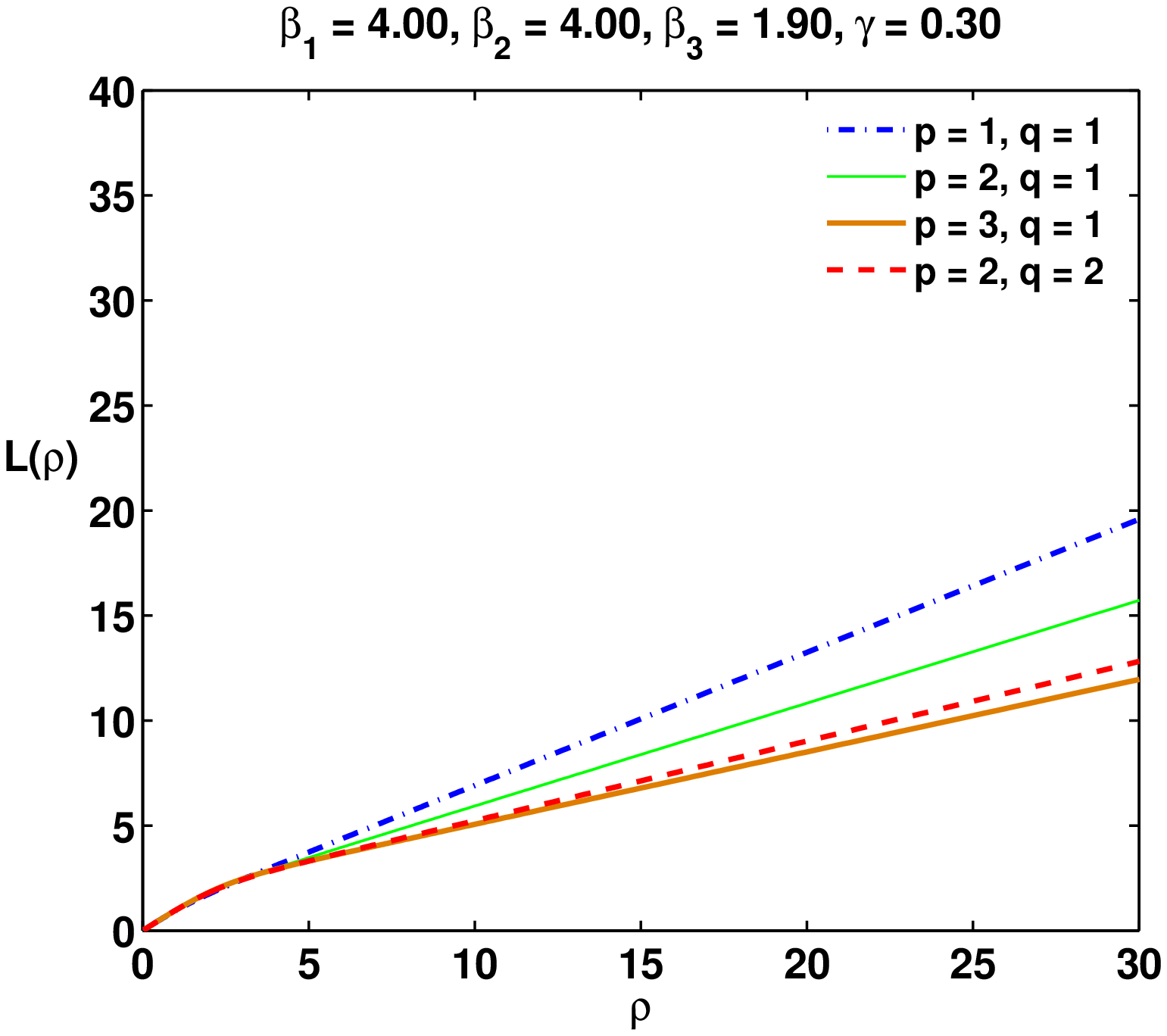}}
    \subfigure[$\beta_1=\beta_2=2$, $\beta_3=0.7$]{\label{LB-B}\includegraphics[scale=0.29]{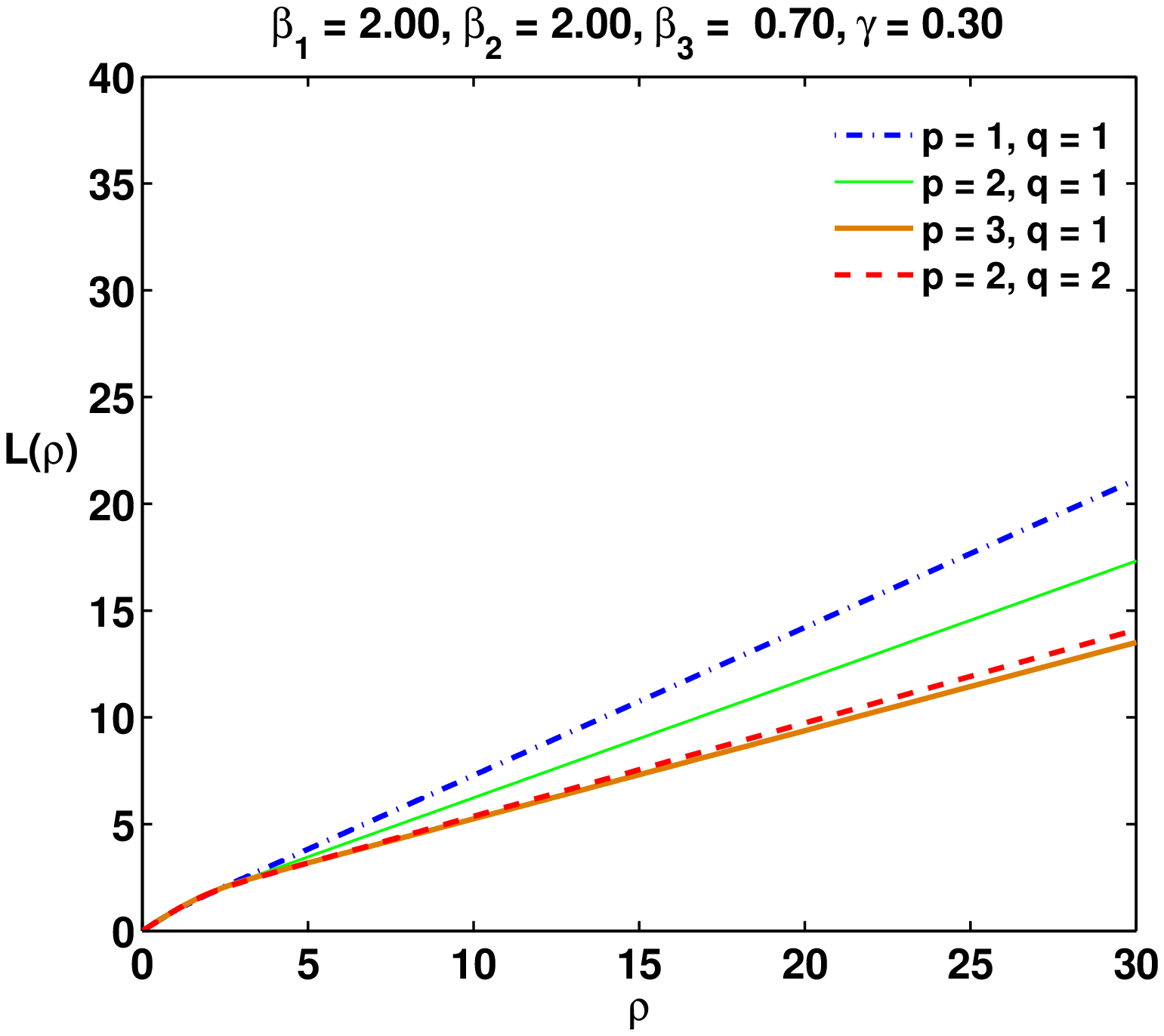}}
    \subfigure[$\beta_1 =\beta_2=0.5$, $\beta_3 =0.24$]{\label{LI-I}\includegraphics[scale=0.29]{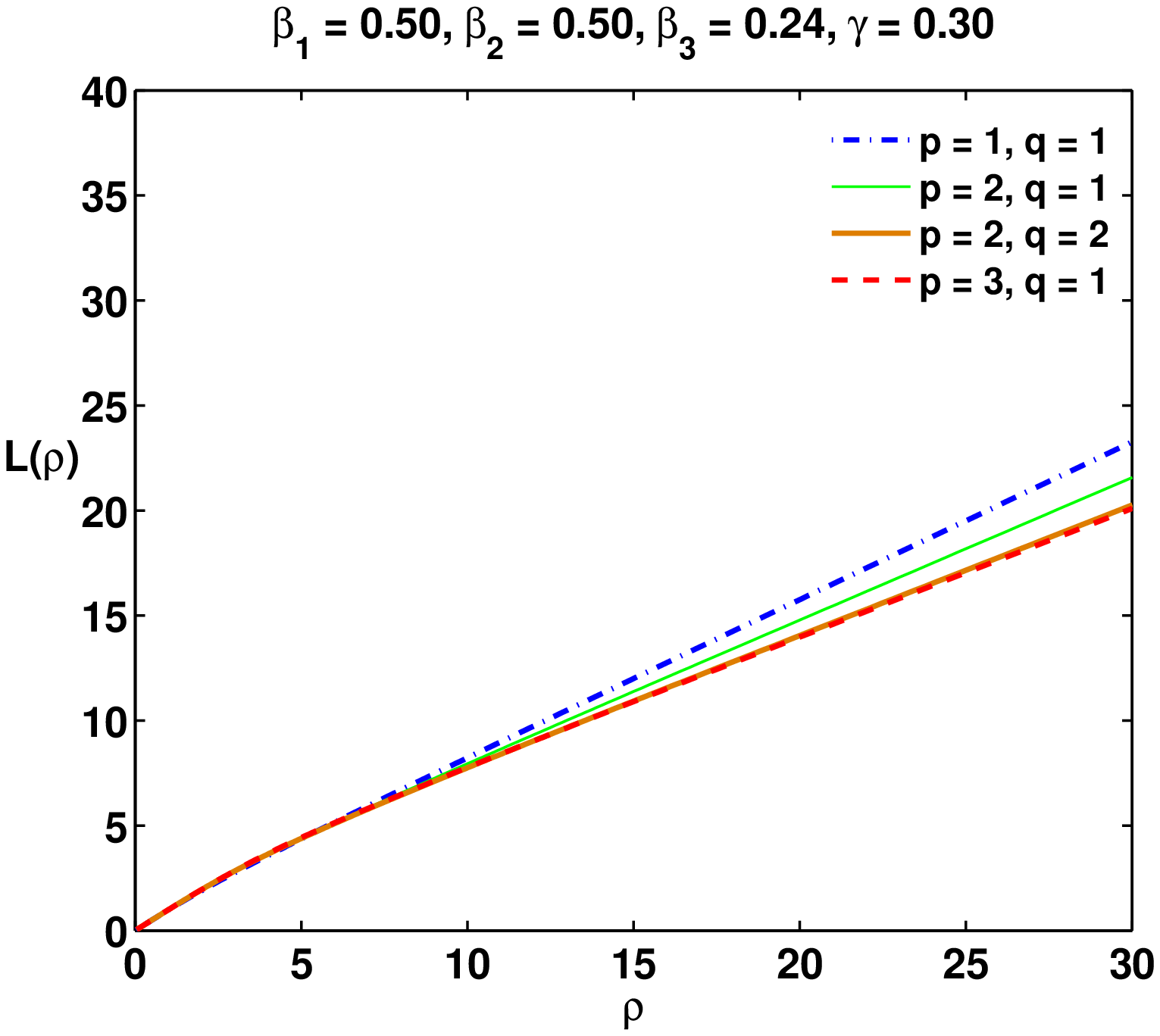}}\\
    \subfigure[$\beta_1 = 1$, $\beta_2 = 4$, $\beta_3=0.99$]{\label{LI-II}\includegraphics[scale=0.29]{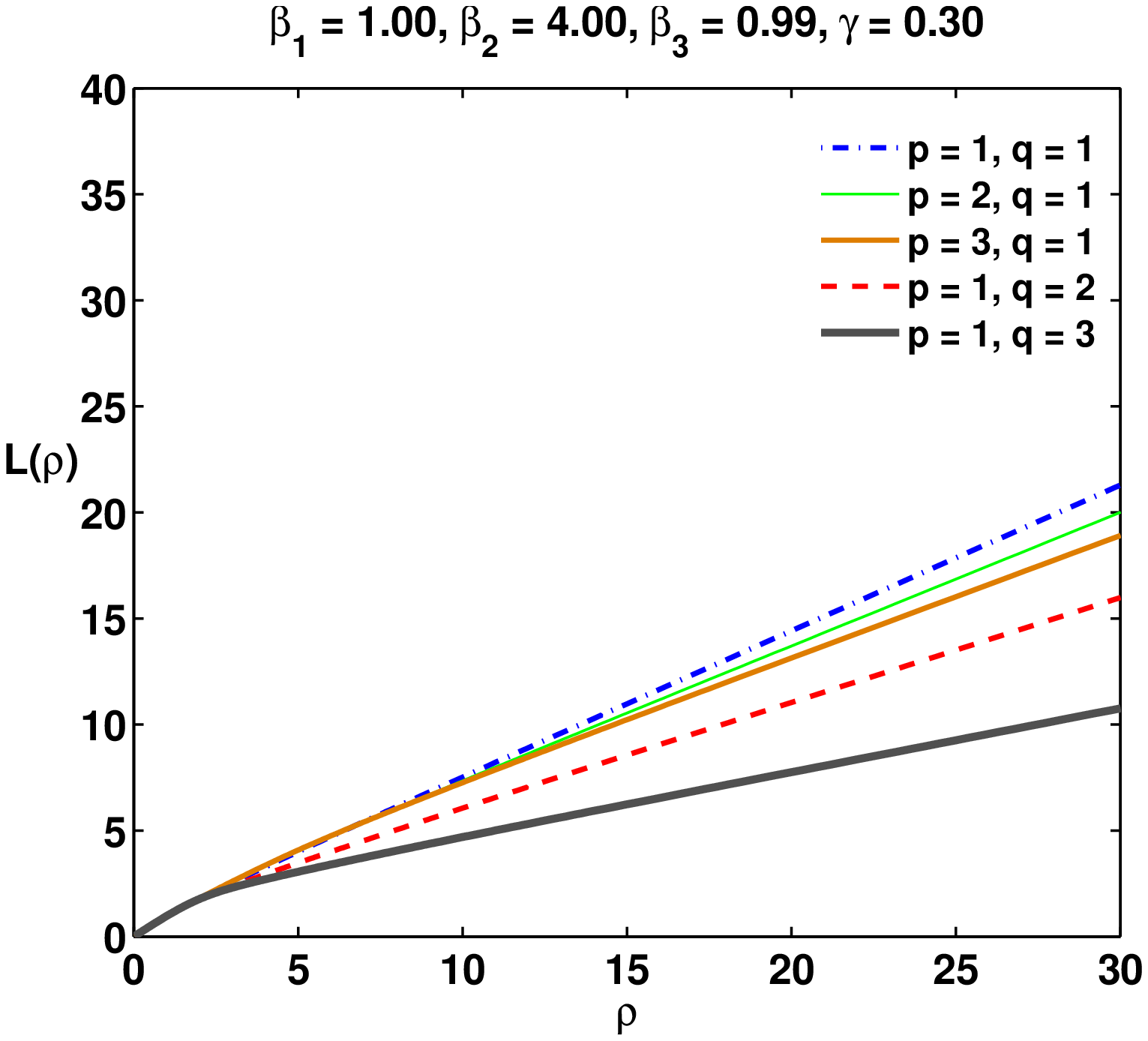}}
    \subfigure[$\beta_1 = 2$, $\beta_2=4.5$, $\beta_3=1.45$]{\label{LII-B}\includegraphics[scale=0.29]{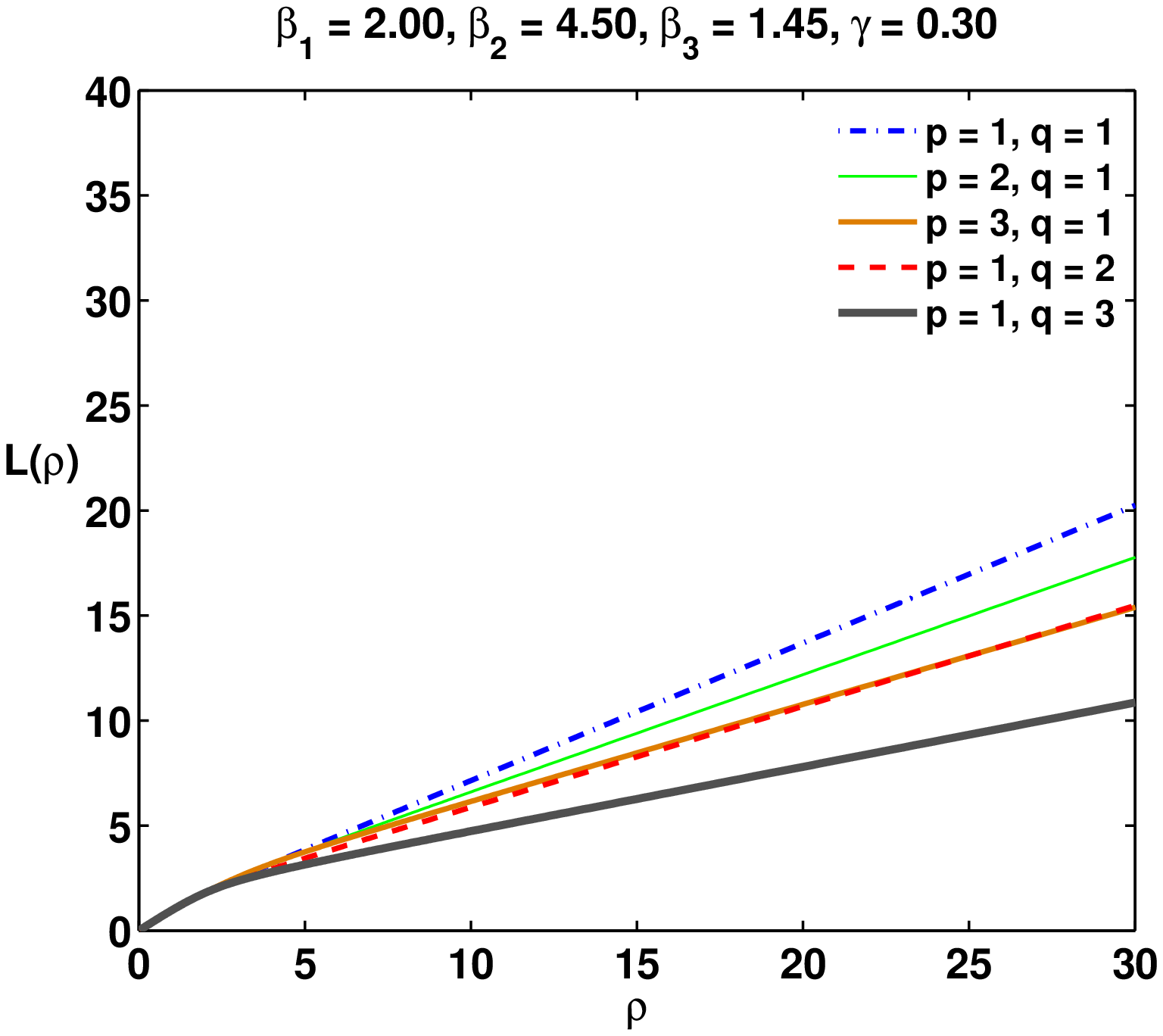}}
     \subfigure[$\beta_1 = 0.5$, $\beta_2=2$, $\beta_3=0.45$]{\label{LI-B}\includegraphics[scale=0.29]{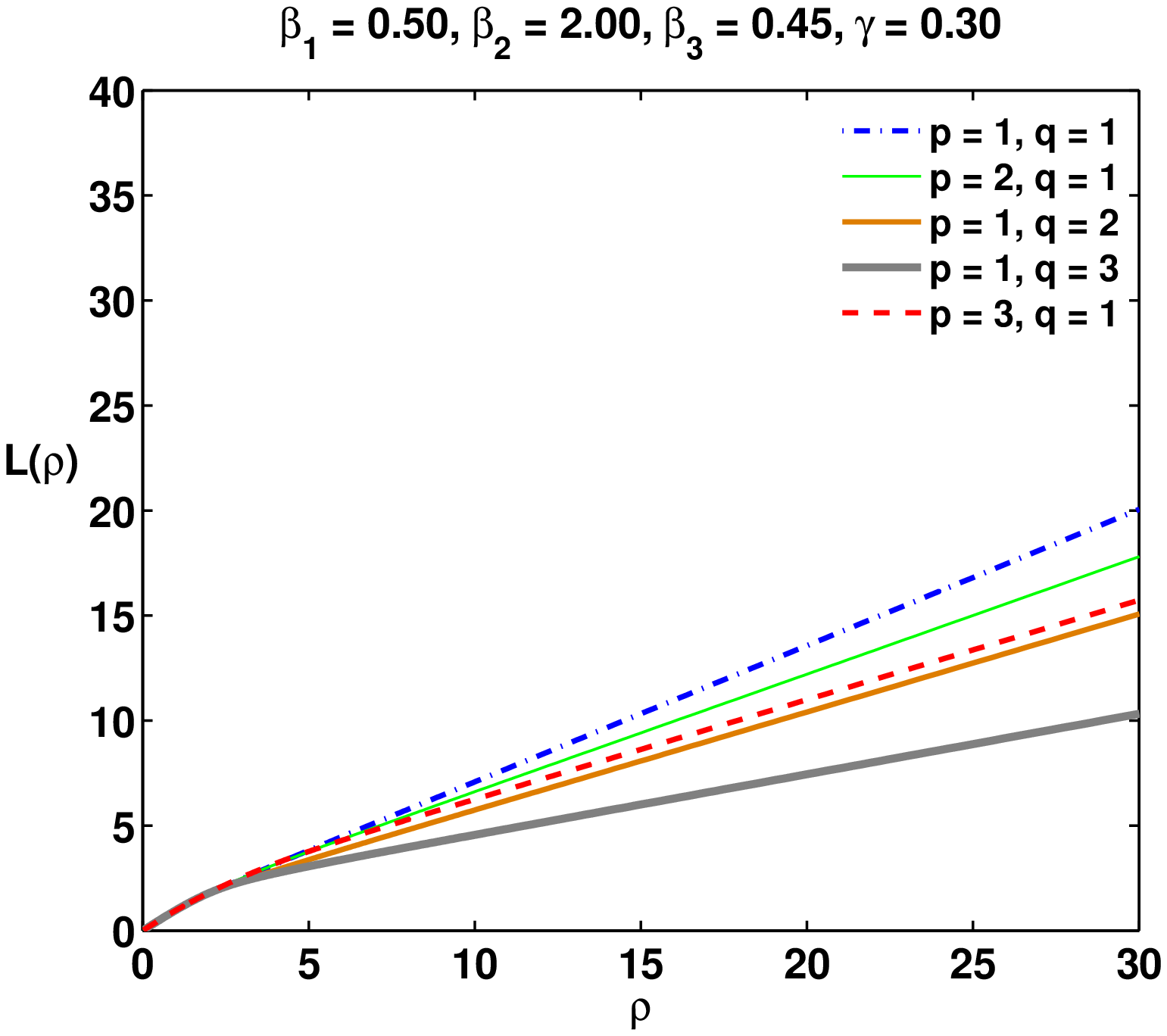}}
   \end{center}
   \caption{The metric function $L(\rho)$ of a (p,q)-string for $\gamma=0.3$, different
choices of $\beta_i$, $i=1,2,3$ and different choices of (p,q)$\equiv (n,m)$.}  \label{Lprofvpq}
  \end{figure}

We observe that the increase in the total magnetic flux along the string increases the asymptotic value of 
the metric function $N(\rho)$ if $N'(\rho) > 0$. The qualitative features do not change. If a minimum of the
metric function exists for $n=m=1$ it exists for all choices of $n$ and $m$ (see Fig.\ref{NI-II})
and if $N'(\rho) > 0$ for $n=m=1$ this will be the same
for other choices of $n$ and $m$.

Note that the profiles of the metric functions $L(\rho)$ for all cases are similar to those for $\beta_3=0$.
The deviation of $L'(\rho\gg 1)=c_2$ from one determines the deficit angle of the space-time and depends on the 
inertial mass per unit length. This is shown for $\gamma=0.3$ and different choices of $n$, $m$ and $\beta_i$ in Fig.\ref{Lprofvpq}.
For solutions with $\beta_1=\beta_2$ the slope of $L(\rho)$ at infinity decreases with increasing sum $n+m\equiv p+q$.
This is natural since an increase in the windings leads to an increase in the mass per unit length and hence to an increase
of the deficit angle. Moreover, for a given sum $p+q$ the solutions with $p=q$ have the biggest slope of $L(\rho)$ at infinity.
This is related to the fact that the solutions with equal winding are bound strongest (see also the results in \cite{hartmann_urrestilla}).

\subsubsection{Classification of solutions} 
The geodesics can be classified according to the test particles
energy $E$, angular momentum $L_z$ and momentum in the direction of the string axis $p_z$. 
Intersection points of $\mathcal{E}$ with the effective potential, i.e. points
where $\mathcal{E}=V_{\rm eff}$ correspond to turning points of the motion.
The maximum and minimum of the effective potential determine the largest, respectively smallest
possible value of $\mathcal{E}$ for bound orbits. The effective potential is determined by the metric functions
$N$ and $L$ as well as the constants of motion. 
Choosing $\beta_i$, $i=1,2,3$, $\gamma$ and $n$, $m$ we find the numerical profiles of $N$ and $L$.
For a given $L_z^2$ (and $\varepsilon=0$ or $\varepsilon=1$) there is an $E^2$ such that the value of $\mathcal{E}$ 
is equal to the maximal value of the effective potential $V_{\rm eff}(\rho)$ and one $E^2$ such that $\mathcal{E}$ is equal to
the minimal value of $V_{\rm eff}(\rho)$. In the former case, the corresponding orbit is an unstable
circular orbit, while in the latter the orbit is a stable circular orbit. 

\begin{figure}[p!]
  \begin{center}
    \subfigure[$\beta_1 =\beta_2 = 2$, $\beta_3=0.1$ , $\gamma= 0.30$]{\label{N2pts}\includegraphics[scale=0.45]{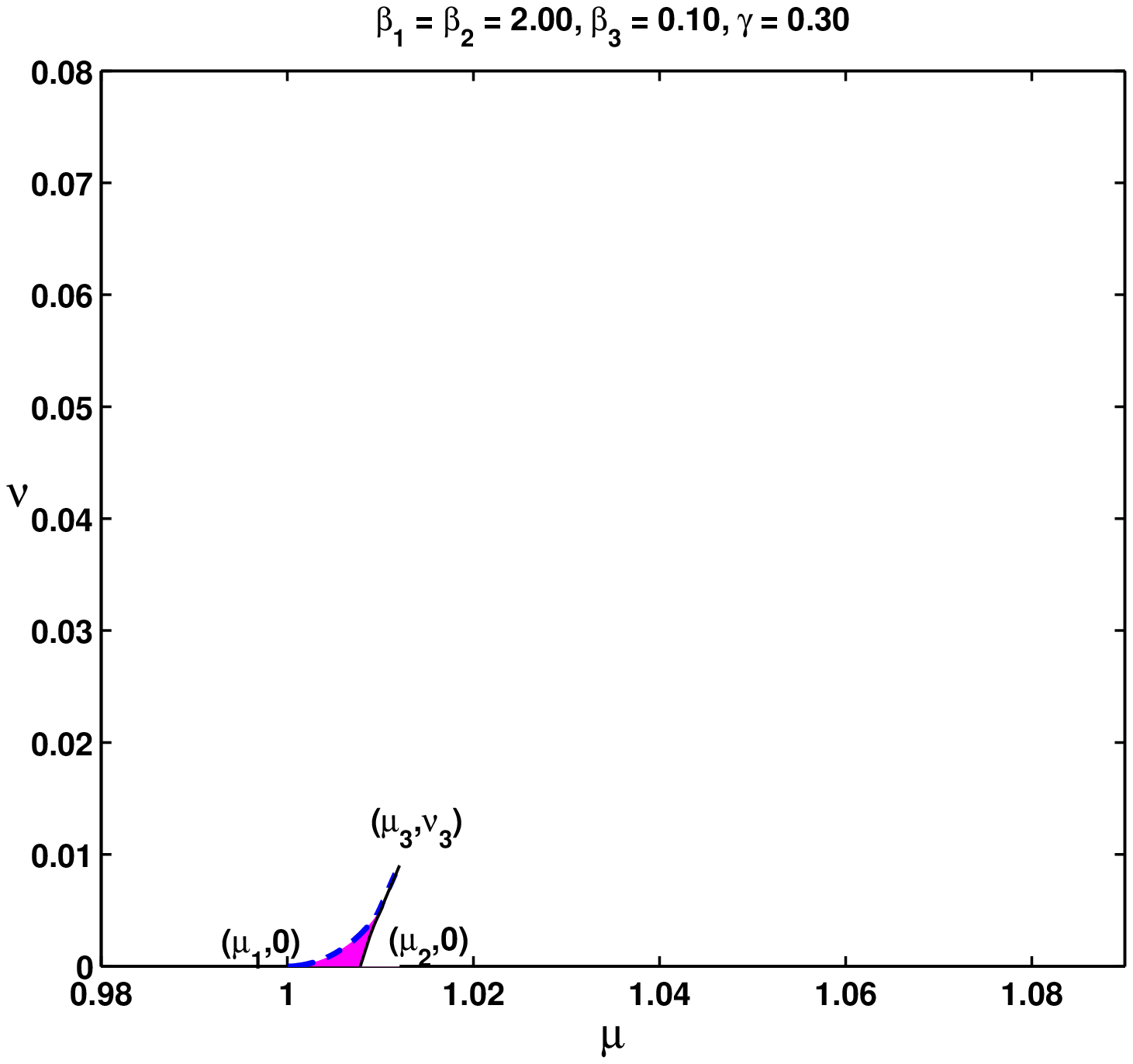}}
    \subfigure[$\beta_1 =\beta_2 = 2$, $\beta_3=0.75$ , $\gamma= 0.30$]{\label{N3pts}\includegraphics[scale=0.45]{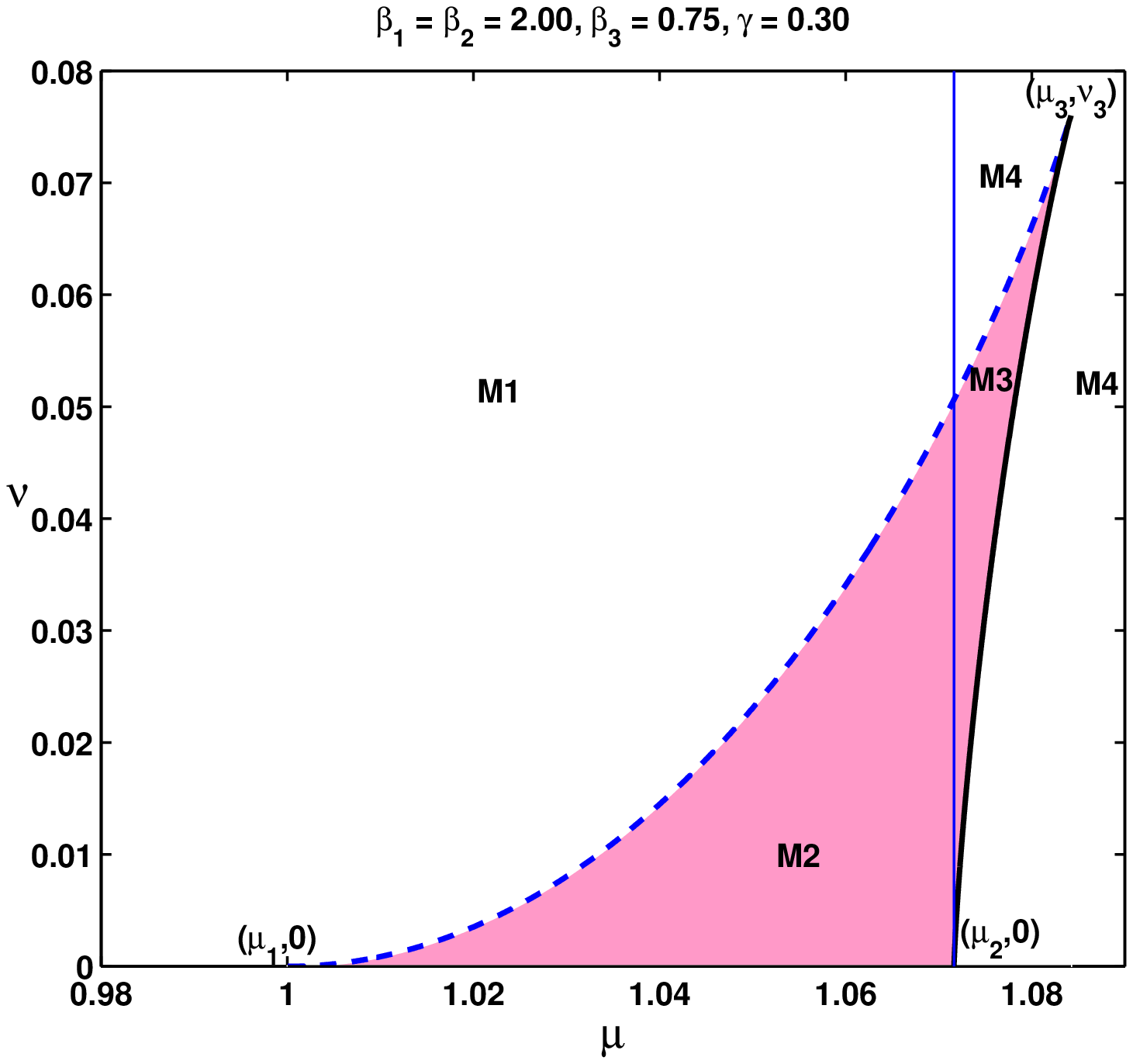}}\\
   \end{center}
  \caption{$\mu$-$\nu$ plot for a (1,1)-string space-time with a monotonically increasing $N(\rho)$. Here $\gamma=0.3$, $\beta_1=\beta_2=2$ 
and $\beta_3=0.1$ (left) and $\beta_3=0.75$ (right), respectively.
The blue dashed and solid black line represent the choice of ($E$, $L_z$, $p_z$) for stable and unstable 
circular orbits, respectively. Here $p_z=0$.}\label{munuplot}
  \end{figure}

\begin{figure}[p!]
  \begin{center}
    \subfigure[$\beta_1= 10$,  $\beta_2 = 3.6$, $\beta_3 = 2.15$, $\gamma = 0.35$]{\includegraphics[scale=0.45]{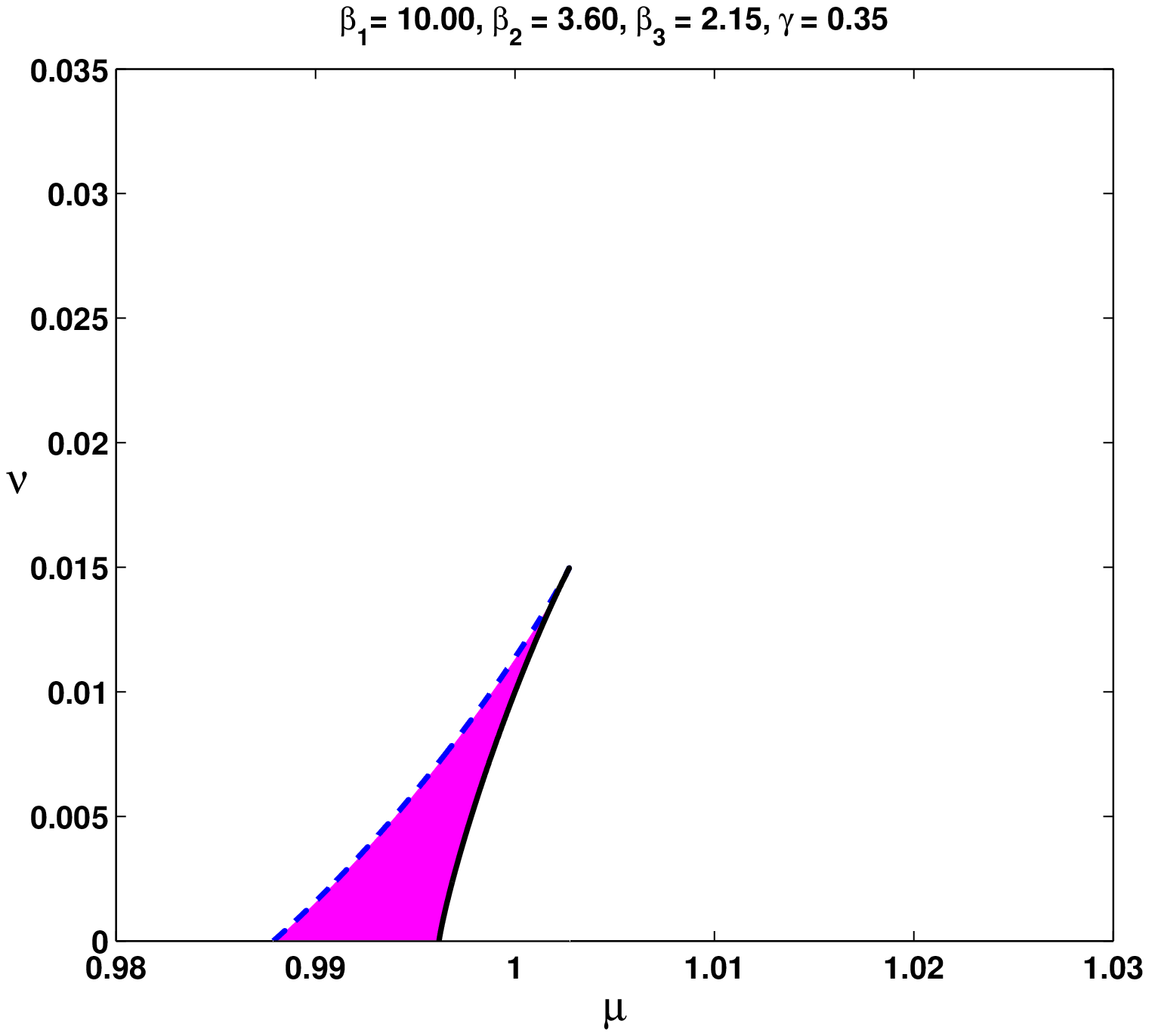}}
    \subfigure[$\beta_1 = 10$, $\beta_2 = 3.6$, $\beta_3 = 2.38$, $\gamma = 0.35$]{\includegraphics[scale=0.45]{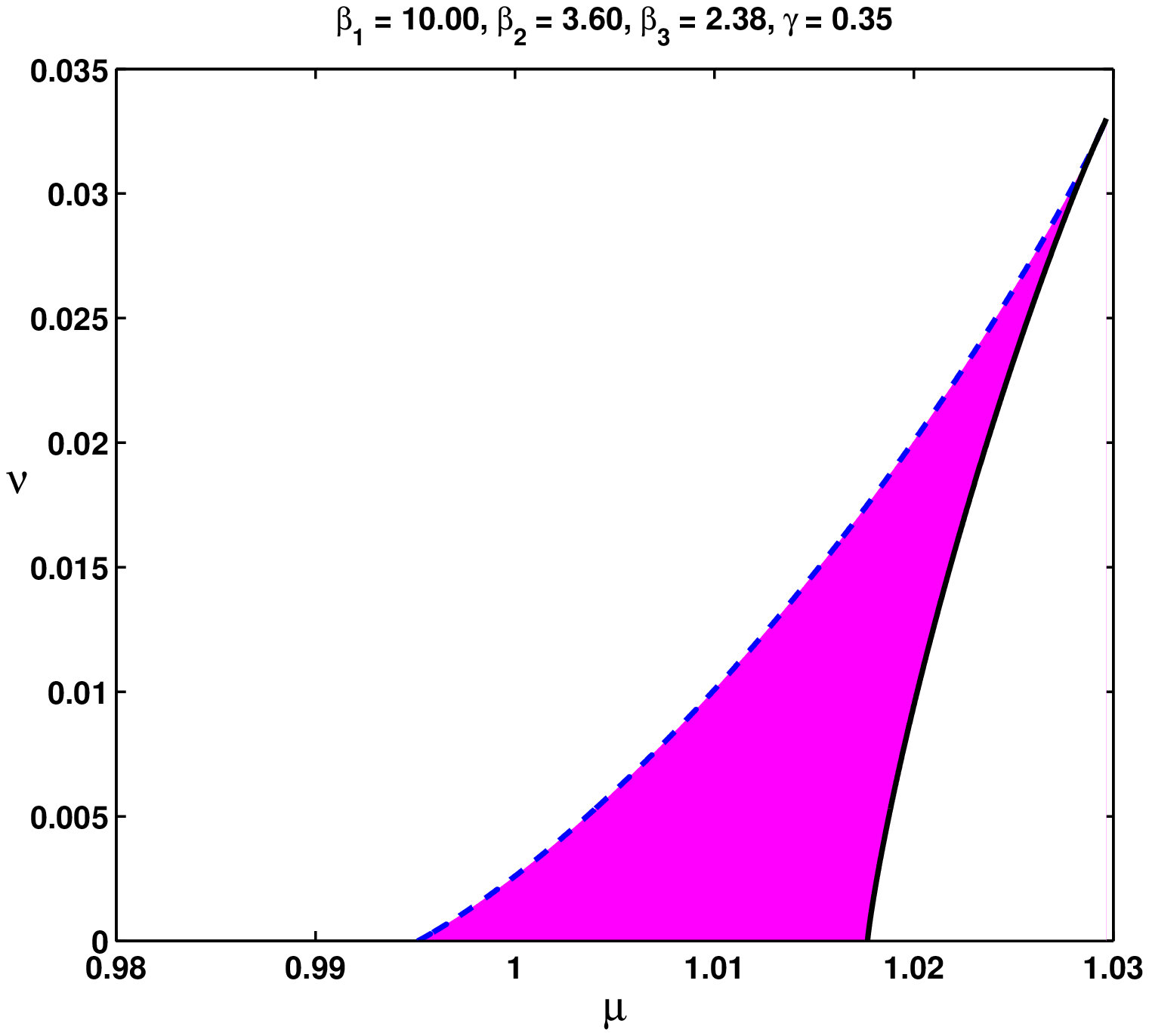}}\\
  \end{center}
  \caption{The $\mu$-$\nu$-plot for a  (1,1)-string space-time which has $N'(\rho)=0$ at some non-vanishing, finite value of $\rho$.
Here $\gamma=0.35$, $\beta_1=10$, $\beta_2=3.6$ and two different values of
$\beta_3=2.15$ (left) and $\beta_3=2.38$ (right), respectively. The blue dashed and solid black line represent the choice of ($E$, $L_z$, $p_z$) for stable and unstable 
circular orbits, respectively. Here $p_z=0$.}
\label{munuplotVlambda3}
  \end{figure}

Defining $\mu$ := $E^2$ and $\nu$ := $L_z^2$ we can then plot the domain of existence of 
bound orbits in the $\mu$-$\nu$-plane.
Our results for massive particles with $p_z=0$ are given in Fig.s \ref{munuplot}, \ref{munuplotVlambda3} for $n=m=1$.
Fig.\ref{munuplot} corresponds to the case
of a (p,q)-string space-time with monotonically increasing $N(\rho)$ and Fig. \ref{munuplotVlambda3} to 
the case of a (p,q)-string space-time which has $N'(\rho)=0$ at some non-vanishing, finite value of $\rho$.
The  blue dashed and solid black line from ($\mu_1$,0) to ($\mu_3$,$\nu_3$) and ($\mu_2$,0) to ($\mu_3$,$\nu_3$) , respectively,
represent the choice of ($E$, $L_z$, $p_z$) for stable and unstable 
circular orbits, respectively, and bound orbits exist in the colored domain between the two bounding curves.
$(\mu_3,\nu_3)$ corresponds to the largest possible values of $\mu$ and $\nu$ for bound orbits.
M1 denotes the domain in the $\mu$-$\nu$-plane in which $\mathcal{E}$ is smaller than the minimum of the effective potential and
hence there are no solutions to the geodesic equation. M4 denotes the domain in which $\mathcal{E}$ is larger than
the maximum of the effective potential and only escape orbits are possible.
In M2 and M3 on the other hand bound orbits are possible. In M2 $\mathcal{E}$ is smaller than the asymptotic value
of the effective potential, but larger than the minimum of $V_{\rm eff}$ and only bound orbits are possible. In M3 
$\mathcal{E}$ is larger than the
asymptotic value of the effective potential but smaller than the maximum of $V_{\rm eff}$. Hence, in M3
there are bound orbits, but escape orbits are also possible.

For $\beta_1=\beta_2=2$ and $\beta_3 > 0$ we find that $\mu_1=1$ for all values of $\beta_3$, while $\mu_2$ as well
as ($\mu_3$,$\nu_3$) increase with increasing $\beta_3$.
While for $\beta_1=\beta_2=2$, $\beta_3=0$ no bound orbits exist at all \cite{hartmann_sirimachan}, bound orbits are possible
for $\beta_3=0.1$ and the domain
of existence of bound orbits in the $\mu$-$\nu$-plane is extending for increasing $\beta_3$ (compare the plots for
$\beta_3=0.1$ and $\beta_3=0.75$). The existence of bound orbits in the limit
where $M_{{\rm H},i}=M_{{\rm W},i}$, $i=1,2$ is new as compared to the $\beta_3=0$ case.

This, however, is not the only difference as compared to the space-time of an Abelian-Higgs string. As stated above we find that
it is possible to have negative valued minima of the effective potential in (p,q)-string space-times. This leads
to the observation that massive test particles with $\mu < 1$ can now move on bound orbits. This is a new feature 
as compared to the $\beta_3=0$ case, where we had to require that $\mu > 1$. 
This means that test particles
with less energy can move on bound orbits in (p,q)-string space-times as compared to the $\beta_3=0$ case, which
corresponds to the space-time of two non-interacting Abelian-Higgs strings. 
This is clearly
seen in Fig. \ref{munuplotVlambda3} for $\gamma=0.35$, $\beta_1=10$, $\beta_2=3.6$ and two different
values of $\beta_3$. While for $M_{{\rm H},i} > M_{{\rm W},i}$, $i=1,2$ and  in the $\beta_3=0$ limit no bound 
orbits exist \cite{hartmann_sirimachan} they
exist in a small domain of the $\mu$-$\nu$-plane for sufficiently large $\beta_3$. The extension 
of the domain in the $\mu$-$\nu$-plane for which
bound orbits exist increases with increasing $\beta_3$, i.e. the values of $\mu_1$, $\mu_2$ and $(\mu_3,\nu_3)$ increase.

\begin{figure}[h!]
  \begin{center}
    \subfigure[$p=q = 1$]{\includegraphics[scale=0.32]{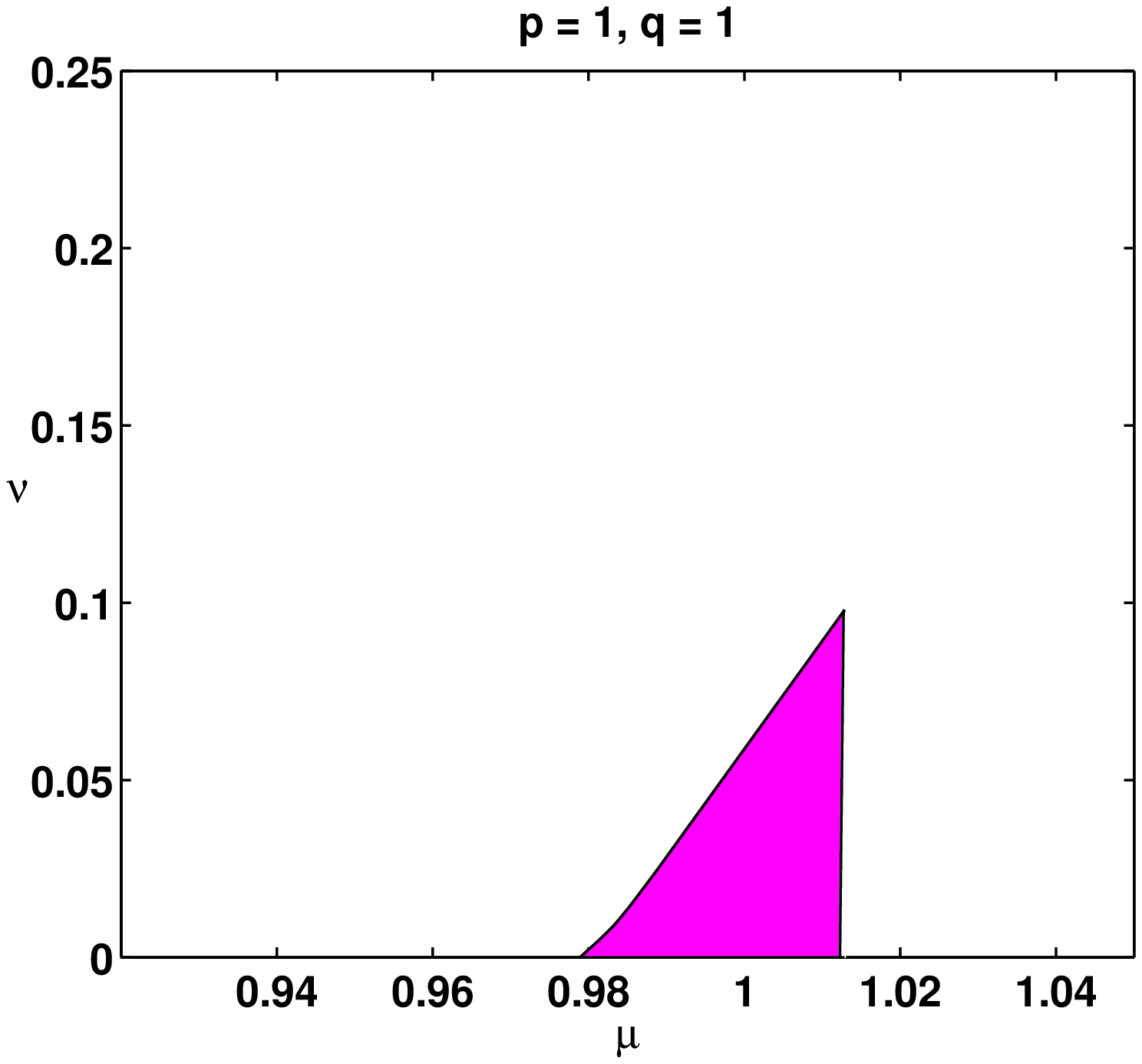}}
    \subfigure[$p=2$, $q = 1$]{\includegraphics[scale=0.32]{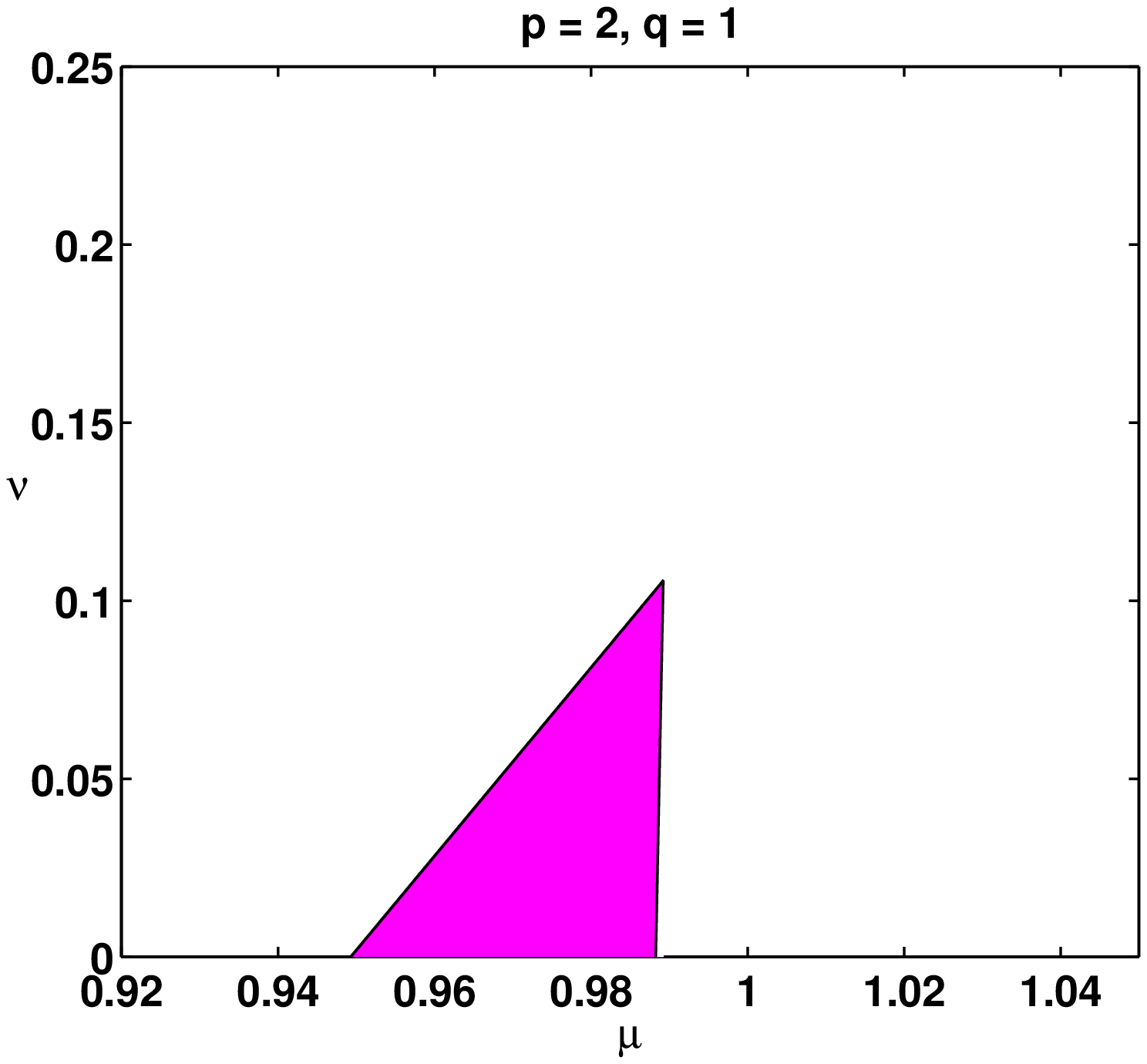}}
    \subfigure[$p = 3$, $q = 1$]{\includegraphics[scale=0.32]{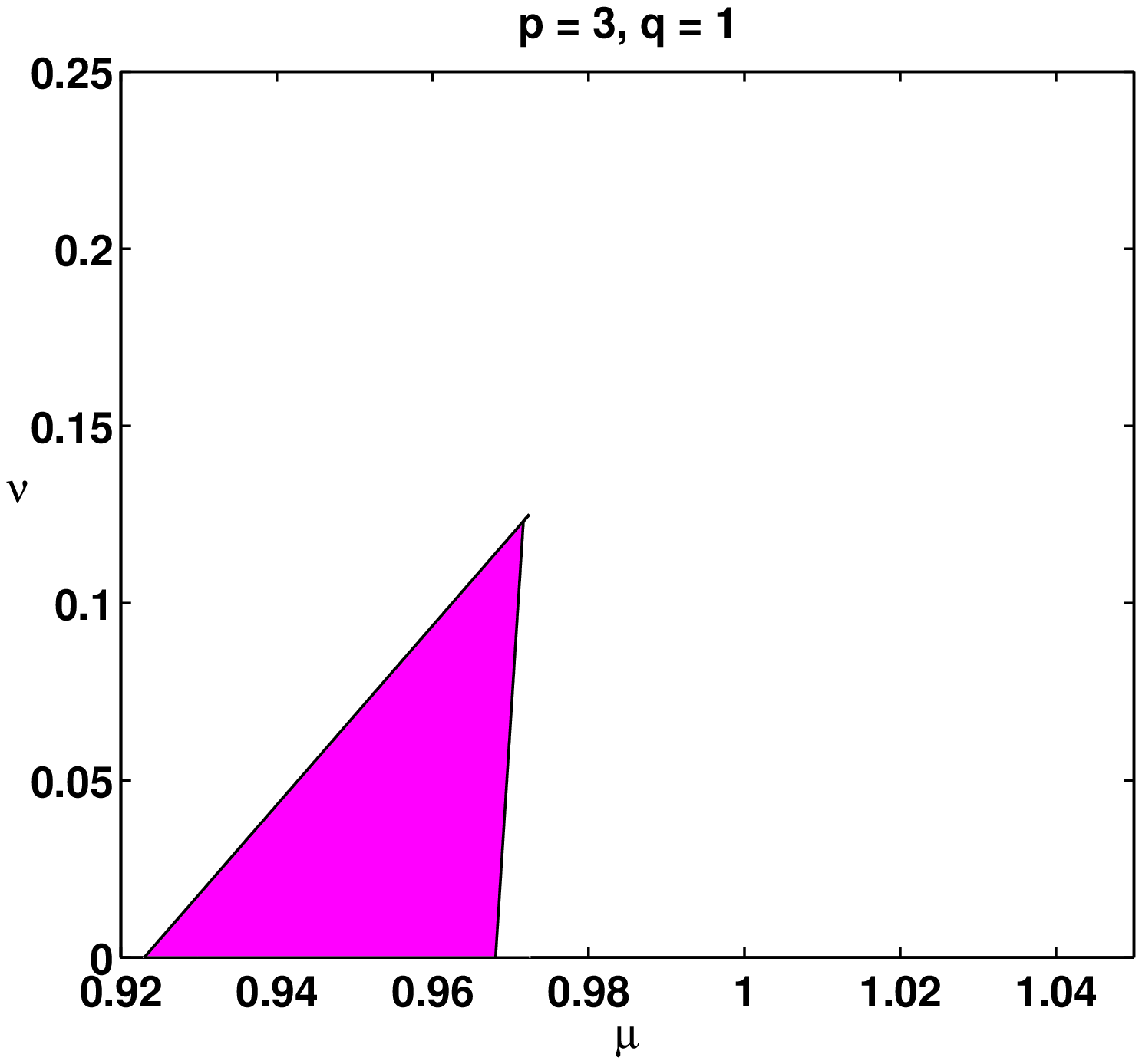}}\\
    \subfigure[$p = 1$, $q = 2$]{\includegraphics[scale=0.32]{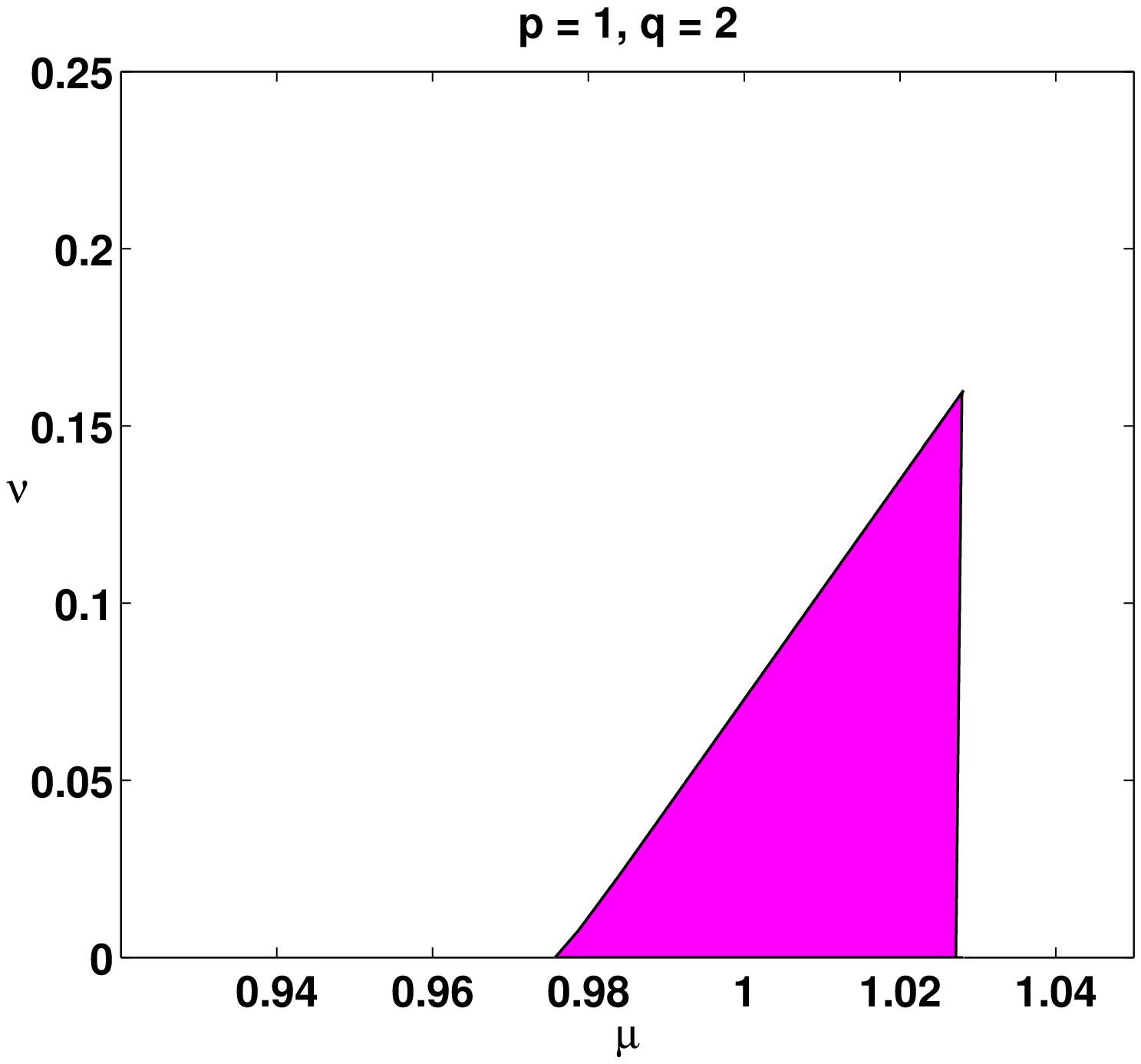}}
    \subfigure[$p = 1$, $q = 3$]{\includegraphics[scale=0.32]{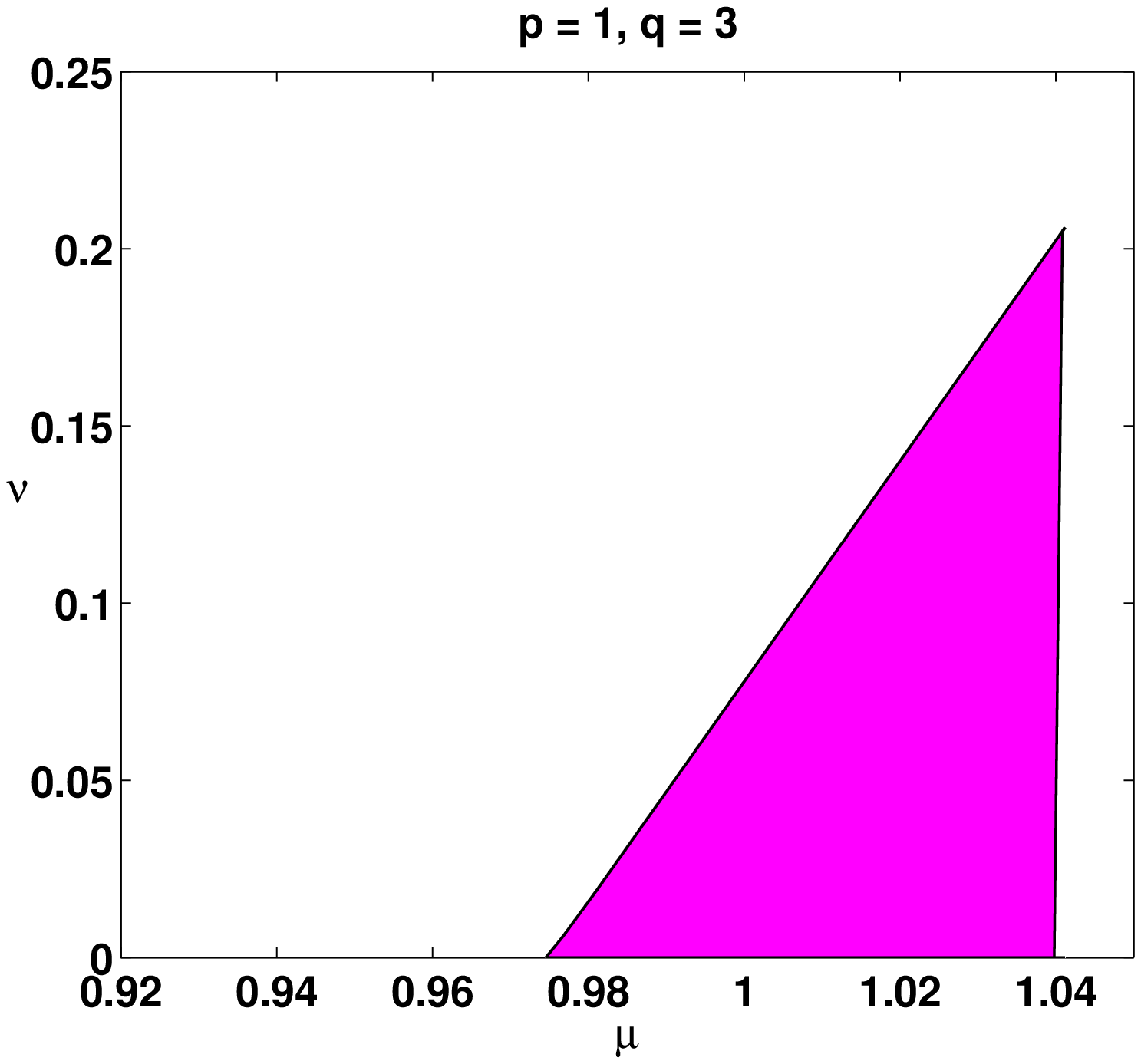}}
    \subfigure[$p = 2$, $q = 3$]{\includegraphics[scale=0.32]{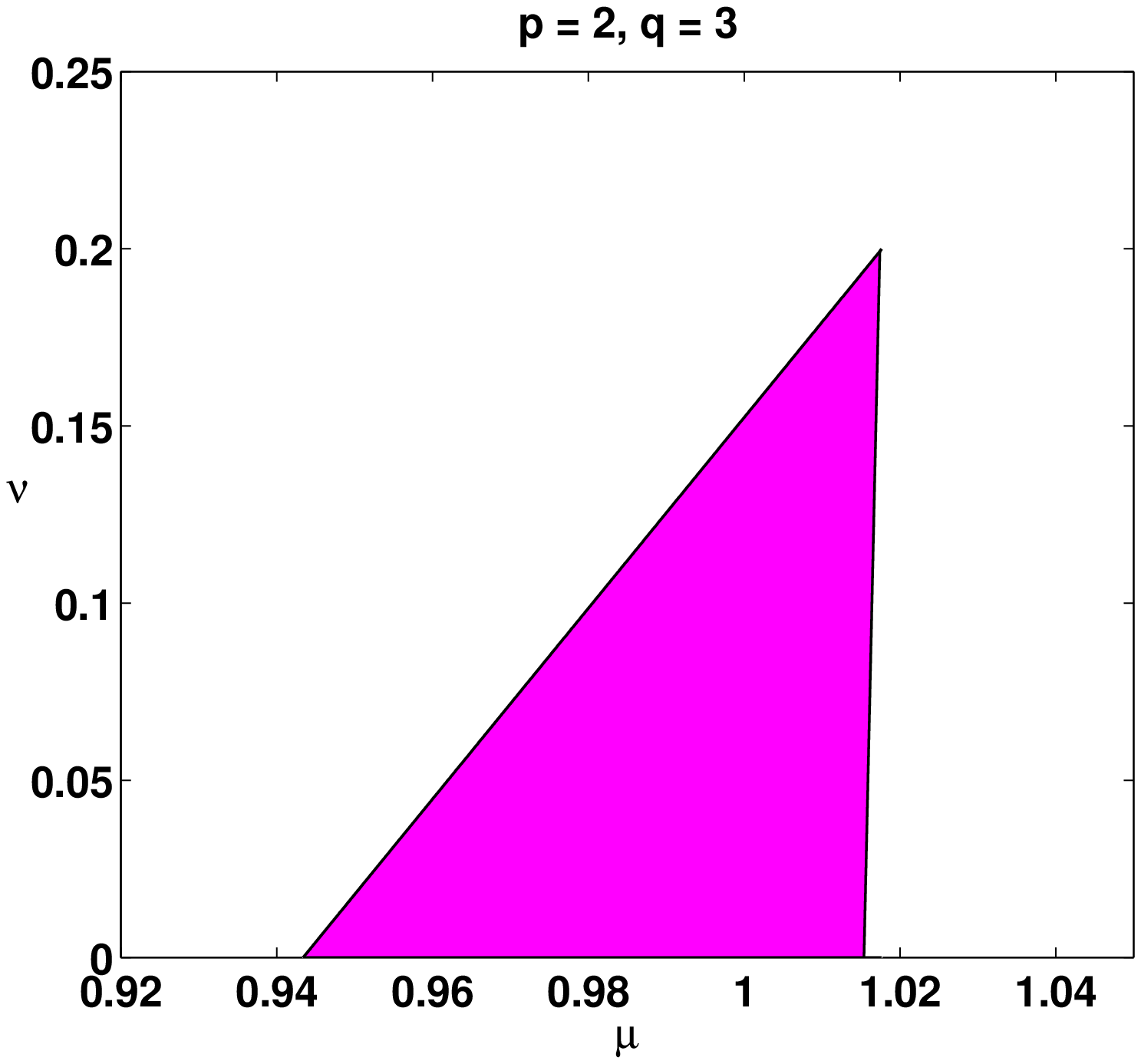}}\\
   \end{center}
  \caption{The $\mu$-$\nu$-plot for a (p,q)-string with $\gamma=0.2$, $\beta_1 = 8$, $\beta_2 = 0.5$, $\beta_3= 0.99$
and different choices of (p,q)$\equiv (n,m)$.}\label{munuvpq}
  \end{figure}

The change of the $\mu$-$\nu$-plot of a (p,q)-string with 
$\gamma=0.2$, $\beta_1 = 8$, $\beta_2 = 0.5$, $\beta_3= 0.99$ resulting from the change of the winding
numbers (p,q)$\equiv (n,m)$ and hence the change of the magnetic fluxes along the 
(p,q)-string are shown in Fig.\ref{munuvpq}. 
Here we concentrate on the case of a string with $\rho_{{\rm H},1} < \rho_{{\rm W},1}$ (the p-string) interacting
with a string that has $\rho_{{\rm H},2} >  \rho_{{\rm W},2}$ (the q-string). 
Increasing the winding of the p-string while keeping the winding of the q-string fixed 
shifts $\mu_1$ and $\mu_2$ to lower values, while
the difference $\mu_2-\mu_1$ slightly increases with increasing p. Hence bound orbits are possible in a 
slightly bigger domain of the $\mu$-$\nu$-plane and in particular test particles need less energy to
be able to move on bound orbits when increasing the winding of the p-string. 
On the other hand, increasing the winding of the q-string while keeping the winding of the p-string fixed
increases the value of $\mu_2$, while $\mu_1$ is nearly constant. Again, the domain of existence of
bound orbits becomes larger when increasing the winding of the q-string.

For $p_z\neq 0$, the qualitative features are the same. We observe, however, that the whole domain
of existence of bound orbits shifts to larger values of $\mu$ when increasing $p_z$. This
is obviously related to the fact the the effective potential $V_{\rm eff}$ is energy-dependent
(see (\ref{effective_pot})).

In contrast to massive test particles, we find that massless particles can only move on escape orbits. This is
very similar to what has been observed in the $\beta_3=0$ limit \cite{hartmann_sirimachan} and 
agrees with the result found in \cite{gibbons} 
which states that for a general cosmic string space--time with topology $\mathbb{R}^2\times \Sigma$
massless test particles must move on geodesics that escape to infinity in both directions, i.e.
closed geodesics are not possible.
The assumption made in \cite{gibbons} is that $\Sigma$ must have positive Gaussian curvature.
To show that $\Sigma$ has positive Gaussian curvature in our case, we rewrite the metric
(\ref{metric}) for massless particles ($ds^2=0$) moving in a plane parallel to the $x$-$y$-plane
as follows
\begin{equation}
 dt^2=\frac{1}{N^2} d\rho^2+ \frac{L^2}{N^2}d\varphi^2 = \tilde{g}_{ij} dx^i dx^j \ \  ,  \ \ i=1,2
\end{equation}
where $\tilde{g}_{ij}$ is the so-called optical metric \cite{acl} of which the spatial
projection of geodesics of massless particles, i.e. light rays are geodesics. 
$\tilde{g}_{ij}$ is the metric of the above mentioned 2-manifold $\Sigma$ and has Gaussian
curvature $K$ given by
\begin{eqnarray}
\label{gauss}
 K=\frac{L'}{L} N' N - \frac{L''}{L} N^2 - (N')^2 + N N''   \ .
\end{eqnarray}

For $\beta_1=\beta_2=2$ and $\beta_3=0$ (the BPS limit) we know that $N\equiv 1$ and the 
Gaussian curvature is obviously positive,
away from the BPS limit one has to use the numerical solution and compute the curvature.
We find that for most values of $\beta_1$, $\beta_2$, $\beta_3$ and $\gamma$
the Gaussian curvature is indeed positive and our result is in agreement with that of \cite{gibbons}. 
However, if $\beta_1$ and $\beta_2$ are sufficiently
large and $\beta_3$ sufficiently small, we find that $K$ can become negative close to the string axis. 
Though the theorem of \cite{gibbons} is not applicable here,
we nevertheless find that bound orbits do not exist.

\subsubsection{Examples of orbits}

\begin{figure}[p!]
\centering

\subfigure[$\beta_3 = 2.1$]{
\includegraphics[scale=0.8]{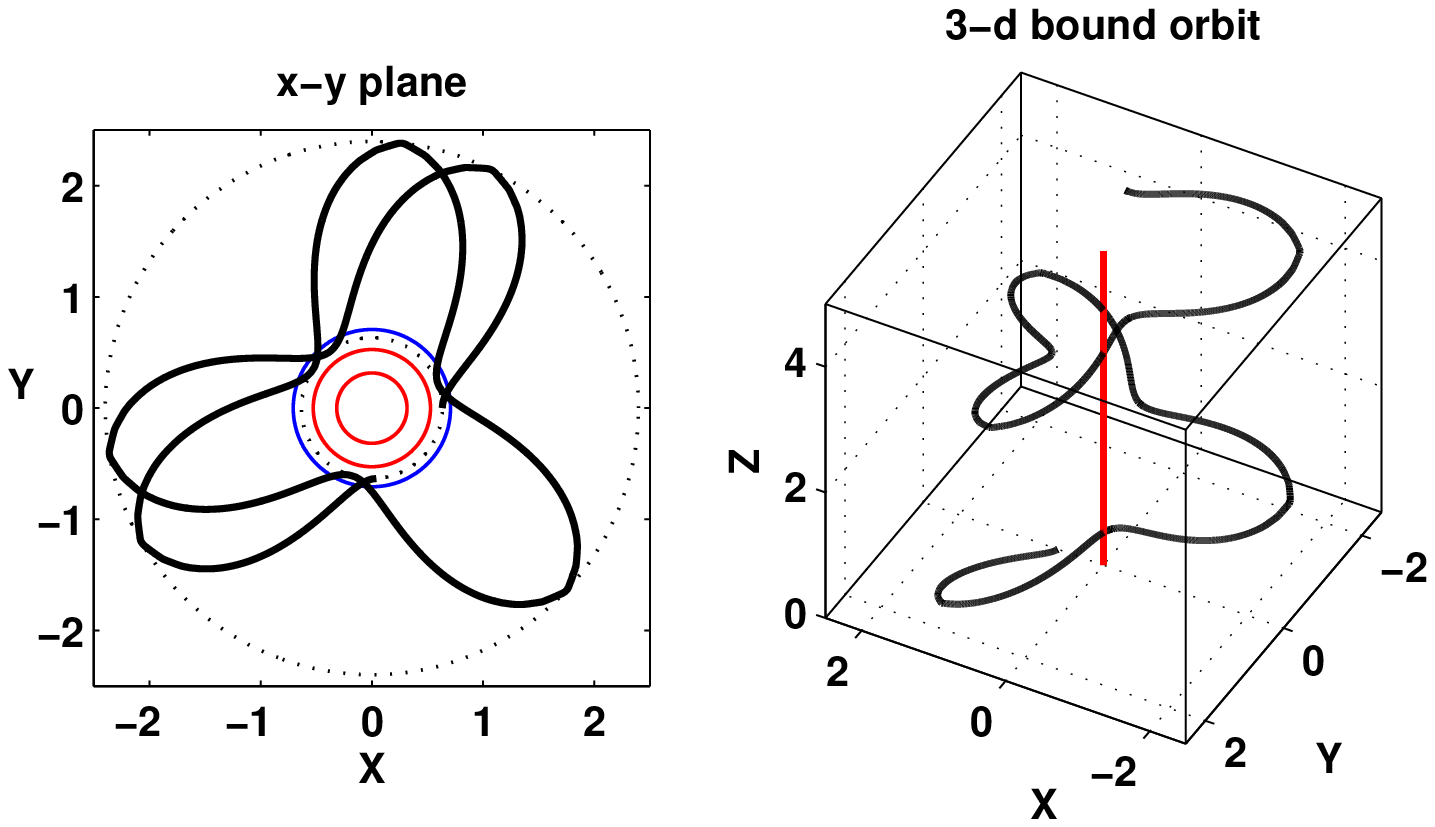}
}

\subfigure[$\beta_3 = 2.15$]{
\includegraphics[scale=0.8]{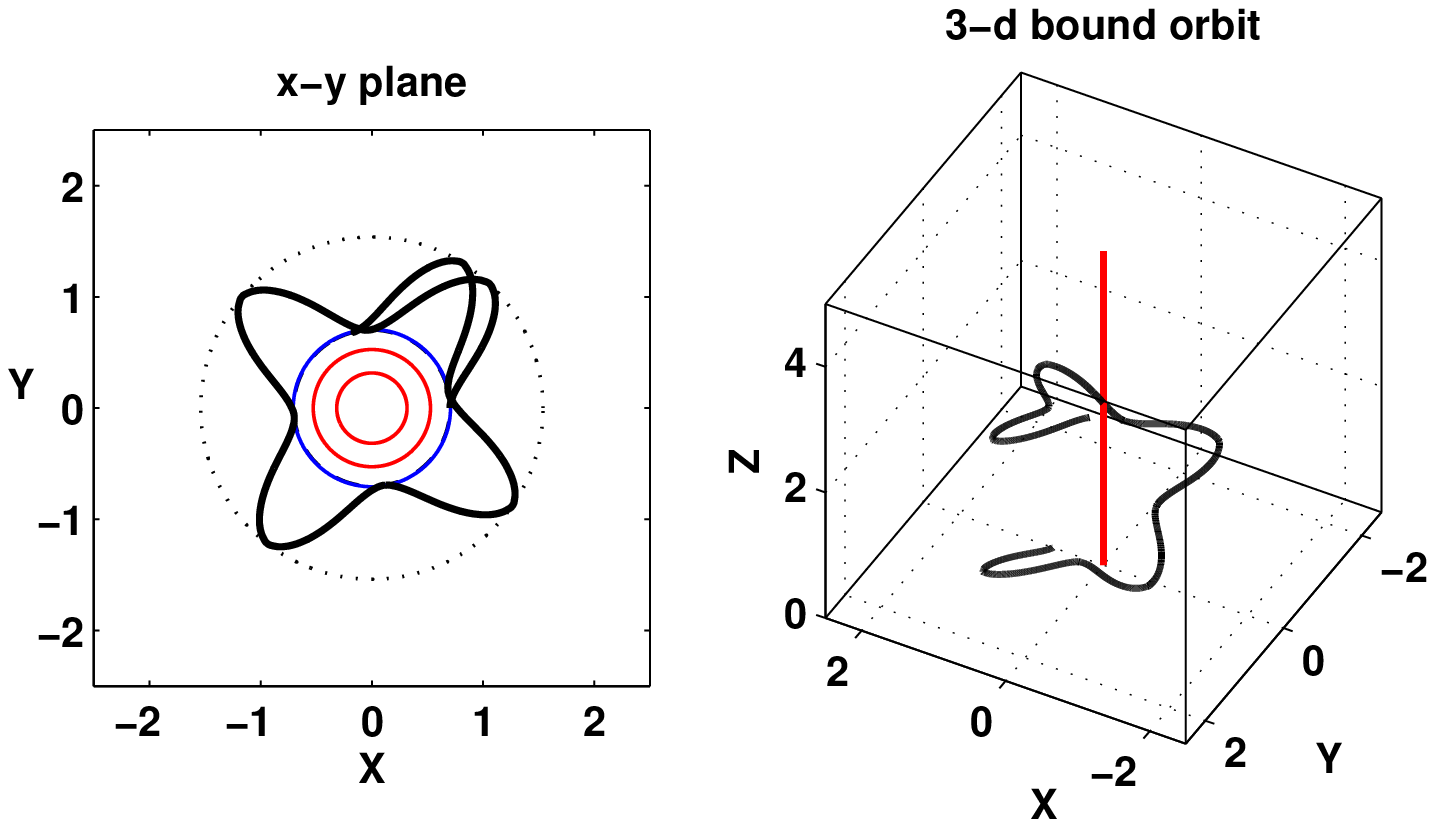}
}
\subfigure[$\beta_3 = 2.2$]{
\includegraphics[scale=0.8]{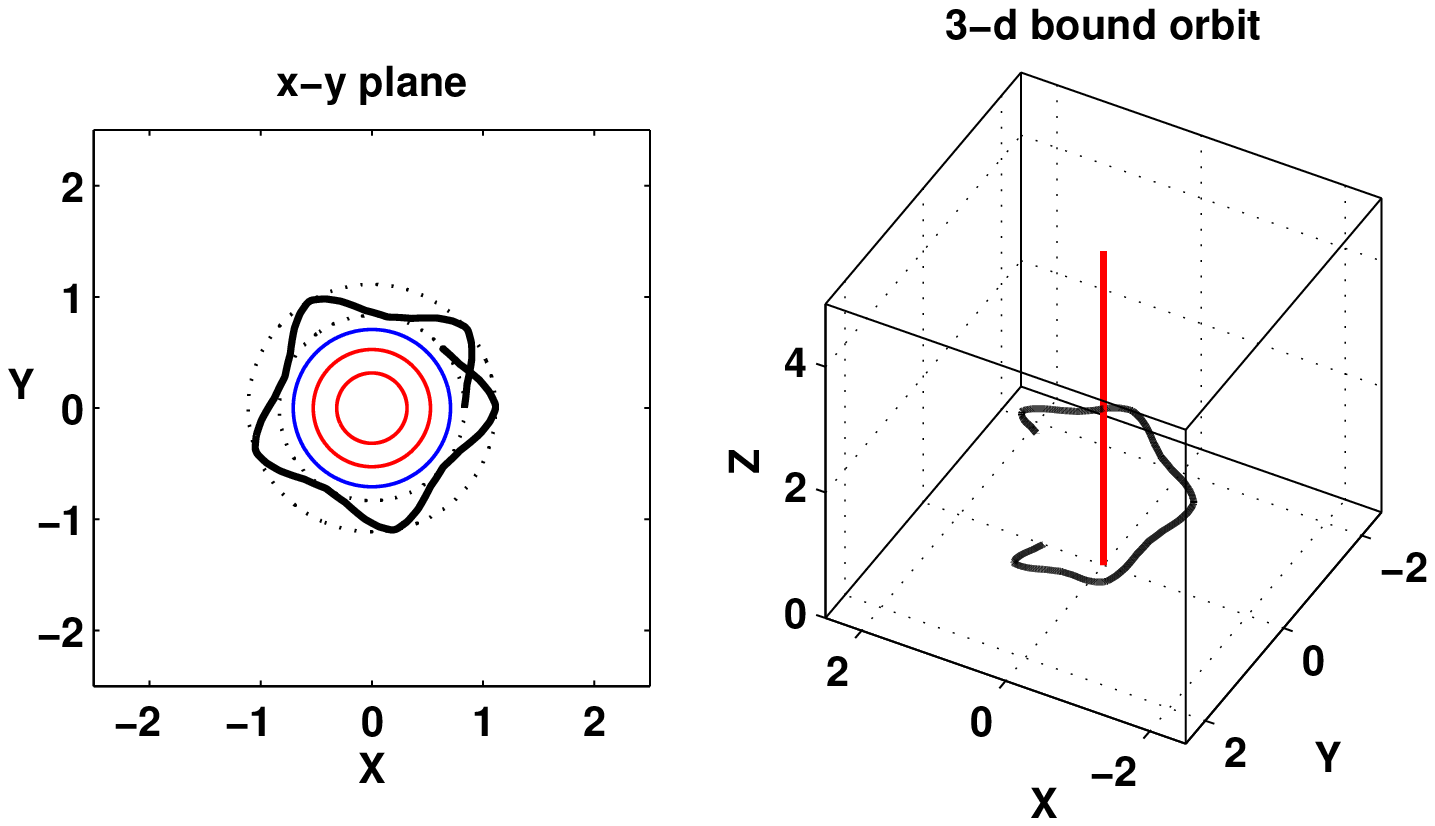}
}

\caption[Optional caption for list of figures]{The bound orbit of a massive
test particle with $E=0.995$, $L_z= 0.022$ and $p_z = 0.011$ in the space-time of a (1,1)-string
with $\gamma=0.3$, $\beta_1 = 10.00$, $\beta_2 = 3.6$ and different choices of $\beta_3$. 
The dotted circles denote the minimal and the maximal radius of the bound orbit, while
the blue and the red circles indicate the radius of the gauge and scalar field cores, respectively. }\label{msbBoundL110L2360G035vL3}
\end{figure}

\begin{figure}[p!]
\centering

\subfigure[$\beta_3 = 2.15$]{
\includegraphics[scale=0.8]{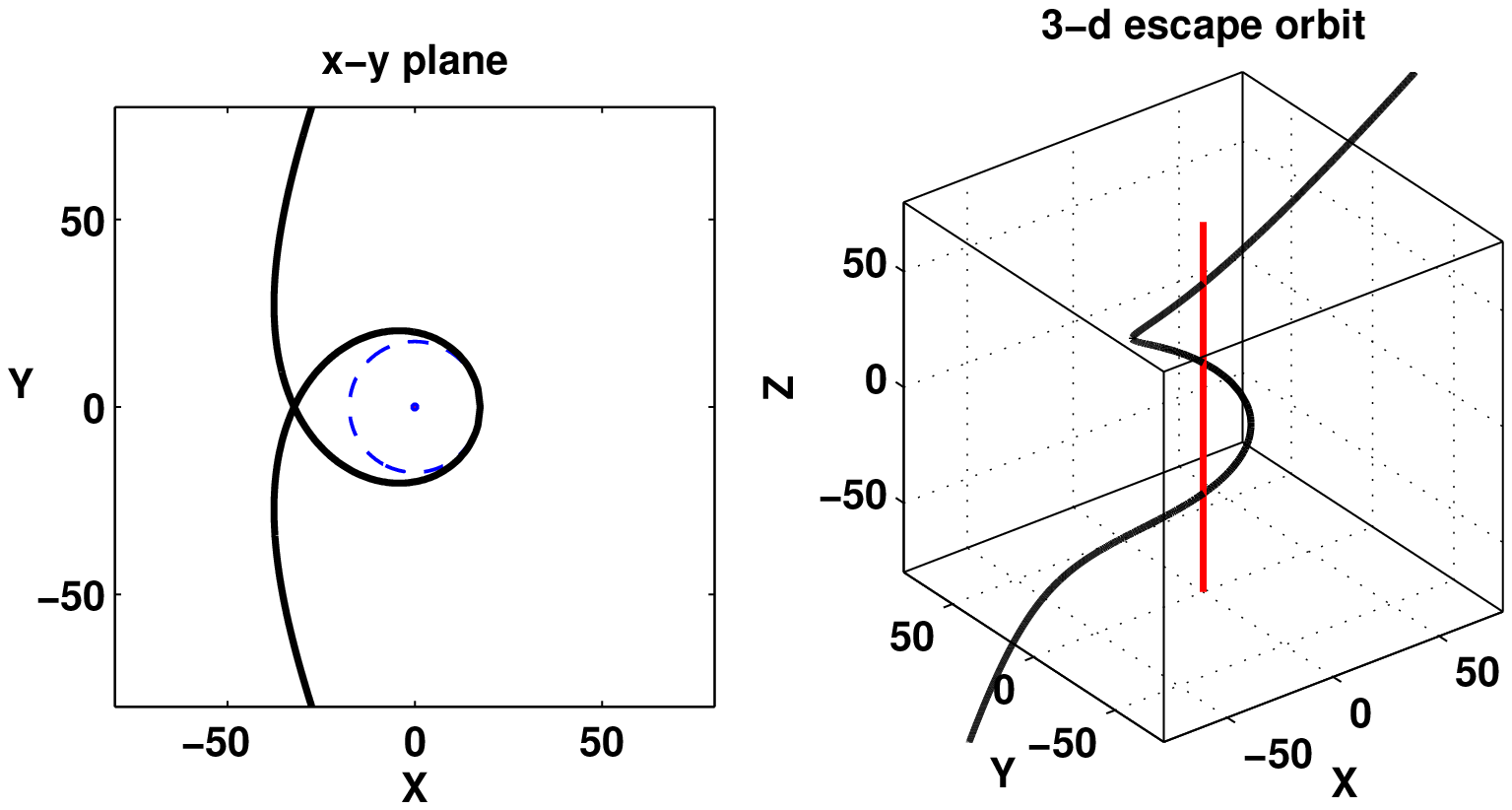}
}

\subfigure[$\beta_3 = 2.48$]{
\includegraphics[scale=0.8]{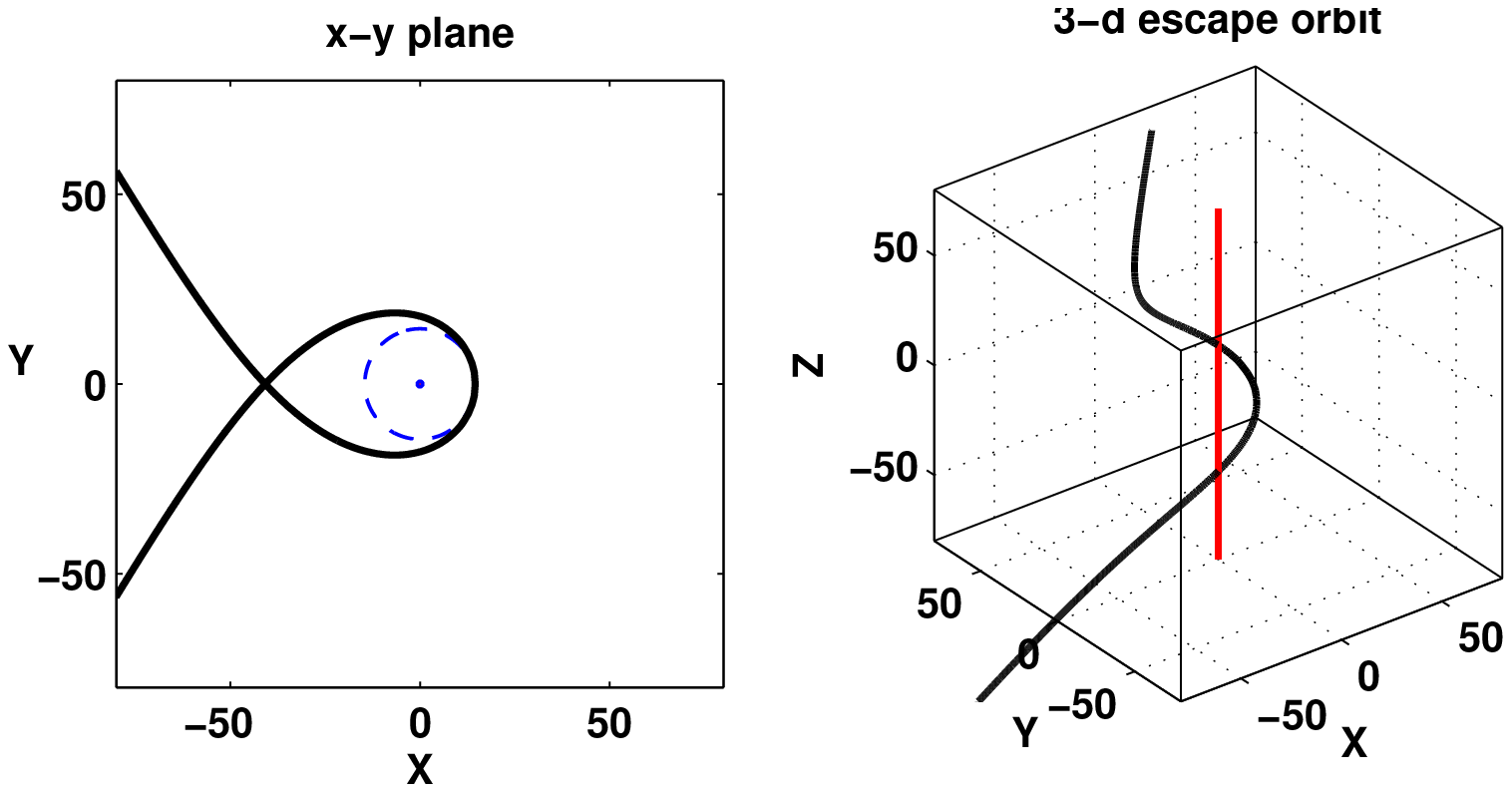}
}
\subfigure[$\beta_3 = 2.96$]{
\includegraphics[scale=0.8]{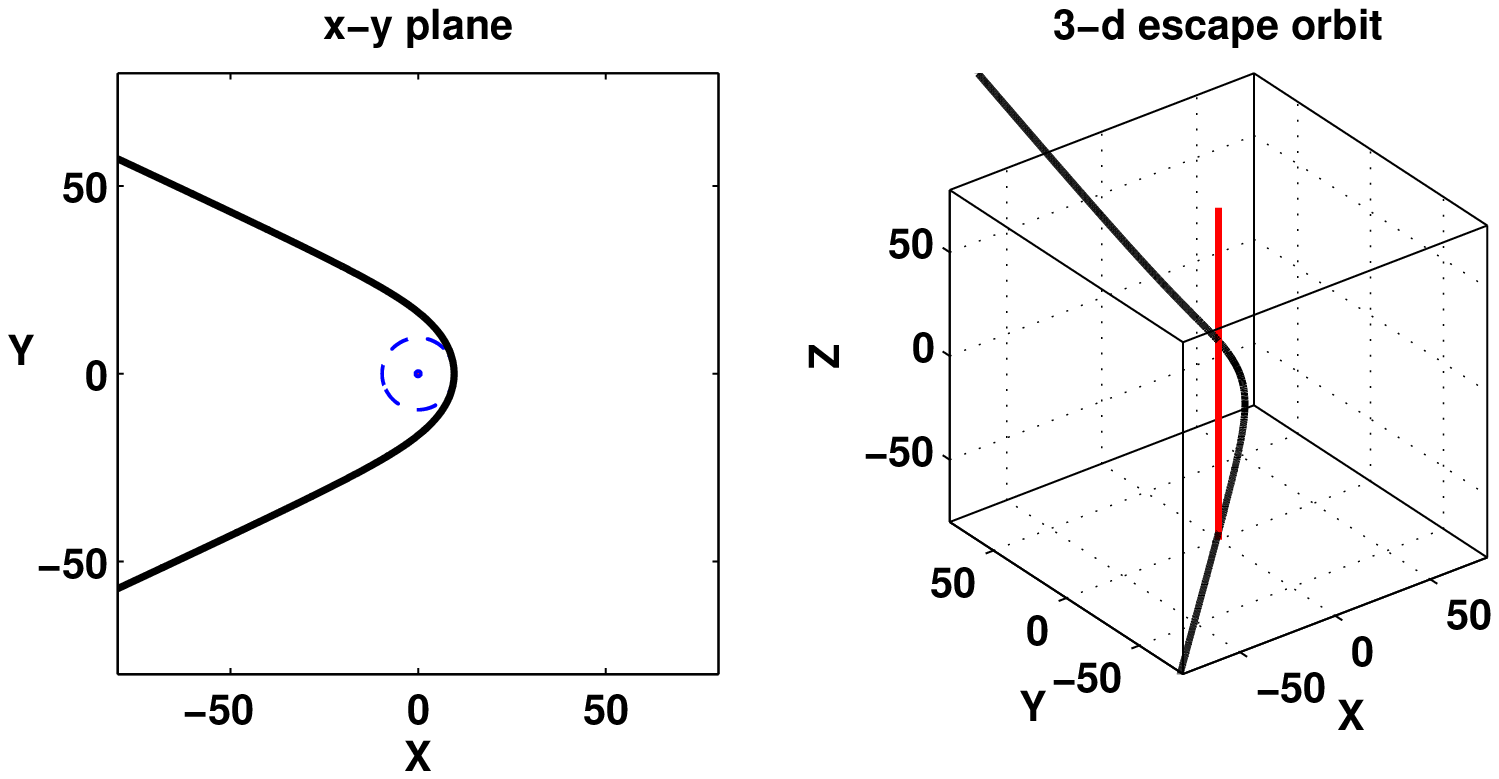}
}

\caption[Optional caption for list of figures]{
The escape orbit of a massive
test particle with $L_z = 0.02$, $E = 0.06$ and $p_z = 0.05$ in the space-time of a (1,1)-string
with $\gamma=0.35$, $\beta_1 = 10$, $\beta_2 = 3.6$ and different choices of $\beta_3$. 
The blue dot denotes the core of the string, while the dashed blue line denotes the circle with minimal
radius of the orbit, i.e. closest approach of the particle to the string.
}\label{mlbBoundL110L2360G035vL3}
\end{figure}

In Fig.\ref{msbBoundL110L2360G035vL3} we show how a massive test particle with $E=0.995$, $L_z=0.022$ and $p_z=0.011$
moves around a (1,1)-string with $\gamma=0.35$, $\beta_1=10$, $\beta_2=3.6$ and different choices of $\beta_3$.
Note that the orbit is not planar due to the fact that the test particle has momentum
in $z$-direction (see (\ref{zint})).
The red and blue circles indicate the width of the scalar cores and the gauge field cores, while the
dotted circles denote the minimal and maximal radius of the orbit. For our choice
of parameters the width of the gauge field cores is larger than that of the scalar cores. We observe that
the larger $\beta_3$ the closer the test particle moves around the string. For $\beta_3=2.1$, the orbit
extends out to roughly four times the radius of the gauge field core, while for $\beta_3=2.2$ the maximal radius is only roughly
twice that of the gauge field core. Stating it differently: for smaller $\beta_3$ the test particle moves mainly
in the exterior vacuum region of the string, while for larger $\beta_3$ it moves mainly close to or inside
the string core, where the matter fields are non-trivial.
As stated above, bound orbits are only possible in a limited domain of the $\mu$-$\nu$-plane. Test particles
with values of $E$ and $L_z$ outside of this domain will not be able to move on
a bound orbit around the string and will escape to infinity. An example of such an escape orbit
of a massive test particle is shown in Fig.\ref{mlbBoundL110L2360G035vL3}. Since the radii of the
scalar and gauge field cores are very small in comparison to the extension of the orbit, we denote the
core by a blue dot. The blue dashed line indicates the minimal radius of the orbit. 
We observe that the test particle comes from infinity, encircles the string core once
and then moves again away to infinity for $\beta_3=2.15$ and $\beta_3=2.48$. For $\beta_3=2.96$ the particle
gets simply deflected by the string without encircling it. In fact, 
the deflection is decreasing for increasing $\beta_3$. This can be explained by the fact that the energy per unit
length and hence the deficit angle decreases with increasing $\beta_3$.

\begin{figure}[htp]
\centering

\subfigure[$p=q=1$]{
\includegraphics[scale=0.8]{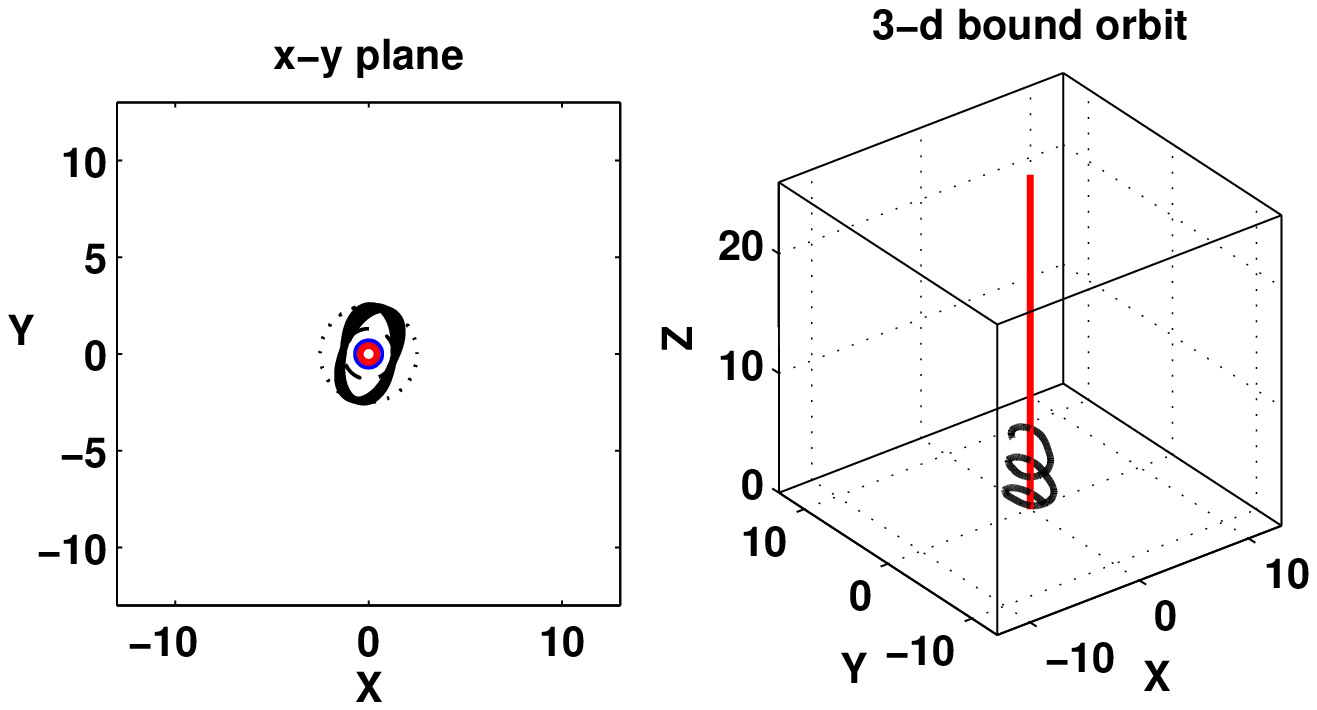}
}

\subfigure[$p = 2$, $q = 1$]{\label{orbitMsvPQ(b)}
\includegraphics[scale=0.8]{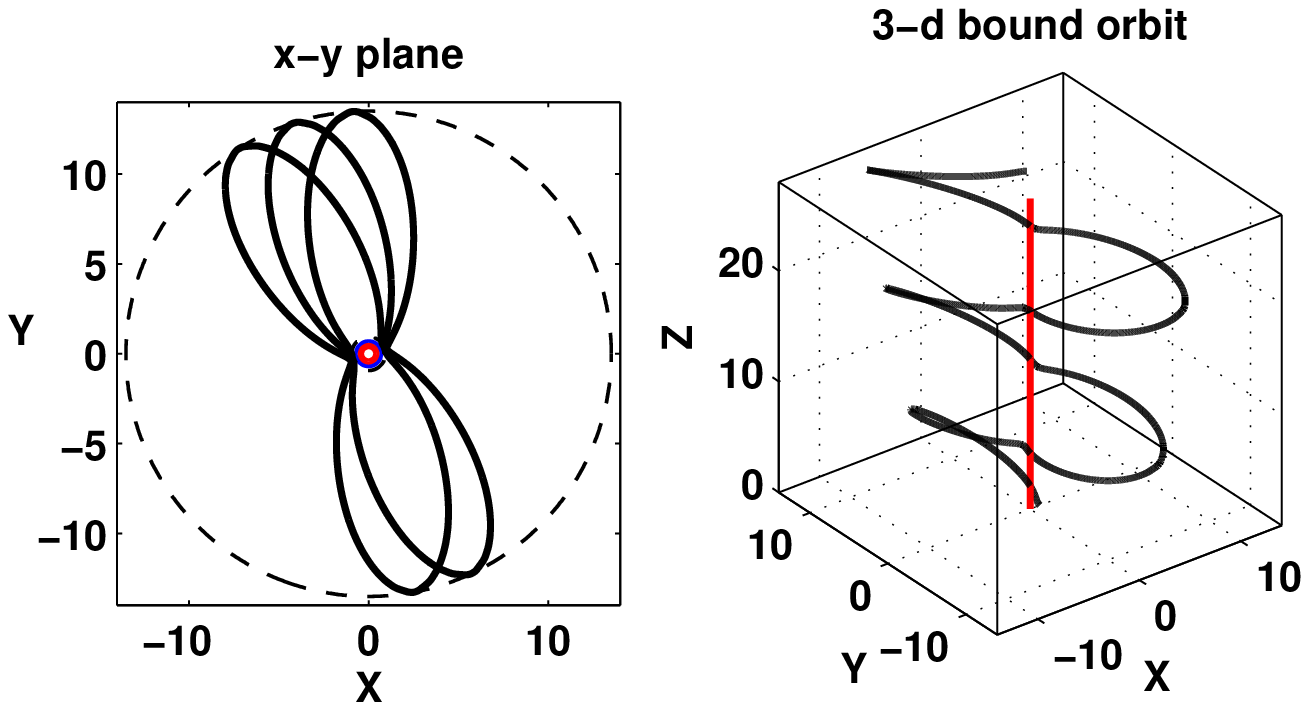}
}
\subfigure[$p = 1$, $q = 2$]{\label{orbitMsvPQ(c)}
\includegraphics[scale=0.8]{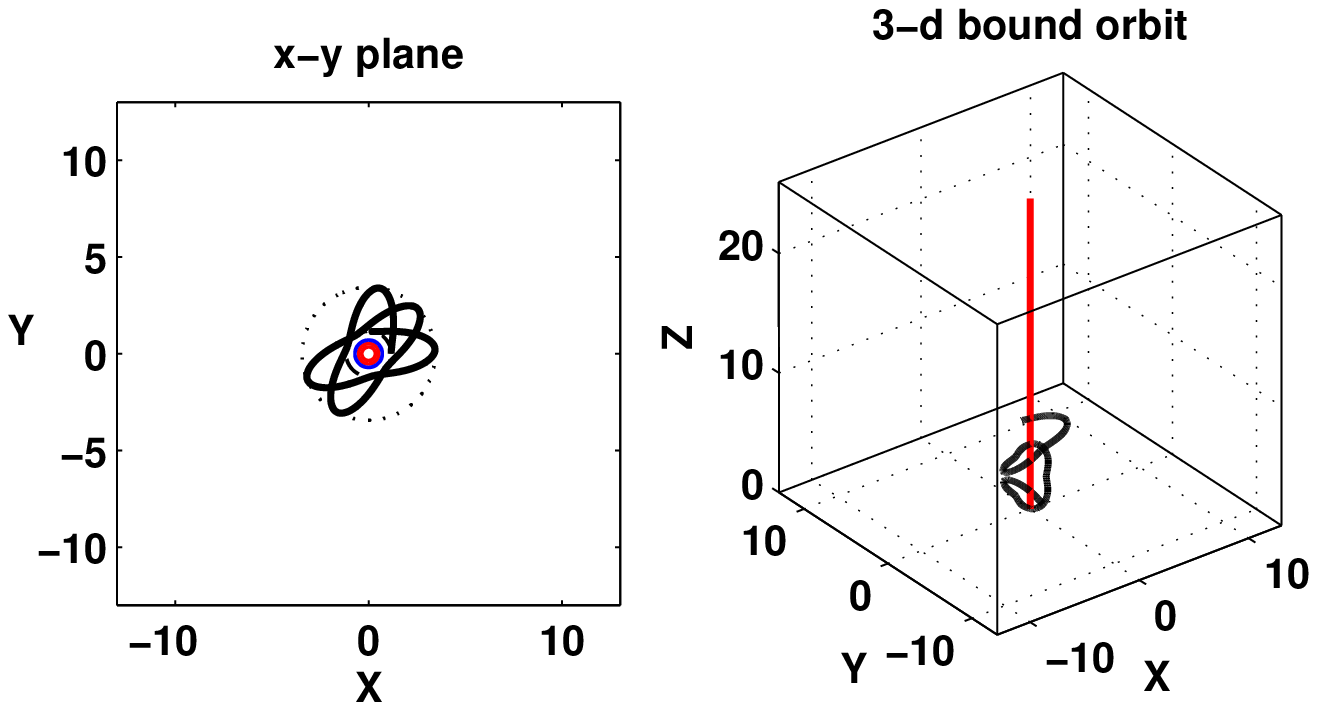}
}

\caption[Optional caption for list of figures]{
The bound orbit of a massive
test particle with $E=0.9931$, $L_z= 0.01$ and $p_z = 0.015$ in the space-time of a (p,q)-string
with $\gamma=0.2$, $\beta_1 = 8$, $\beta_2 = 0.5$, $\beta_3=0.99$ and different choices of (p,q)$\equiv (n,m)$. 
The dotted circles denote the minimal and the maximal radius of the bound orbit, while
the blue and the red circles indicate the radius of the gauge and scalar field cores, respectively.
}\label{orbitMsvPQ}
\end{figure}

We have also studied how the orbits change when changing the winding numbers (p,q)$=(n,m)$ and hence the magnetic fluxes.
This is shown in Fig. \ref{orbitMsvPQ} for the bound orbit of 
a massive test particle
with $E=0.9931$, $L_z= 0.01$ and $p_z = 0.015$ in the space-time of a (p,q)-string 
with $\gamma=0.2$, $\beta_1 = 8$, $\beta_2 = 0.5$, $\beta_3=0.99$. While for
(p,q)$=(1,1)$ and (p,q)$=(1,2)$ the test particle moves close to the core of the string, it can extend considerably into the
vacuum region for (p,q)$=(2,1)$. Apparently, the change of the winding 
of the q-string which has $\rho_{{\rm H},2} > \rho_{{\rm W},2}$ mainly
influences the perihelion shift of the orbit, which increases with increasing winding. On the other hand the increase
of the winding of the p-string which has $\rho_{{\rm W},1} > \rho_{{\rm H},2}$ allows the test particle to move further away from the string core.  

\begin{figure}[htp]
\centering

\subfigure[$p$ = 1, $q$ = 1]{
\includegraphics[scale=0.8]{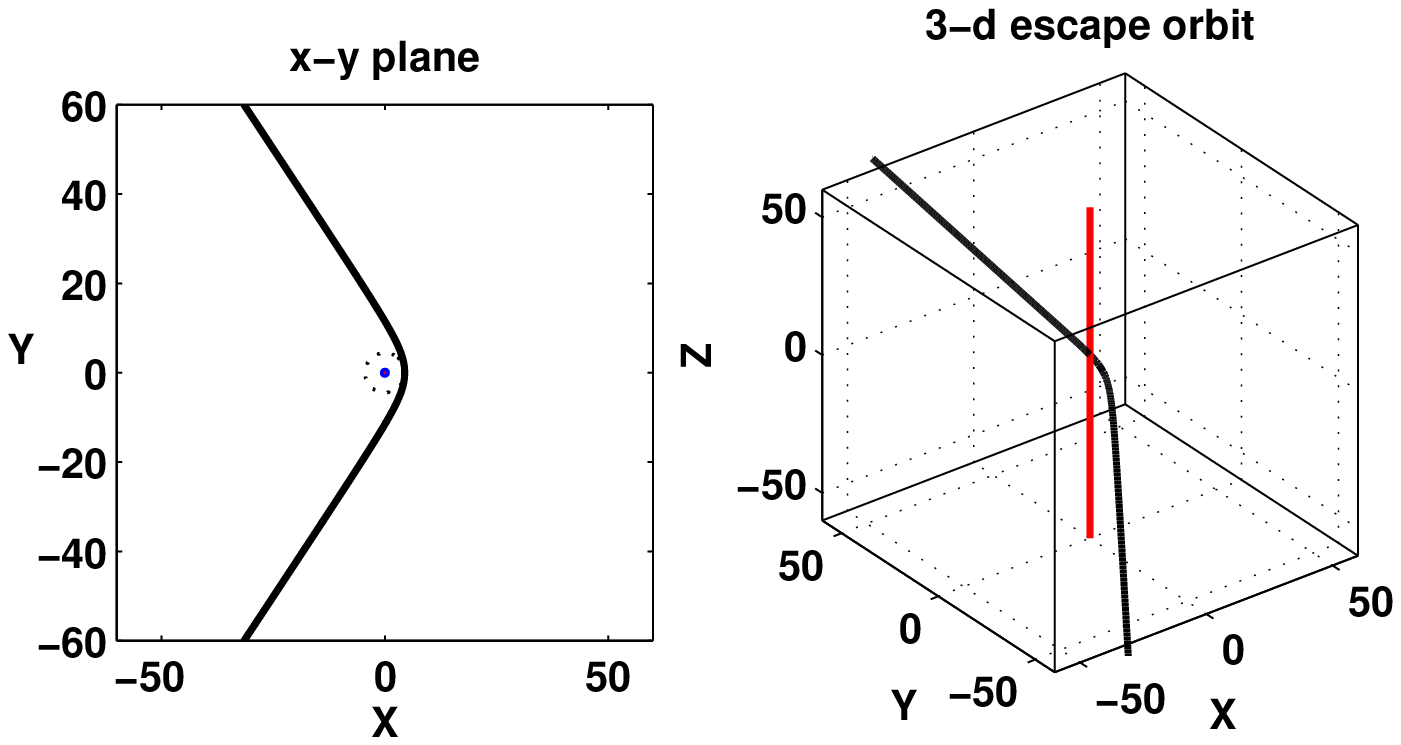}
}

\subfigure[$p$ = 1, $q$ = 3]{\label{orbitMlvPQ(b)}
\includegraphics[scale=0.8]{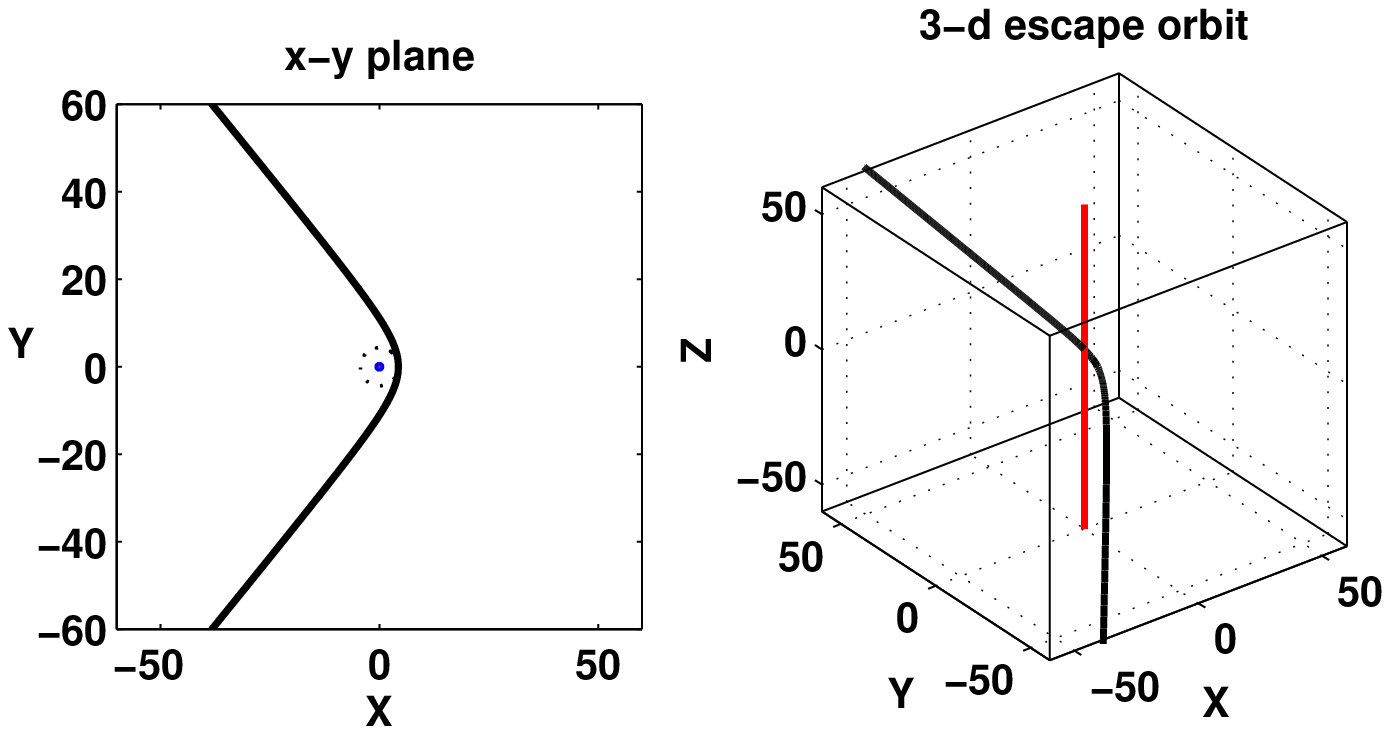}
}

\subfigure[$p$ = 3, $q$ = 1]{\label{orbitMlvPQ(c)}
\includegraphics[scale=0.8]{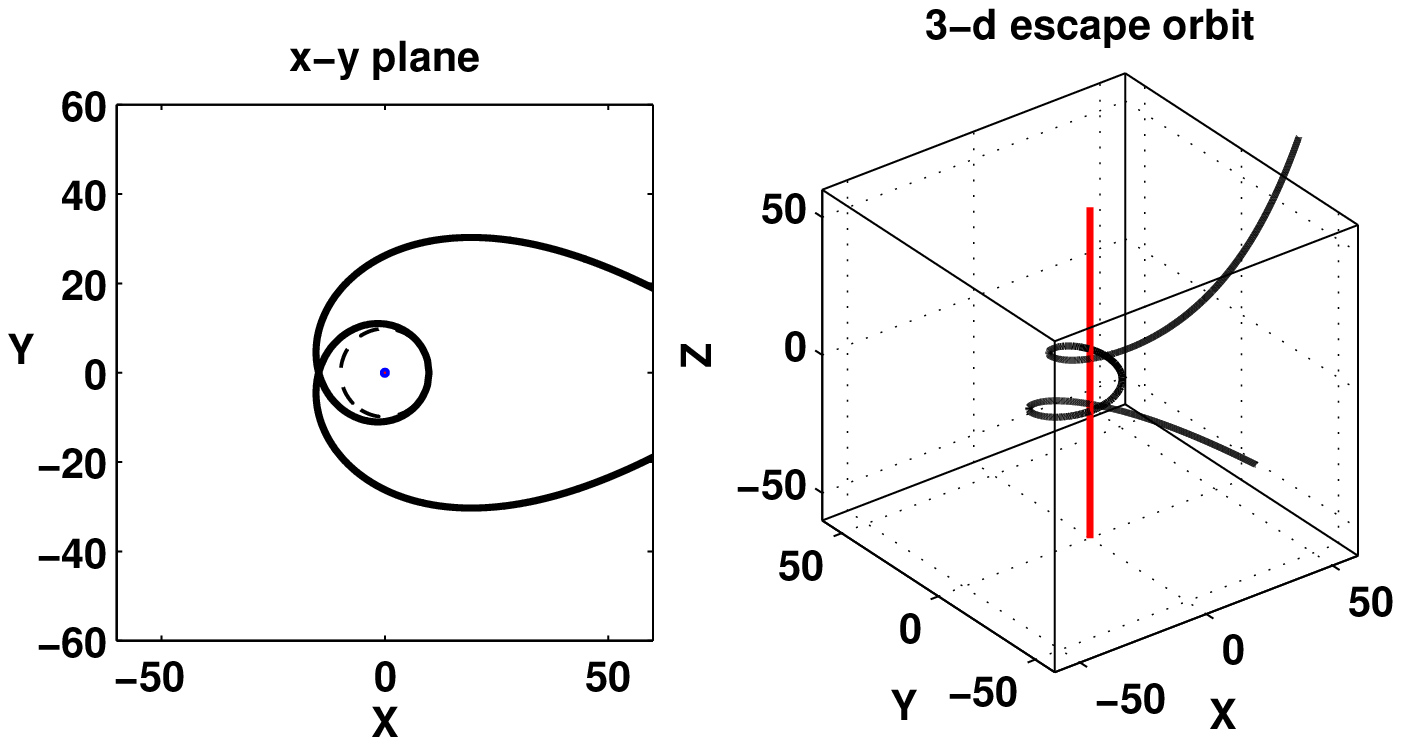}
}

\caption[Optional caption for list of figures]{
The escape orbit of a massless
test particle with $L_z = 0.2236$, $E = 0.08$ and $p_z = 0.05$ in the space-time of a (p,q)-string
with $\gamma=0.2$, $\beta_1 = 8$, $\beta_2 = 0.5$, $\beta_3=0.99$ and different choices of (p,q)$\equiv (n,m)$. 
The blue dot denotes the core of the string, while the dotted line denotes the circle with minimal
radius of the orbit, i.e. closest approach of the particle to the string.
}\label{orbitMlvPQ}
\end{figure}

The change of an escape orbit of a massless test particle with the change of the windings is shown in 
Fig.\ref{orbitMlvPQ}. While for
(p,q)$=(1,1)$ and (p,q)$=(1,3)$ the test particle gets simply deflected by the string it encircles the string
before escaping to infinity for (p,q)$=(3,1)$. Apparently, the change of the winding of the q-string which has 
$\rho_{{\rm H},2} > \rho_{{\rm W},2}$ influences the deflection only slightly. On the other hand the increase
of the winding of the p-string which has $\rho_{{\rm W},1} > \rho_{{\rm H},2}$ leads to an encirclement of the string.

\subsection{Geodesic motion in Melvin space-times}
\begin{figure}[h!]
\begin{center}
\resizebox{4in}{!}{\includegraphics{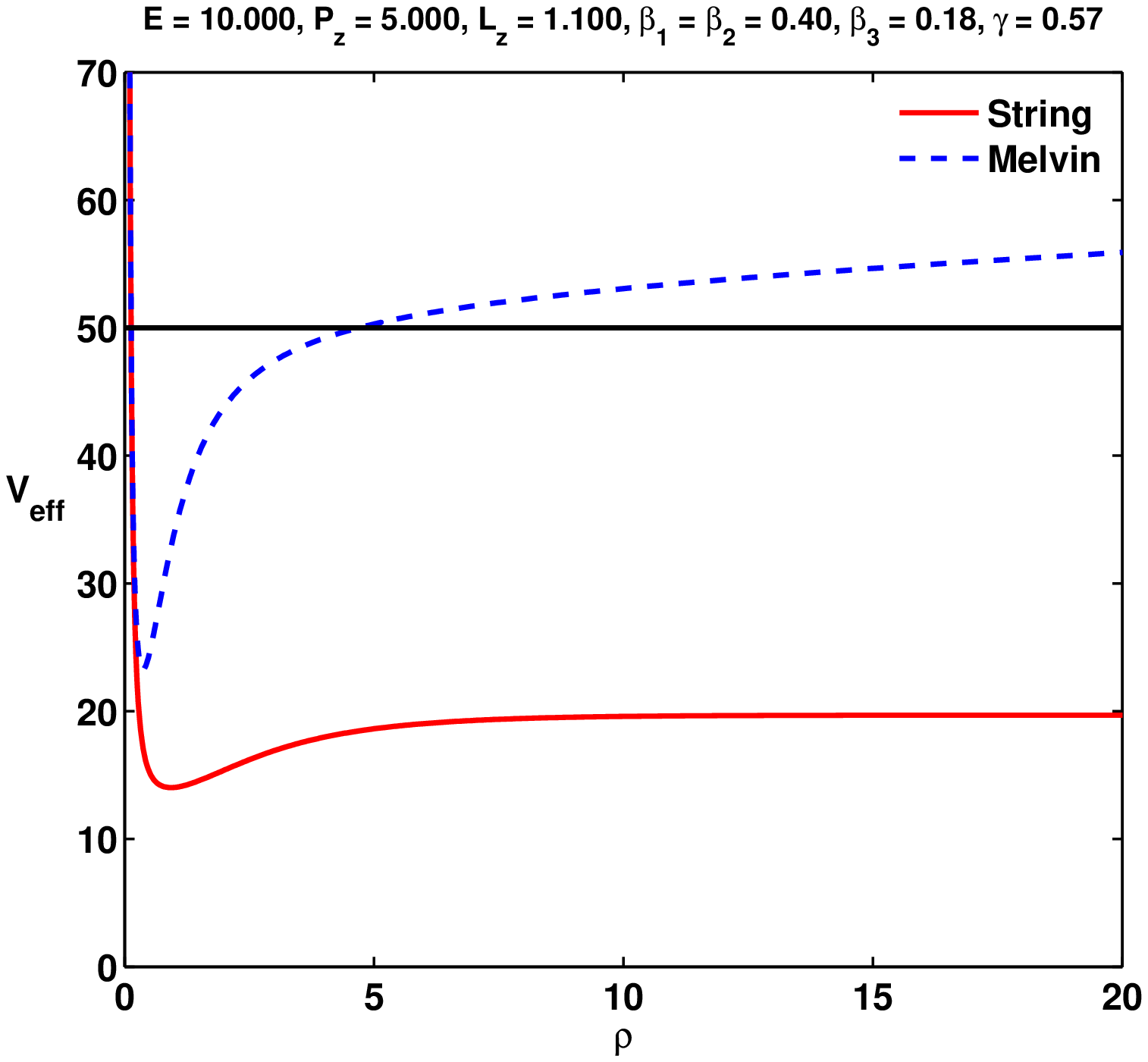}} 
\caption{The effective potential $V_{\rm eff}(\rho)$ is shown for a Melvin solution (dashed)
and in comparison for a string solution (solid) for $E=10$, $p_z=5$, $L_z=1.1$ and
coupling constants $\beta_1=\beta_2=0.4$, $\beta_3=0.18$ and $\gamma=0.57$. The horizontal
solid line denotes the value of $\mathcal{E}$.}
\label{melvin_potential}
\end{center}
\end{figure}

\begin{figure}[htp]
\centering
\includegraphics[scale=0.8]{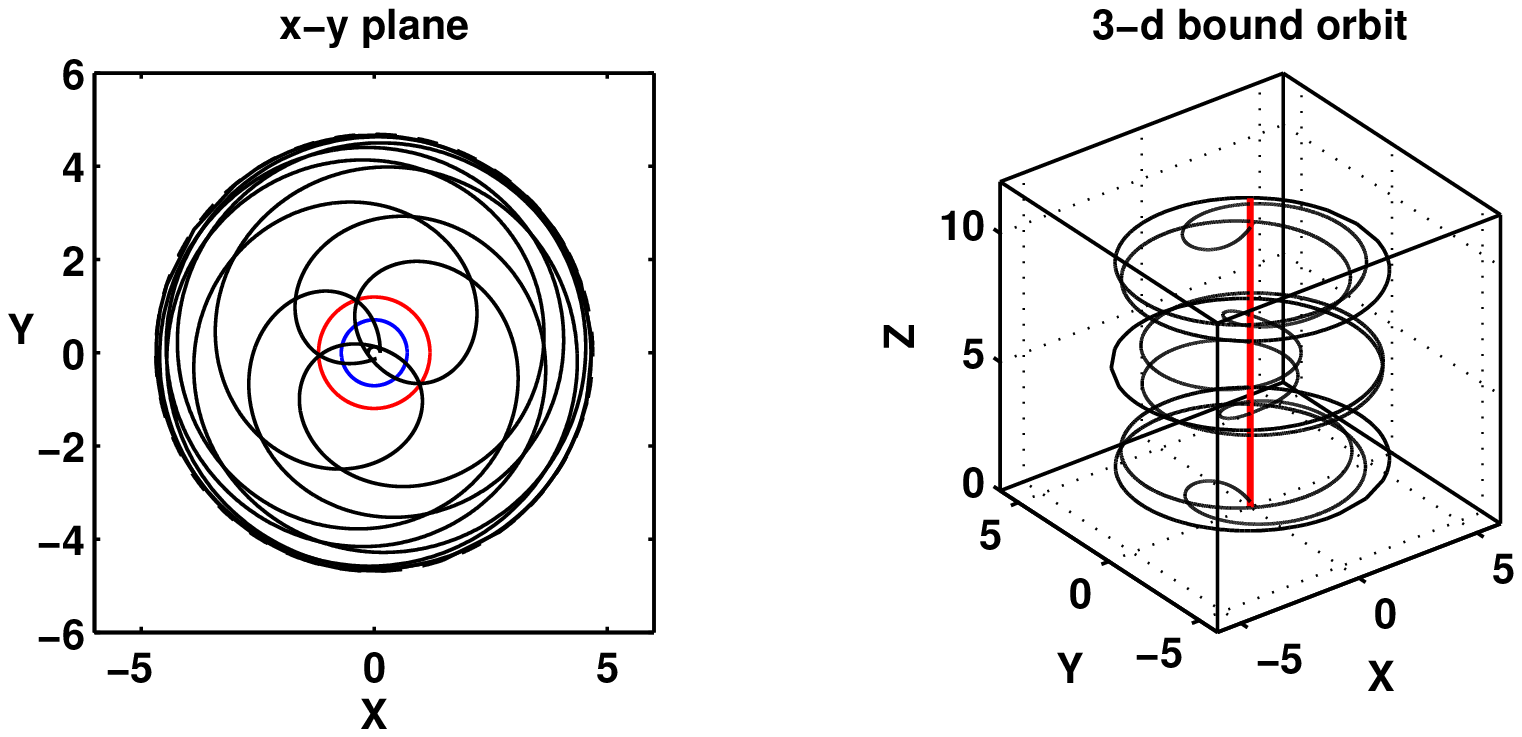}
\caption[Optional caption for list of figures]{The bound orbit of a massless test particle 
with $E=10$, $L_z=1.1$, $p_z=5$ 
moving in a Melvin space-time 
with $\beta_1=\beta_2=0.4$, $\beta_3=0.18$, $\gamma=0.57$. 
The blue and the red circle indicates the radius of the gauge and scalar field cores, respectively.}\label{orbit_melvin}
\end{figure}

In Fig.\ref{melvin_potential} we show the effective potential for a Melvin space-time with
$\beta_1=\beta_2=0.4$, $\beta_3=0.18$ and $\gamma=0.57$ and test particle
parameters $E=10$, $p_z=5$, $L_z=1.1$. In comparison, we also give the effective
potential of the corresponding string space-time. Close to the $z$-axis the effective potential
of the Melvin space-time is equivalent to that of a string space-time. In contrast to the
string space-time the effective potential in a Melvin space-time will always have a
local minimum (see also (\ref{dv0})) and 
$V_{\rm eff}(\rho\rightarrow \infty)\rightarrow \infty$ for $L_z\neq 0$. Hence there will be only bound orbits, while escape 
orbits do not exist. As already mentioned this is related to the fact that the space-time is not asymptotically
flat and particles can never reach infinity. 
This is interesting since in string space-times a massless test particle will 
always escape from the string,
while a massive test particle can move on a bound orbit provided both energy $E$ and angular momentum $L_z$ are not too large.
In Melvin space-times test particles will always move on bound orbits.
In Fig. \ref{orbit_melvin} we show the bound orbit for a massless test particle with $E=10$, $L_z=1.1$, $p_z=5$ 
moving in a Melvin space-time 
with $\beta_1=\beta_2=0.4$, $\beta_3=0.18$, $\gamma=0.57$.

\section{Observables}
Since we believe the string space-time to be the physically relevant case, we compute all observables in the
space-time with asymptotic behaviour (\ref{stringinfinity}).
\subsection{Perihelion shift}
The perihelion shift of a bound orbit of a massive test particle ($\varepsilon=1$) can be calculated by the following expression
\begin{eqnarray}
\delta\varphi_{\varepsilon=1}&=&2\int_{\rho_{\rm min}}^{\rho_{\rm max}}\frac{L_z d\rho}{L(\rho)^2\left(\frac{E^2-p_z^2}{N(\rho)^2}
-\frac{L_z^2}{L(\rho)^2}-1\right)^{1/2}}-2\pi\label{perihel}  \ ,
\end{eqnarray}
where $\rho_{\rm min}$ and $\rho_{\rm max}$ are the minimal and the maximal radius of the bound orbit, respectively.
The dependence of the perihelion shift $\delta\varphi_{\varepsilon=1}$ of a bound planar orbit of a massive test particle 
with $E = 1.01$, $L_z = 0.02$, $p_z=0$
on the binding parameter $\beta_3$  is shown 
in Fig.\ref{periplot}. 
For this particular case, the perihelion shift is negative which means 
that the test particle moves from the minimal to the maximal radius and returns back 
to the minimal radius 
under an angle less than $2\pi$. 
In asymptotically flat black hole space-times and even asymptotically flat space-times of black holes pierced
by infinitely thin cosmic strings (see \cite{hhls1,hhls2}) the perihelion shift is
positive. In a Schwarzschild--(Anti)--de Sitter black hole space-time the positive (negative) cosmological
constant gives a positive (negative) contribution to the perihelion shift. 
Since all observations point to a positive cosmological constant, we would expect that
the perihelion shift is positive for astrophysically relevant black hole solutions.
Note that in the space-time of an infinitely thin cosmic string alone
no bound orbits exist (see discussion above) and hence it makes no sense to 
calculate the perihelion shift. On the other hand, the presence of an infinitely
thin cosmic string in black hole space-times enhances the (positive) perihelion shift \cite{hhls1}. The fact that
the perihelion shift can become negative in the case of finite width cosmic strings 
is hence related to the fact that the space-time is that of a smoothed cone close to the string
axis.
In fact, the absolute value of the perihelion shift increases with increasing $\beta_3$, i.e. for increasing $\beta_3$ 
the change of the $\varphi$ coordinate from the first to the second minimal radius decreases. 
This can be understood when considering the influence of 
$\beta_3$ on the effective potential. In fact, the potential becomes steeper when increasing
$\beta_3$ and hence the difference between the
minimal and the maximal radius for fixed values of $E$, $L_z$ and $p_z$ decreases.
Moreover, we observe that the perihelion shift for a (p,q)-string with $M_{{\rm H},i} > M_{{\rm W},i}$, $i=1,2$ has the largest 
negative value, while a (p,q)-string with $M_{{\rm H},i} < M_{{\rm W},i}$, $i=1,2$ has the smallest negative value.

We have also studied whether the perihelion shift is always negative and find that it becomes positive
for cosmic string space-times with coupling constants chosen such that the deficit angle is close to $2\pi$ 
and the values of $E$ and $L_z$ are large, i.e. close to the boundary of the $\mu$-$\nu$-domain in which bound orbits
exist (see Fig.s \ref{munuplot}-\ref{munuvpq}). In this case, the difference between $\rho_{\rm min}$ and $\rho_{\rm max}$ is
quite large and the test particle shows mainly in the vacuum region outside the string. We find e.g. for a p-q-string space-time with $\beta_1=\beta_2=4$, $\beta_3=1.02$, $\gamma = 0.499963$ 
and resulting deficit angle $\delta/(2\pi)\approx 0.9943$ that the perihelion shift of a bound orbit of a massive test particle
with $E=1.0018$, $L_z=0.007$ and $p_z=0$ is positive and has value $\delta\varphi_{\varepsilon=1}\approx 2.3695$ rad.

\subsection{Light deflection}
Is it very important for gravitational lensing to understand how massless test particles move on escape orbits.

The deflection of light (massless test particle, i.e. $\varepsilon=0$) by a (p,q) string can be calculated by the following equation
\begin{eqnarray}
\delta\varphi_{\varepsilon=0}&=&\int_{\rho_{\rm min}}^{\infty}\frac{L_zd\rho}{L(\rho)^2\left(\frac{E^2-p_z^2}{N(\rho)^2}-
\frac{L_z^2}{L(\rho)^2}\right)^{1/2}}-\pi
\end{eqnarray}
where $\rho_{\rm min}$ is the minimal radius of the escape orbit of the massless test particle. 

\begin{figure}[h!]
  \begin{center}
    \subfigure[$E = 1.01$,  $L_z = 0.02$, $p_z = 0$, $\varepsilon = 1$]{\label{periplot}\includegraphics[scale=0.45]{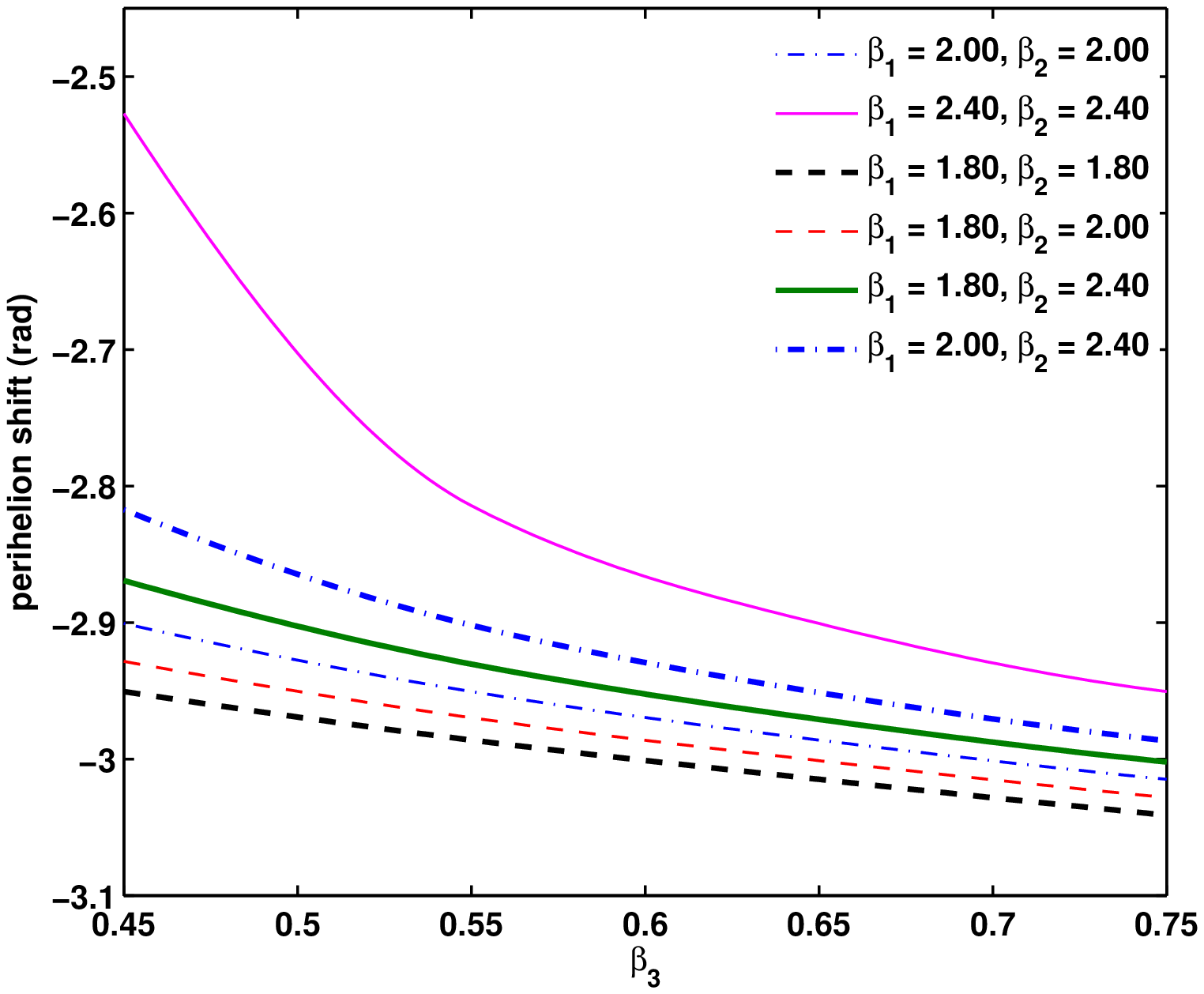}}
    \subfigure[$E = 1.04$,  $L_z = 0.28$, $p_z = 0$, $\varepsilon = 0$]{\label{lightdefplot}\includegraphics[scale=0.45]{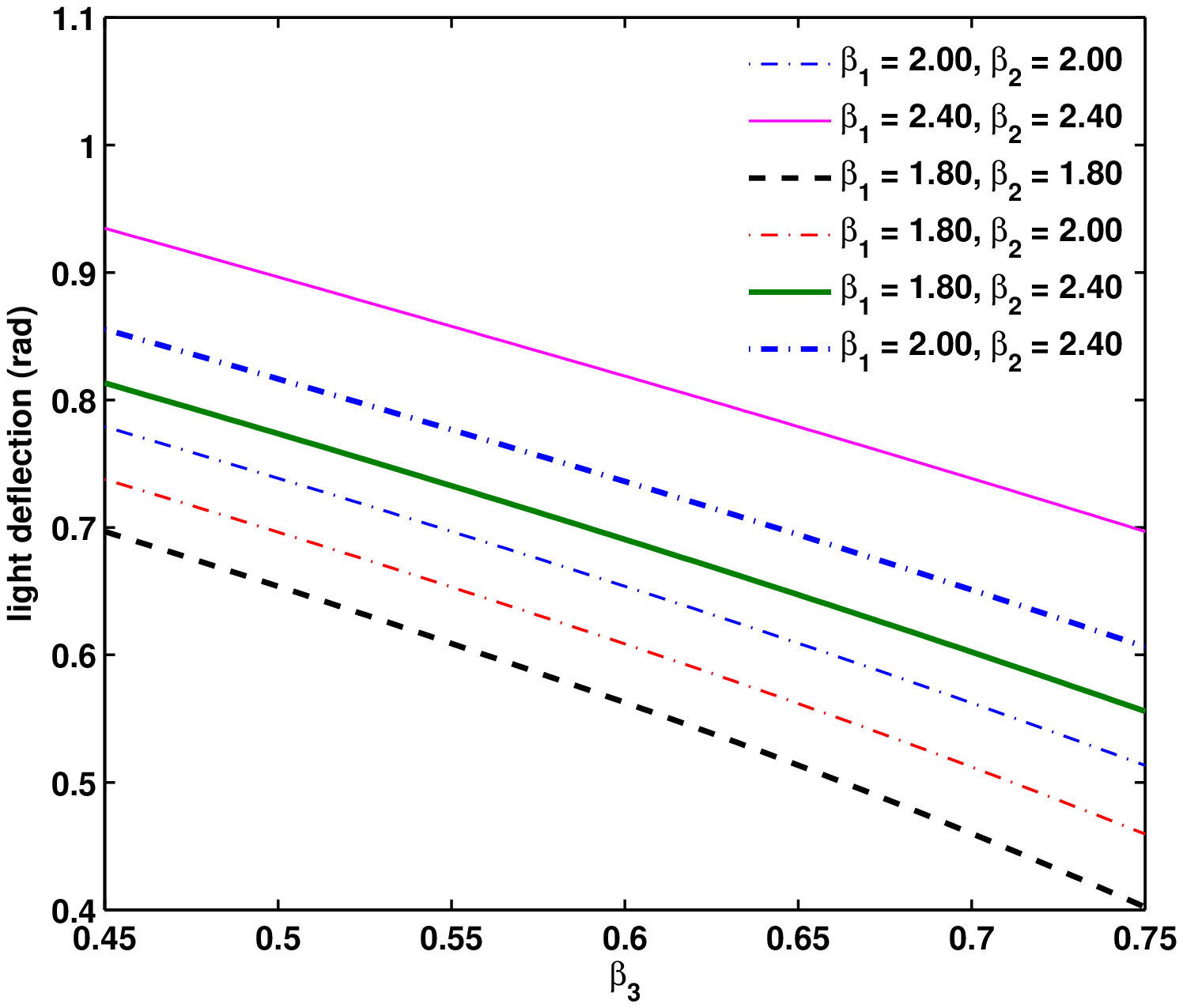}}
   \end{center}
  \caption{The dependence of the perihelion shift of a planar bound orbit of a massive test particle with $E = 1.01$, $L_z = 0.02$ on $\beta_3$ and
(left) and the dependence of the deflection angle of a planar escape orbit of a massless test particle 
with  $E = 1.04$, $L_z = 0.28$ on $\beta_3$ (right). }\label{pqphshiftlightdef}
  \end{figure}

The dependence of the deflection $\delta\varphi_{\varepsilon=0}$ of a planar escape orbit of a massless test particle with
$E = 1.04$, $L_z = 0.28$, $p_z=0$ on $\beta_3$ is shown in Fig.\ref{lightdefplot}.  
The light deflection decreases  
when increasing $\beta_3$. This is not surprising since the energy per unit length of the (p,q)-string and with it
the deficit angle decrease with increasing $\beta_3$. 

Moreover, we observe that the light deflection for a (p,q)-string with $M_{{\rm H},i} > M_{{\rm W},i}$, $i=1,2$ has the smallest
value, while a (p,q)-string with $M_{{\rm H},i} < M_{{\rm W},i}$, $i=1,2$ has the largest value. This is related to the
fact that if the scalar (gauge) field cores dominate strings tend to attract (repel) each other, hence lowering (increasing) the total
energy per unit length as compared to the BPS limit. This leads to a decrease (increase) of the deficit angle.  

\section{Conclusions}
In this paper we have studied test particle motion in the space-time of a cosmic superstring that consists of
p D-strings and q F-strings. We have studied the asymptotically conical 
string space-time as well as the Melvin space-time that has vanishing circumference of a circle
at infinity. We observe that the binding between the strings has important effects
on the motion of test particles in string space-times. In the $\beta_3=0$ limit which corresponds to the space-time
of two non-interacting Abelian-Higgs strings and is qualitatively similar to that studied in \cite{hartmann_sirimachan}
massive test particles can only move on bound orbits if the scalar core width of the string is larger
than that of the gauge field core. For $\beta_3 >0$ massive test particles can now
move on bound orbits if the scalar core width is smaller than the gauge field core width and need
less energy than in the $\beta_3=0$ limit to do so. The perihelion shift can become negative due to 
the smoothed conical nature of the
space-time close to the string axis and the absolute value of the perihelion shift increases with increasing $\beta_3$. 
The fact that the perihelion shift can become negative seems to be a characteristic of
the space-time of a finite width cosmic string that -- to our knowledge -- has not be noticed in any other astrophysically 
relevant space--time yet.
Massless particles can only move on escape orbits and the deflection by the string decreases with increasing
binding between the p- and the q-string.
In Melvin space-times, on the other hand, massless and massive particles cannot escape to infinity and must move
on bound orbits.

The deflection of light by cosmic strings
should be detectable. Though the identification of cosmic strings due to their gravitational
lensing effects has been discussed extensively \cite{csl1} no such cosmic string lens has been detected
to date. Moreover, there might also be other sources of gravitational lensing and when
identifying cosmic strings through gravitational lensing it has to be made sure that no 
``standard'' matter distributions are the source of the lensing. On the other
hand, the negative perihelion shift seems to be generic to finite width cosmic string
space-times. To state it differently: if a negative perihelion shift would be observed this
would be a strong evidence for the existence of cosmic strings.
The main characteristic of cosmic superstrings is that they can form bound states and our field theoretical 
solutions describe such bound states. The fact that bound states can form so effectively 
alters the set of possible orbits 
considerably in comparison to standard field theoretical cosmic string models. 
In particular, the mass ratios $\gamma$ and $\beta_i$ have an important impact.
E.g. the perihelion shift can become positive or negative depending on the choice of these parameters
and its absolute value can vary considerably. 
\\ 
\\

{\bf Acknowledgments}
The work of PS was supported by DFG grant HA-4426/5-1.


\begin{thebibliography}{34}
\bibitem{remark1} Note that bound orbits exist when
combining the space-time of an infinitely thin cosmic string with that of black hole solutions, see e.g.
\cite{hhls1,hhls2}.
\end{thebibliography}
\end{document}